\documentclass[a4paper,fleqn]{cas-sc}

\usepackage[authoryear]{natbib}

\usepackage{amssymb}
\usepackage{amsthm}
\usepackage{amsmath}
\usepackage{color}
\usepackage{multirow}
\usepackage{graphicx}
\usepackage{subcaption}
\usepackage{lineno}
\usepackage{nomencl}
\makenomenclature
\usepackage{hyperref}

\def\tsc#1{\csdef{#1}{\textsc{\lowercase{#1}}\xspace}}
\tsc{WGM}
\tsc{QE}

\begin{document}
\let\WriteBookmarks\relax
\def\floatpagepagefraction{1}
\def\textpagefraction{.001}

\shorttitle{On applicability of von Karman’s momentum theory in predicting the water entry load}    

\shortauthors{Yujin Lu and Alessandro Del Buono}  

\title [mode = title]{On applicability of von Karman’s momentum theory in predicting the water entry load of V-shaped structures with varying initial velocity}  



%

\author[inst1,inst2]{Yujin Lu}[type=author]





\credit{Conceptualization, Methodology, Software, Investigation, Data Curation, Visualization, Writing - Original Draft}


\affiliation[inst1]{organization={Nanjing University of Aeronautics and Astronautics},
            addressline={Yudao Street 29}, 
            city={Nanjing},
            postcode={210016}, 
            state={Jiangsu},
            country={People’s Republic of China}}
\affiliation[inst2]{organization={National Research Council–Institute of Marine Engineering (CNR-INM)},
            addressline={Via di Vallerano 139}, 
            city={Roma},
            postcode={00128}, 
            state={Lazio},
            country={Italy}}

\author[inst2]{Alessandro Del Buono}[type=author]
\cormark[1]

\ead{alessandro.delbuono@inm.cnr.it}


\credit{Validation, Investigation, Formal analysis, Writing - Review \& Editing}

\author[inst1]{Tianhang Xiao}[type=author]
\cormark[2]
\ead{xthang@nuaa.edu.cn}
\credit{Conceptualization, Writing - Review \& Editing, Supervision}

\author[inst2]{Alessandro Iafrati}[type=author]
\credit{Project administration, Writing - Review \& Editing, Supervision}

\author[inst1]{Shuanghou Deng}[type=author]
\credit{Funding acquisition, Writing - Review \& Editing, Supervision}

\author[inst1]{Jinfa Xu}[type=author]
\credit{Resources, Supervision}

\cortext[1]{Corresponding author}
\cortext[2]{Corresponding author}



\begin{abstract}
The water landing of an amphibious aircraft is a complicated problem that can lead to uncomfortable riding situation and structural damage due to large vertical accelerations and the consequent dynamic responses. The problem herein is investigated by solving unsteady incompressible Reynolds-averaged Navier-Stokes equations with a standard $k-\omega$ turbulence closure model. The theoretical solutions established by the \textcolor{black}{von Karman's} momentum theory are also employed. In order to validate the relationships between the initial vertical velocity and the peak value of vertical acceleration, free fall test cases of 2D symmetric wedge oblique entry and 3D cabin section vertical entry are presented first. The other parameters at which the maximum acceleration occurs, such as time, penetration depth, velocity, are also evaluated. Hence, the quantitative relations are investigated to water landing event for amphibious aircraft. Detailed results in terms of free surface shape and pressure distribution are provided to show the slamming effects. The results show that a linear dependence of the maximal acceleration from the square of initial vertical velocity can be derived for two-dimensional wedge, three-dimensional cabin section and seaplane with V-shaped hull. Moreover, the ratio between the corresponding velocity and the initial vertical velocity tends to a constant threshold value, 5/6, derived from the theoretical solution, when increasing the initial vertical velocity in all three cases.
\\

\noindent
© 2022. This manuscript version is made available under the CC-BY-NC-ND 4.0 license
https://creativecommons.org/licenses/by-nc-nd/4.0/
\end{abstract}


\begin{highlights}
\item The maximal acceleration is proportional to the square of the initial velocity for the V-shaped body
\item The theoretical ratio of the corresponding velocity to the initial velocity is valid for large impact velocity
\item Gravity effect should be considered with slow impact speed
\item A coupled relation among $a_{z\mathrm{max}}$,$\upsilon_z^*$ and $z^*$ is found
\end{highlights}

\maketitle

\nomenclature{$a_x, a_z$}{non-dimensional acceleration in $x$- and $z$-direction}%
\nomenclature{$a_\mathrm{max}$}{non-dimensional maximal acceleration}%
\nomenclature{$b$}{intercept}%
\nomenclature{$C_{p}$}{pressure coefficient}%
\nomenclature{$F_\mathrm{hd}^{*}, F_\mathrm{hs}^{*}$}{maximal vertical hydrodynamic and hydrostatic force, N}%
\nomenclature{$k$}{slope}%
\nomenclature{$\boldsymbol L$}{tensor of the moments of inertia, kg$\cdot\mathrm{m}^2$}%
\nomenclature{$L$}{length of the cabin and the fuselage, m}%
\nomenclature{$\boldsymbol M$}{resultant moment acting on the object, N$\cdot\mathrm{m}$}%
\nomenclature{$M$}{mass, kg}%
\nomenclature{$m_\mathrm{added}$}{added mass, kg}%
\nomenclature{$\upsilon, \upsilon_0$}{velocity and the initial velocity, m/s}%
\nomenclature{$V,V_\mathrm{w}$ }{the volume of the cell and the volume of water in the cell, $\mathrm{m}^3$}%
\nomenclature{$W$}{width, m}%
\nomenclature{$x_{shift}$}{shifted coordinate in x-axis, m}%
\nomenclature{$z$}{penetration depth, m}%
\nomenclature{$\alpha_\mathrm{w}$}{volume fraction of water}%
\nomenclature{$\alpha$}{velocity angle, $^\circ$}%
\nomenclature{$\beta$}{deadrise angle, $^\circ$}%
\nomenclature{$\theta$}{the heel angle, $^\circ$}%
\nomenclature{$\kappa$}{the ratio of the corresponding velocity to the initial velocity}%
\nomenclature{$\boldsymbol \omega$}{angular velocity of the object, rad/s}%
\nomenclature{$\zeta$}{resultant displacement, m}%
\nomenclature{w, a}{water and air}%
\nomenclature{aero}{aerodynamic}%
\printnomenclature

\linenumbers

\section{Introduction}
\label{sec:introduction}
Amphibious aircraft is a special flight vehicle that is capable of taking off and landing both on water and conventional runways \citep{qiu2013efficient}. The amphibious aircrafts have drawn considerable attentions by the nations with maritime supremacy due to their potential military and civilian applications. In the flight operational envelope of amphibious aircraft, landing on water is regarded as the most dangerous phase where the hydrodynamic impact load significantly influences the occupants survivability and structural integrity \citep{hughes2013from}. In terms of the design and analysis of water entry load, full scale tests are regarded as the most straightforward and reliable way. Investigating the hydrodynamics of the water landing of an amphibious aircraft with full scale test are highly expensive and time demanding and, may be challenged by a low repeatability level. In order to derive reliable estimates of the hydrodynamic loads acting on the aircraft during water landing, another practicable way is to perform scaled-model experiments in water basins. As an example, experimental studies on the water entry problems have been conducted at  NACA Langley Memorial Aeronautical Laboratory, resulting in extensive and valuable archived test data and recommendations in industrial applications \citep{benson1945bibliography}. The study provides interesting information about the effects on performances of design parameters such as deadrise angle, depth of step, configuration of hull body, hydrofoils, etc. In general, hydrodynamics of water impacting can be demonstrated commendably by scaled model water tank tests.  In the case of seaplanes, both hydrodynamics and aerodynamic aspects play the same key roles in the dynamic behavior and there is however a difficulty in achieving the correct scaling for the air and water domains \citep{duan2019numerical}. Froude ($Fr$) scaling guarantees the correct reproduction of the ratio between the inertia and gravity force in the water domain but it do not allow to preserve the  Reynolds ($Re$) similarity and thus the correct scaling of the viscous effects which are important in both water and, especially, in air for the aerodynamic lift and drag \citep{terziev2022scale,iafrati2019cavitation}. Depending on the full-scale speed, other phenomena like cavitation and ventilation might be also relevant in the water domain that would not be properly reproduced in scaled model tests based on Froude similarity only \citep{iafrati2019cavitation}.

As an alternative to expensive experimental campaign, the recent developed computational approaches allow to simulate the hydro- and aero- dynamics and kinematic motion of amphibious aircraft in full scale. Different phases during the whole process, such as takeoff/landing, skiing, and other serious situations were investigated recently by numerical simulation. For the takeoff process, \citep{qiu2013efficient} proposed a decoupled algorithm to investigate the kinematic characteristics, whereby the aerodynamic forces of the full configuration and the hydrodynamic forces of the hull body were computed separately. The whole process was divided into a number of small time-step, and the forces were calculated at each time step. In \citep{duan2019numerical} evaluated the porpoising motion, an unstable oscillation phenomenon that threatens the flying safety of amphibious aircrafts, by using a two-phase flow solver in OpenFOAM. Both slipstream caused by the propeller and external forces, viz. thrust and elevator forces, were taken into consideration as well. Results highlighted the important role played by the hydrodynamic force on the heaving and pitching oscillations, while the aerodynamic forces have a rather marginal effect. Similar to the water landing scenarios of amphibian aircraft, ditching events of conventional aircrafts show the same fluid dynamics phenomena, and have been numerically studied widely. The effects of initial pitching angle and velocity \citep{xiao2021effect,guo2013effect,qu2016numerical,zheng2021numerical}, fluid-structure interaction \citep{hughes2013from,siemann2017advances,yang2020crashworthy}, wave conditions \citep{woodgate2019simulation,xiao2021hydrodynamic} and various numerical strategies \citep{bisagni2017modelling,siemann2017coupled,xiao2017development} on the kinematic characteristics and fluid dynamics phenomena have attracted most of the attention. The vertical acceleration, and its peak value in particular, is even more relevant than other kinematic characteristics as it may be responsible for possible comfort and safety problems occur on crew members, besides, of course, the effects in terms of structural integrity of the fuselage once it strikes the water \citep{neuberg2017fatigue}.

The ditching event, it is usually distinguished in four phases: approach, impact, landing, and flotation \citep{siemann2017advances}. The impact phase is the most important one in terms of complex fluid-structure interaction. Von Karman \citep{karman1929impact} first proposed an analytical estimation method based on a wedge-shaped water impact and introduced the method to settle the impact loads on seaplanes. Subsequently, a number of researches related to water impact have been carried out based on theoretical, computational or experimental approaches \citep{wagner1932phenomena,zhao1993water,scolan2001three,korobkin2004analytical,korobkin2006three,wu2014similarity,breton2020experimental,zekri2021gravity}. It has been shown that, in the case of free-fall, the structure experiences a rapid change of vertical acceleration and velocity, which is similar to what happens in the impact phase of the water landing \citep{wang2015experimental}. Several studies have focused on the relationship between the maximum acceleration and initial parameters on free-fall water entry. Among these studies, \citep{gong2009water} simulated a series of cases with various initial entering velocity of the wedge through a Smoothed Particle Hydrodynamics (SPH) model, and relations for the maximum force on the wedge and the corresponding time in terms of the initial entering velocity of the wedge have been directly expressed by fitting formulas for Froude number greater than 2. In the work of \citep{abraham2014modeling}, the drag-coefficient of a sphere impacting the water surface was found to be independent of some investigated quantities, like the sphere velocity, surface tension, flow regime (laminar or turbulent) and Reynolds number. Hence, algebraic expressions of the drag coefficient versus the dimensionless depth have been established by two fitted polynomials. Effects of parametric variation, such as impact velocity, radius, and mass of the sphere on the impact force and the acceleration, have also been analyzed by \citep{yu2019parametric}. The peak value of the non-dimensional impact force has been found to be independent of the velocity and the radius, whereas it depends on the mass of sphere. In parallel, simplified expressions for the maximal force and acceleration have been obtained through fitting the relations between the peak value of the non-dimensional force and the non-dimensional mass. The relationships derived in \citep{yu2019parametric} have also been mentioned by other researchers’ work \citep{iafrati2019cavitation,iafrati2016experimental,wen2020impact,wang2021cfd,sheng2022acfd}. However, it is worth noting that only fitting functions of force and acceleration were discussed in the previous studies, whereas the detailed theoretical basis with related relationships have not been derived yet.

The present study is dedicated to numerical simulations of a two-dimensional symmetric wedge and a three-dimensional cabin section in free fall water entry in order to investigate and build up parametric relations, based on the transformation of \textcolor{black}{the von Karman's} momentum theory, that can provide the maximal vertical acceleration and the corresponding vertical velocity, penetration depth and time. Particular attention is paid at the effects of horizontal velocity, and three-dimensional flow. The relations are then used to predict the load acting on amphibious aircraft during the water landing. The present work is organized as follows. Section \ref{sec:method} presents the methodology for the theoretical and numerical approaches, and describes the models and the computational setup; the main results are reported and discussed in Sec.~\ref{sec:results}; final conclusions are drawn in Sec.~\ref{sec:conclusion}.

\section{Methodology and Computational Setup}
\label{sec:method}
\subsection{Von Karman's theoretical method and transformation}

\begin{figure}[hbt!]
\centering
\includegraphics[width=.4\textwidth]{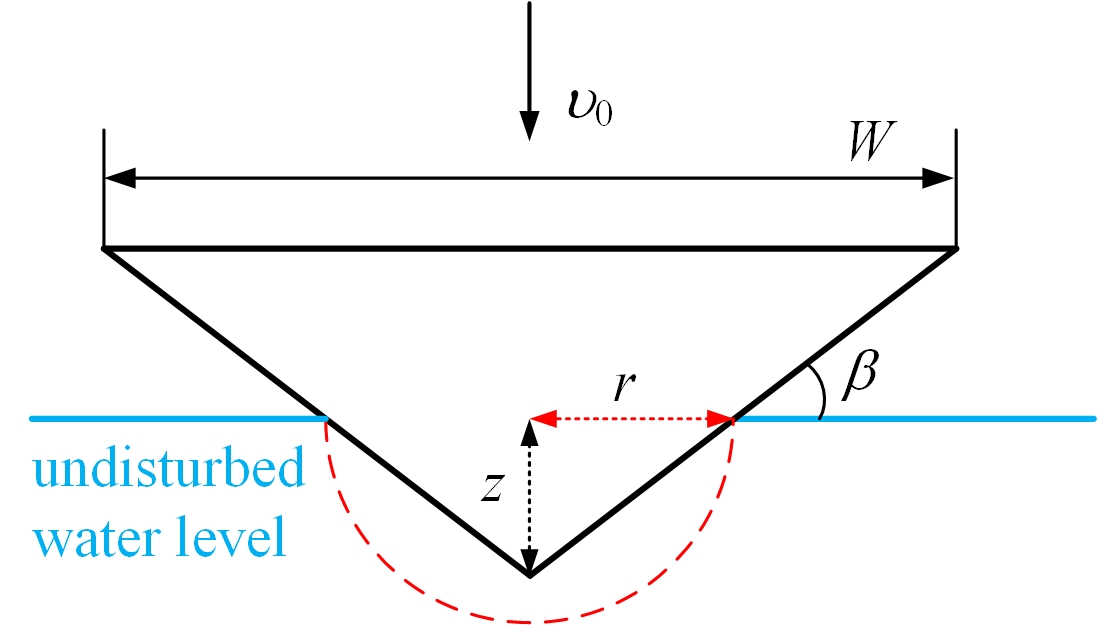}
\caption{Von Karman’s momentum approach.}
\label{fig:figVon}
\end{figure}

Pioneer research in water entry problem has been conducted by von Karman \citep{karman1929impact}, based on momentum theorem and the added mass for the prediction of the hydrodynamic load during the water entry of a V-shaped body penetrating into the water. By applying the momentum conservation at the beginning of the impact and the generic time $t$, it is obtained,
\begin{equation}
\label{eq:momentum}
M\upsilon_0=(M+m_\mathrm{added})\cdot\upsilon(t)
\end{equation}
where $M$ is the mass of the wedge per unit length, $\upsilon_0$ is the initial vertical impact velocity, $\upsilon(t)$ is the instantaneous velocity during the impact. In equation \eqref{eq:momentum}, $m_\mathrm{added}$ is the added mass which is computed by using the flat-plate approximation (see Fig.~\ref{fig:figVon}). It is assumed that the added mass is equal to the mass of a half disk of water of radius $r(t)$, which results into $m_\mathrm{added}=(\pi\rho r^2(t))/2$ \citep{mei1999on}. In such approximation the effect of the water pile-up is ignored.

With such an assumption, the velocity of the body can be retrieved as:
\begin{equation}
\label{eq:upsilon}
\upsilon(t)=\dfrac{M\upsilon_0}{M+m_\mathrm{added}}
=\dfrac{M\upsilon_0}{M+\dfrac{\pi\rho z^2(t)}{2\tan^2(\beta)}}
=\dfrac{2M\tan^2(\beta)\upsilon_0}{2M\tan^2(\beta)+\pi\rho z^2(t)}
\end{equation}

Based on what is provided in the \ref{sec:appendix} and  differentiating Eq.~\eqref{eq:upsilon}, it is possible to analytically derive the instantaneous acceleration as follows \citep{panciroli2013dynamic}:
\begin{equation}
\label{eq:a}
a(t)=\dfrac{\pi\rho z(t)}{M \upsilon_0 \tan^2(\beta)} \cdot \upsilon^3(t)
\end{equation}
which takes a peak of magnitude:
\begin{equation}
\label{eq:amax}
a^*=\upsilon_0^2 \left( \dfrac{5}{6} \right)^3 \dfrac{1}{\tan(\beta)} \sqrt{\dfrac{2\pi\rho}{5M}}
\end{equation}
when the corresponding penetration depth and velocity are:
\begin{equation}
\label{eq:threepara}
\left\{
\begin{aligned}
    z^* &= \sqrt{\dfrac{2M}{5\pi\rho}} \tan(\beta)\\
    \upsilon^* &= \dfrac{5}{6} \upsilon_0
\end{aligned}
\right.
\end{equation}
It should be noticed that we define the positive direction of acceleration upwards, while the vertical velocity and penetration depth are positive downwards. Moreover, according to \citep{panciroli2013dynamic} and \citep{iafrati2000hydroelastic}, the corresponding time $t^*$ can be expressed as:
\begin{equation}
\label{eq:threepara_t}
    t^* = \dfrac{1}{\upsilon_0}\dfrac{16}{15} \sqrt{\dfrac{2M}{5\pi\rho}} \tan(\beta)
\end{equation}
Note that the superscript * indicates the values the different quantities take when the acceleration reaches its peak. It is interesting to notice that $a^*$, $\upsilon^*$ and $t^*$, are proportional to $\upsilon^2_0$, $\upsilon_0$ and $\upsilon_0^{-1}$ respectively, implying that the initial vertical velocity governs those parameters, except $z^*$.

\subsection{Numerical method}

\textcolor{black}{In order to numerically simulate the problem, the commercial package Star CCM+ is utilized herein as the two-phase flow solver.} In the present study the unsteady incompressible Reynolds-averaged Navier-Stokes equations with a standard $k-\omega$ two-equation turbulence model are solved by the finite volume method. The Semi-Implicit Pressure Linked Equations (SIMPLE) algorithm is employed to achieve an implicit coupling between pressure and velocity, and the gradient is reconstructed with the Green-Gauss Node Based method. The modified High Resolution Interface Capturing (HRIC) scheme is adopted for volume fraction transport. The convection terms, as well as diffusion terms, are turned into algebraic parameters using second-order upwind and second-order central methods, respectively. The unsteady terms are discretized in the time domain by applying a second-order implicit scheme.

Volume of fluid (VOF) scheme, originally proposed by Hirt and Nichols \citep{hirt1981volume}, is used in the present computational scheme to capture the water-air interface by introducing a variable, $\alpha_\mathrm{w}$, called the volume fraction of the water in the computational cell, which varies between 0 (air) and 1 (water) and is defined as:
\begin{equation}
\label{eq:alphaw}
\alpha_\mathrm{w}=V_\mathrm{w}/V,
\end{equation}
where $V_\mathrm{w}$ is the volume of water in the cell and $V$ is the volume of the cell. The volume fraction of the air in a cell can be computed as:
\begin{equation}
\label{eq:alphawa}
\alpha_\mathrm{a}=1-\alpha_\mathrm{w}.
\end{equation}
The effective value $\varphi_\mathrm{m}$ of any physical properties, such as density, viscosity, etc., of the mixture of water and air in the transport equations is determined by:
\begin{equation}
\label{eq:varphi}
\varphi_\mathrm{m}=\varphi_\mathrm{w}\alpha_\mathrm{w}+\varphi_\mathrm{a}(1-\alpha_\mathrm{w}).
\end{equation}

To accurately capture the dynamic behavior as well as the load characteristics of water landing process, the motion of the body in response to the fluid forces and moments at the surface is determined via a six degree-of-freedom (6DOF) model. The 6DOF model solves the equations for the rotation and translation of the center of mass of the object. The equation for the translation in the global inertial coordinate system is formulated as:
\begin{equation}
\label{eq:Ftranslation}
M \cdot \dfrac{\mathrm{d} \boldsymbol\upsilon}{\mathrm{d}t}=\boldsymbol{F},
\end{equation}
and the rotation of the object is solved in the body local coordinate system by:
\begin{equation}
\label{eq:Mrotation}
\boldsymbol{L} \dfrac{\mathrm{d} \boldsymbol\omega}{\mathrm{d}t}+ \boldsymbol\omega \times \boldsymbol{L} \boldsymbol\omega=\boldsymbol{M}.
\end{equation}

Subsequently, a dynamic mesh strategy \citep{xiao2021hydrodynamic}, which moves the entire mesh rigidly along with the object at each time step according to the solution of the 6DOF model, is employed to deal with the relative motion between the fluid and the rigid body with on single grid domain. As neither mesh distortion nor mesh reconstruction occurs, the high quality of the initial mesh remains unchanged during the whole simulation, and thus, the solution accuracy of both flow field and water-air interface capture is not degraded for such unsteady problems with large relative motion. It should be mentioned that the water surface level is kept stationary regardless of the translation or rotation of the mesh. To achieve this goal, at the beginning the function of $\alpha_\mathrm{w}$ needs to be implemented on the boundary condition where the water volume fraction of each grid cell was assigned according to its global inertial coordinates. Specifically, the volume fraction is one for the cells located below the interface, and zero for the cells above. The same treatment of pressure function on the boundary condition also should be defined as a part of the initial condition of the fluid field. For the air field, the pressure is assumed as constant at the beginning, while the water pressure varies gradually depending on the depth in the water domain.

\subsection{Models and computational setup}

\begin{figure}[hbt!]
\centering
\includegraphics[width=.3\textwidth]{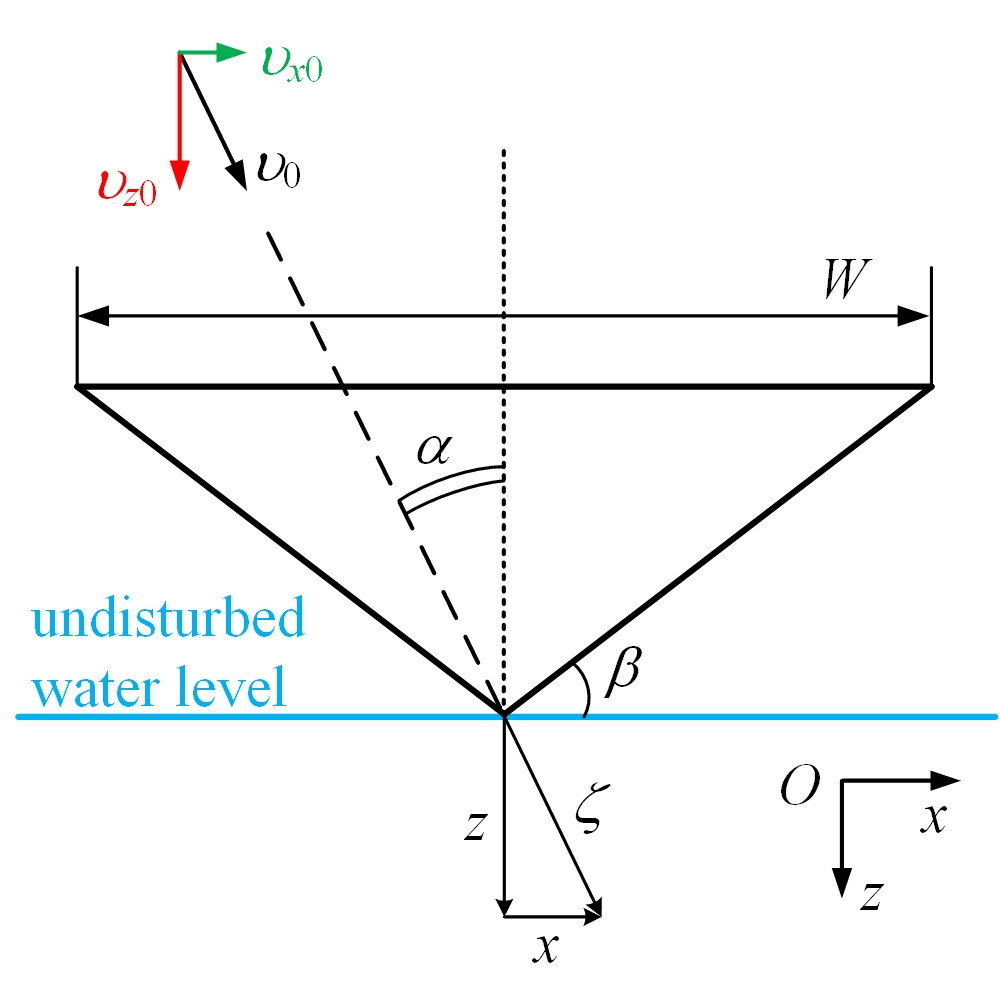}
\caption{Sketch of the wedge at the onset of the entry along with relevant geometric and dynamic parameters.}
\label{fig:Oblique}
\end{figure}

\begin{figure}[hbt!]
\centering
\includegraphics[width=.4\textwidth]{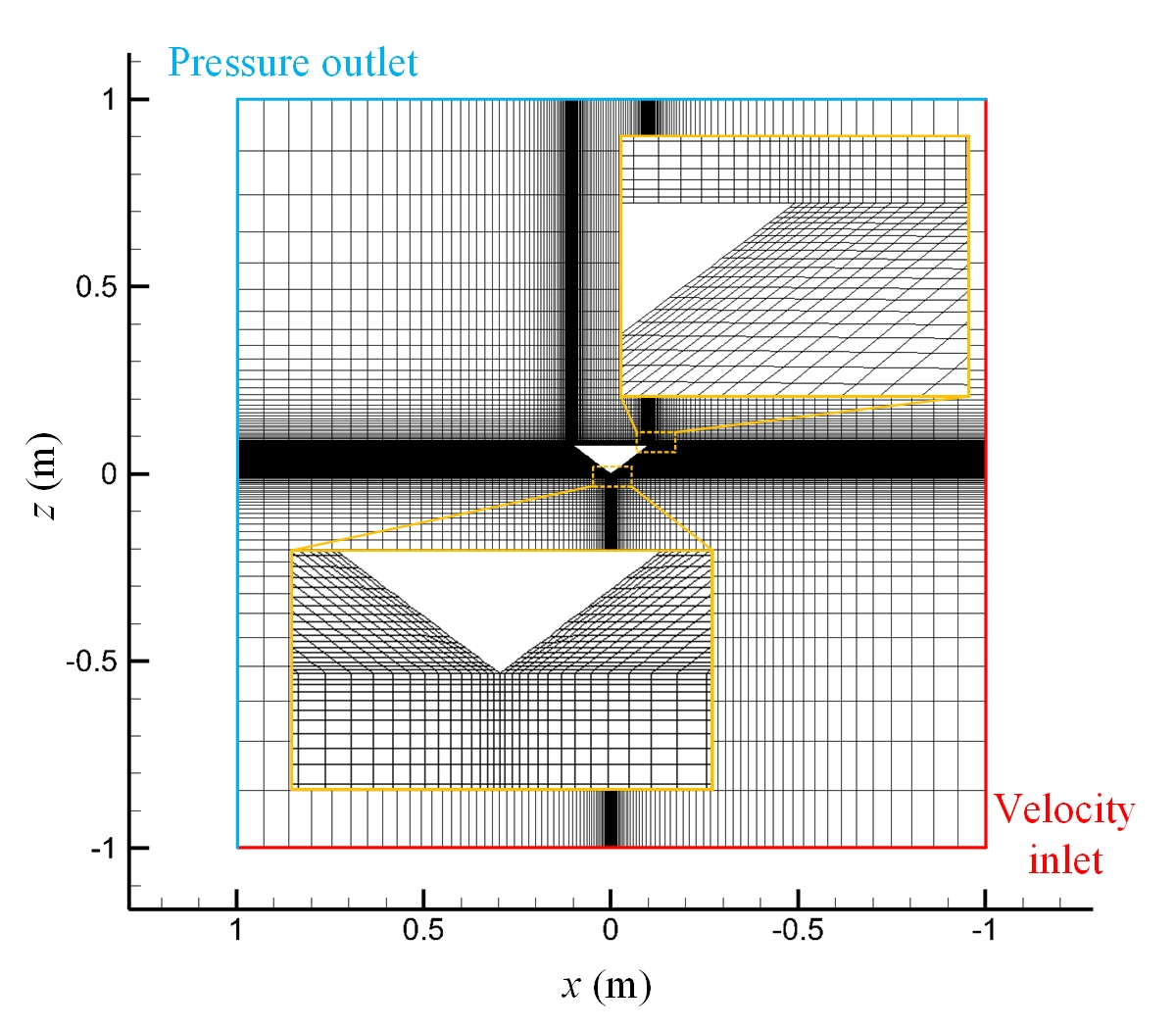}
\caption{Grid topology and density of the wedge.}
\label{fig:ObliqueMesh}
\end{figure}

The theory governing the vertical water entry of wedges and expressed by equations \eqref{eq:amax}, \eqref{eq:threepara} and \eqref{eq:threepara_t} is here validated for the case of oblique entry of a symmetric wedge first,  mainly focusing on the vertical load characteristic. The oblique water entry has been chosen as the motion of the body resembles that of amphibious aircraft during landing and allows to study the effect of varying both the vertical and horizontal components. In \citep{russo2018experimental}, the oblique impact of the wedge has been studied by systematically varying the velocity angle $\alpha$, with the vertical and horizontal motions. The wedge has a width $W$ =0.2 m and a deadrise angle $\beta$ =37$^{\circ}$ and it is impacting with the symmetry axis oriented vertically, as seen in Fig.~\ref{fig:Oblique}. The same configuration is simulated numerically. Besides, in order to carry out a two-dimensional numerical simulation, \textcolor{black}{only 1 cell is set in the y-direction (spanwise direction) with a cell size of 0.002 m. }  
The front and back boundary conditions are defined as symmetry. Fig.~\ref{fig:ObliqueMesh} shows the details of the mesh topology and the grid density with two zoom-in views in the $x-z$ plane. The length of the square boundary is 10 times the width of the wedge. The computational domain is discretized with structured quadrilateral grids and the minimum size of mesh is 0.0005 m. The right hand and bottom sides were set as velocity inlet, when the boundary condition of pressure outlet was specified on the top and the left sides (see Fig.~\ref{fig:ObliqueMesh}).

\begin{figure}[hbt!]
\centering
\includegraphics[width=.5\textwidth]{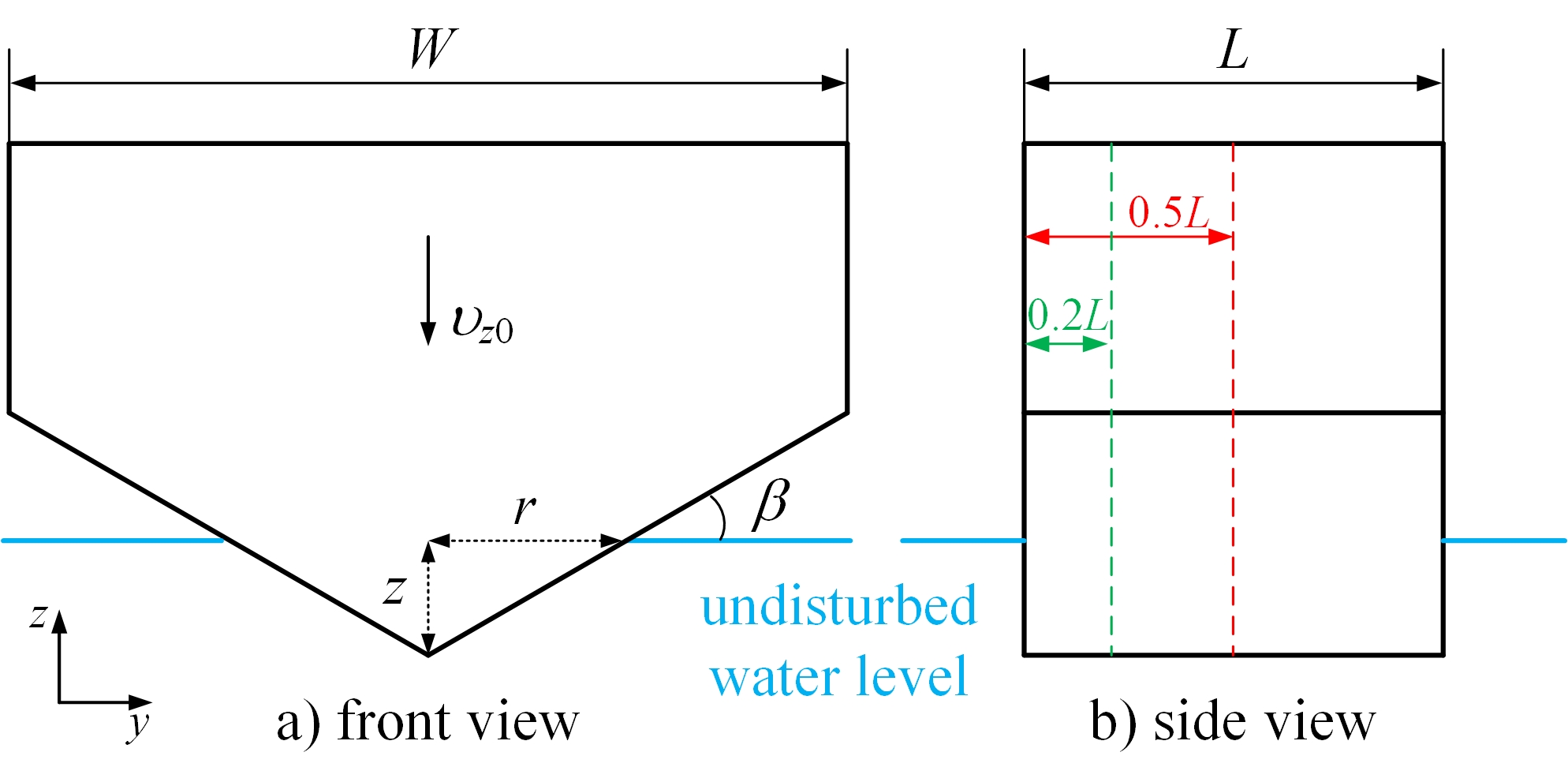}
\caption{Sketch of the cabin section along with relevant geometric parameters.}
\label{fig:CabinSection}
\end{figure}

\begin{figure}[hbt!]
\centering
\includegraphics[width=.5\textwidth]{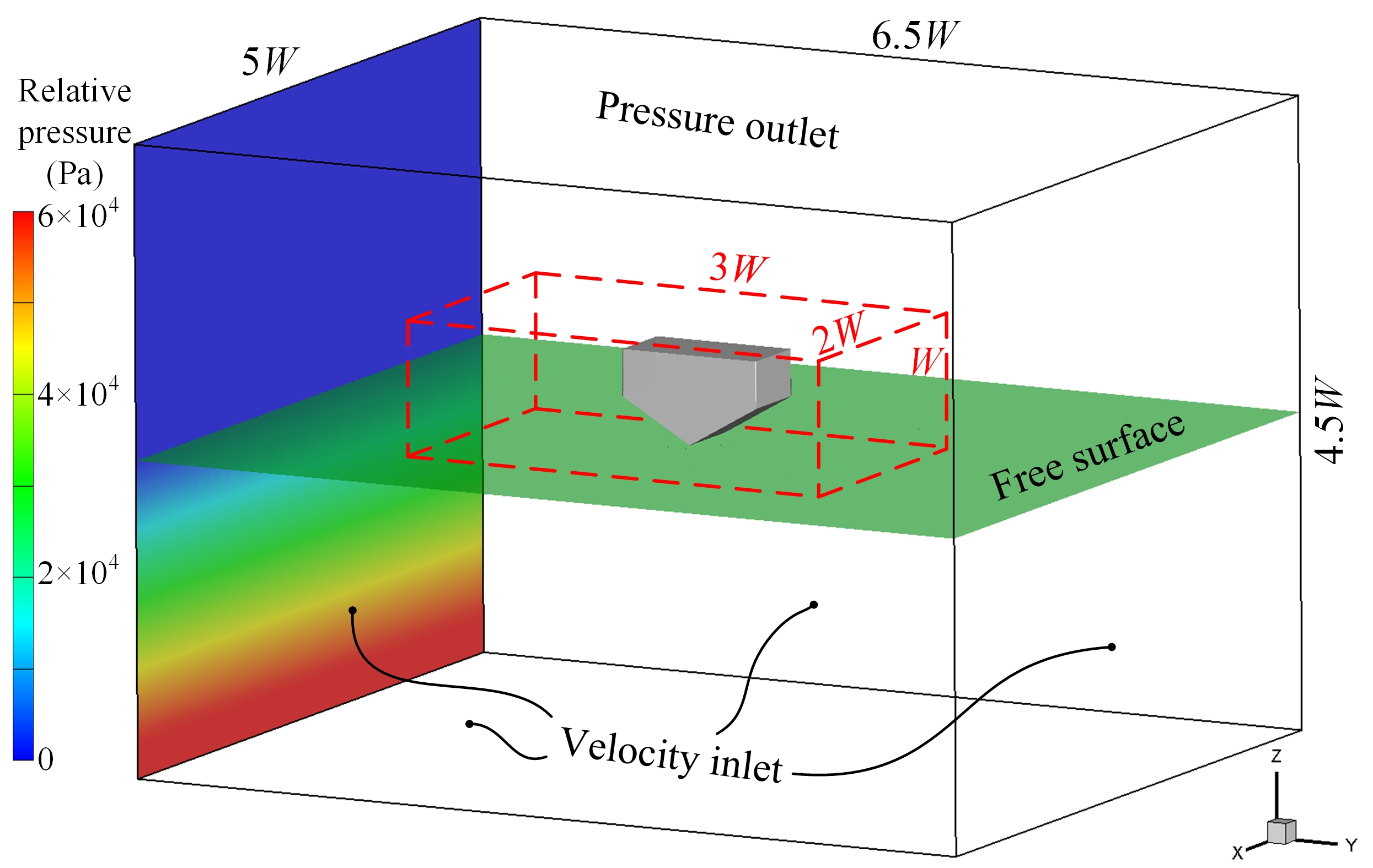}
\caption{Boundary conditions and the initial flow fields \textcolor{black}{of the cabin section.}}
\label{fig:CabinSectionMesh}
\end{figure}

As a second step of test, a cabin section, that is a part of the seaplane, is investigated numerically to examine the quantitative relations, referring to Eq.~\eqref{eq:amax}, \eqref{eq:threepara} and \eqref{eq:threepara_t}, since the 3D effects affect the slamming force during water impact \citep{wang2021assess}. The geometry parameters of the cabin section are shown in Fig.~\ref{fig:CabinSection} with length $L$=1.61 m, width $W$=3.27 m, deadrise $\beta$=30$^{\circ}$ and mass $M$=600 kg. The test condition represent that of the experiments in \citep{chen2022numerical}, where the section is manually lifted to the desired height and released for freely fall. In the simulation, as depicted in Fig.~\ref{fig:CabinSectionMesh}, the cabin is initially released near the water surface with different initial impact velocity to study the effect of velocity on the acceleration. Fig.~\ref{fig:CabinSectionMesh} also shows the boundary conditions and the initial relative pressure field on the left side boundary. A dashed red cuboid was created surrounding the cabin with refined meshes to capture the water surface more accurately.

\begin{figure}[hbt!]
\centering
\includegraphics[width=.8\textwidth]{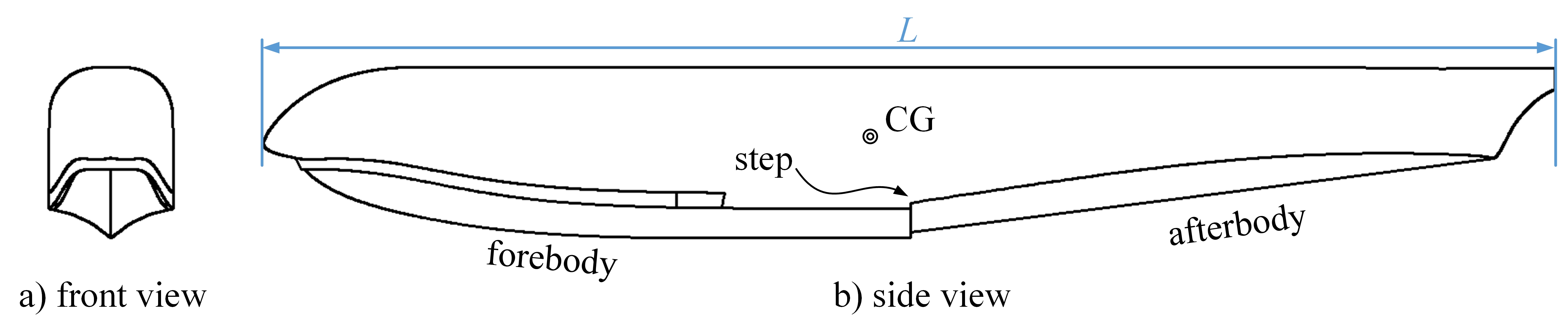}
\caption{V-shaped hull configuration features of amphibious aircraft.}
\label{fig:SeaplaneConfig}
\end{figure}

\begin{figure}[hbt!]
\centering
\includegraphics[width=.45\textwidth]{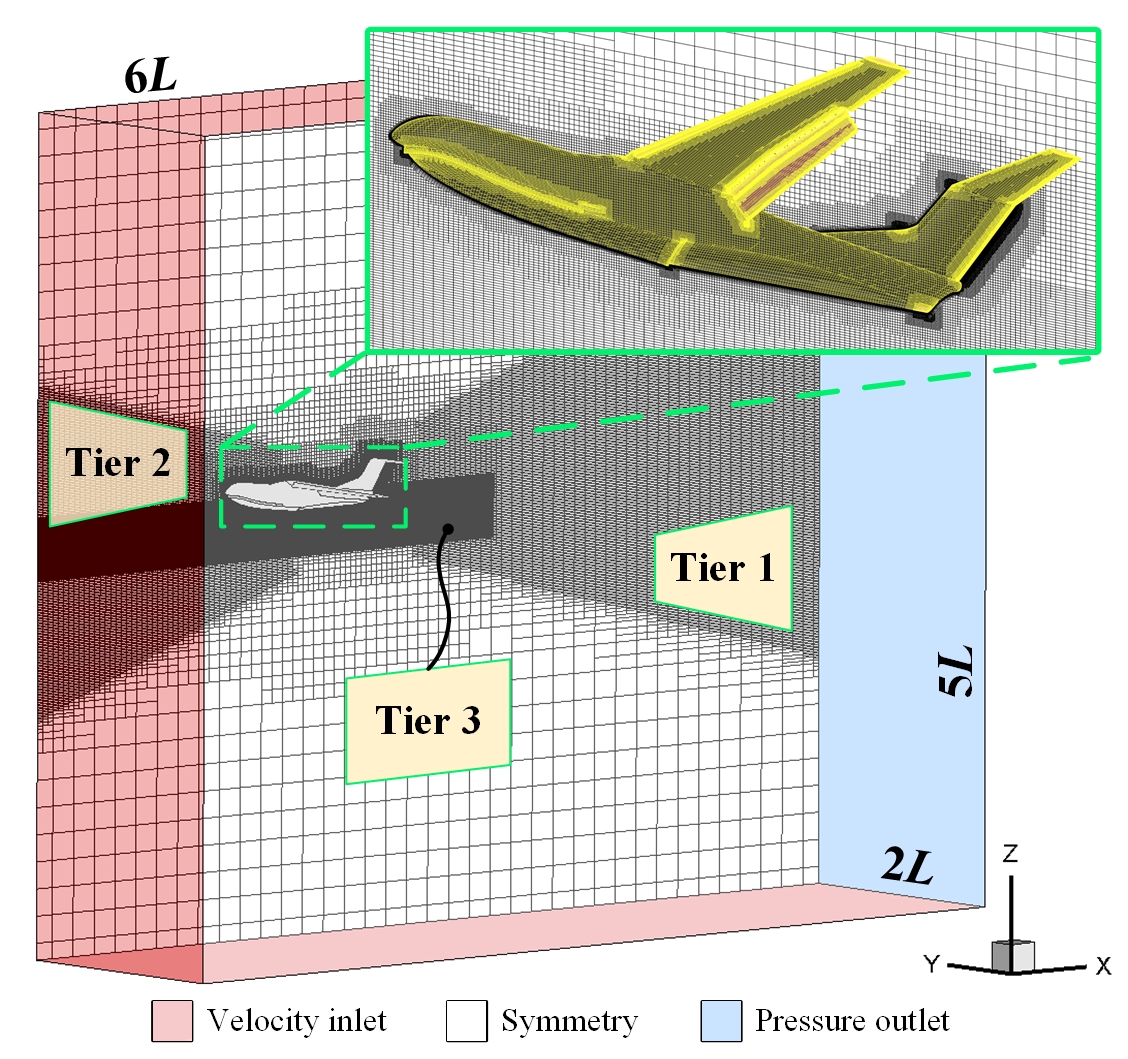}
\caption{Computational domain and boundary conditions of the amphibious aircraft.}
\label{fig:SeaplaneMesh}
\end{figure}

Eventually, the water landing of V-shaped hull on amphibious aircraft is studied to check the capability of the theoretical relations (Eq.~\eqref{eq:amax}, \eqref{eq:threepara} and \eqref{eq:threepara_t}) to deal with complex problems and to verify to which extent they are reliable for engineering applications. A conventional configuration of the fuselage of amphibious aircraft is shown in Fig.~\ref{fig:SeaplaneConfig}. The bottom of hull is divided into two parts, forebody and afterbody, by the step, making it easier to take off on water. The computational domain was created by a cuboid with size of $6\times2\times5L$ in length, width and height, respectively (see Fig.~\ref{fig:SeaplaneMesh}), and is regarded large enough for the present study. The whole domain was discretized with Cartesian cells and prismatic boundary layer grids surrounding the model and moving rigidly without deforming. Three tiers for refining meshes were assigned to the entire domain as follows: tier 3 for the accurate description of the hydrodynamics about the hull; tier 2 and tier 1 fan-shaped regions to enable the large range of pitch motion. The cell height in these tiers is 0.005$L$, 0.01$L$ and 0.015$L$, respectively. The total number of grid cells in the whole domain is almost 12 million. Note that the wing and tail wing are taken into consideration.

\section{Results and Discussion}
\label{sec:results}
\subsection{2D symmetric wedge}

\begin{figure}[hbt!]
\centering
\begin{subfigure}{0.49\textwidth}
\includegraphics[width=\linewidth]{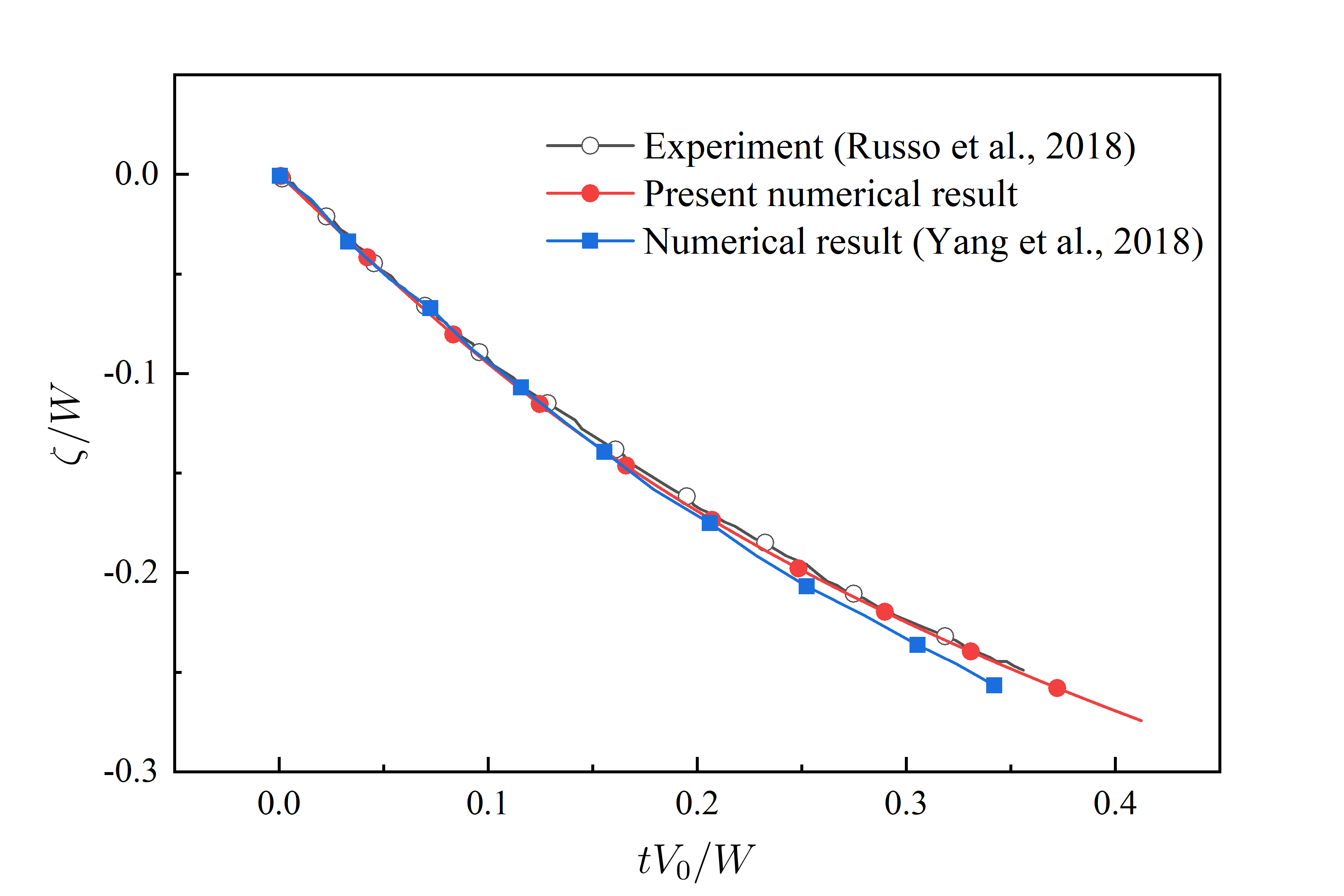} 
\caption{}
\label{fig:CompaPreExpa}
\end{subfigure}
\begin{subfigure}{0.49\textwidth}
\includegraphics[width=\linewidth]{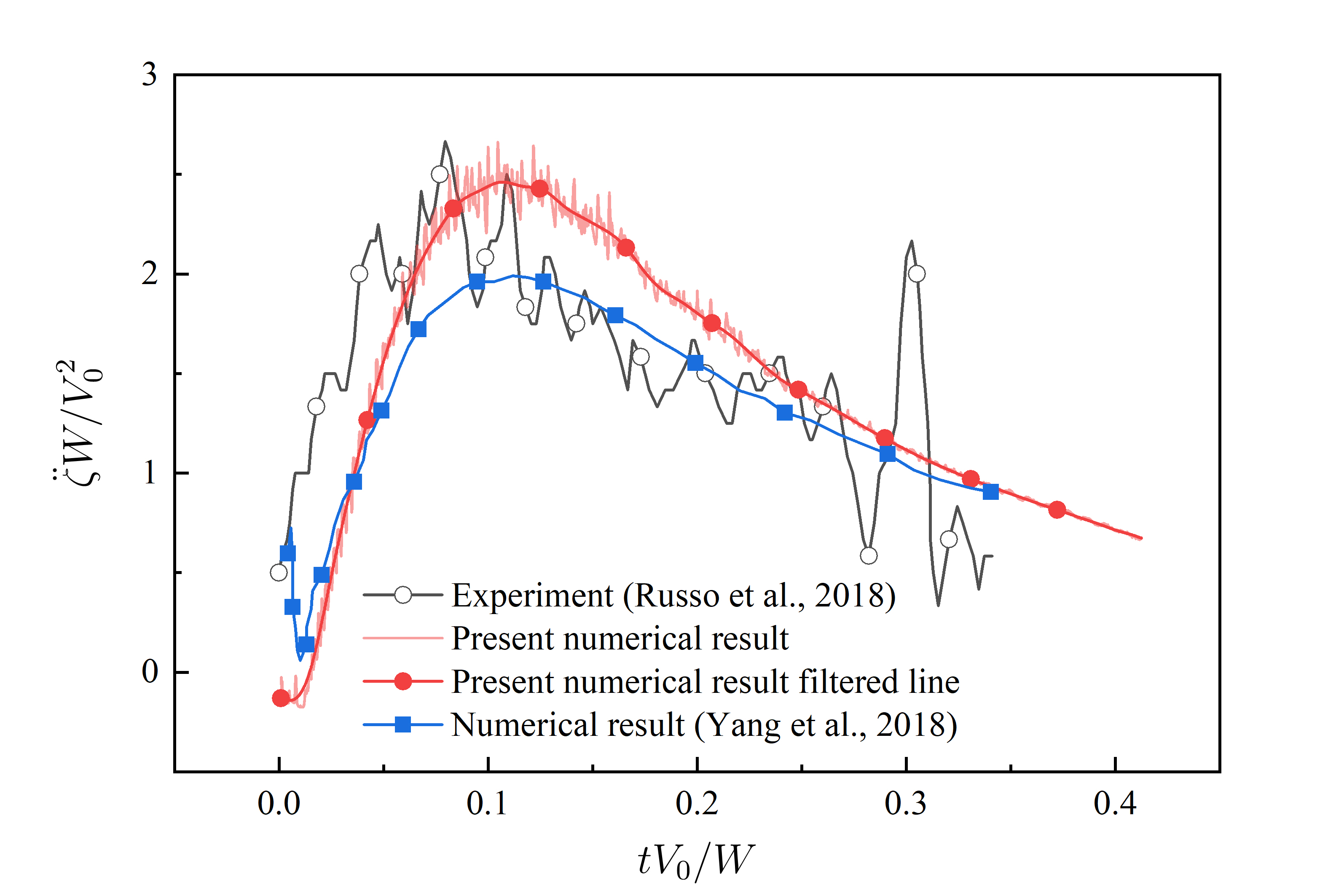}
\caption{}
\label{fig:CompaPreExpb}
\end{subfigure}
\caption{\textcolor{black}{Comparison among the present study, experimental data and numerical results on the oblique water entry of a wedge}: (a) normalized resultant displacement; (b) normalized resultant acceleration.}
\label{fig:CompaPreExp}
\end{figure}

First, the accuracy and efficiency of the numerical method have been validated for a symmetric wedge. In the simulation, at $t$=0.001 s, the wedge is dropped freely against calm water from a small distance at 0.002m, entering the free surface with an initial resultant velocity $\upsilon_0$ = 2.75 m/s and velocity angle $\alpha$ = 20$^{\circ}$ (see Fig.~\ref{fig:Oblique}). Fig.~\ref{fig:CompaPreExp} shows the comparison between \textcolor{black}{the numerical results of the} present study and experimental data \citep{russo2018experimental} in terms of the normalized resultant displacement $\zeta$ and acceleration $\Ddot{\zeta}$. It can be seen, the results are in good agreement with experiments, aside from a little discrepancy occurs at the early stage of the acceleration. Theoretically, at the beginning the acceleration should be close to $-g$, like numerical results show, whereas in the experimental data the acceleration is immediately positive, probably due to measurement problems in the initial phases \citep{russo2018experimental}. \textcolor{black}{Also, a good comparison with another CFD numerical result \citep{yang2018numerical} can be observed in Fig.~\ref{fig:CompaPreExp}.} Overall, numerical results exhibit a satisfactory agreement with the experimental data.

\subsubsection{Effect of vertical velocity}
\label{sec:effectvertical}
\begin{figure}[hbt!]
\centering
\begin{subfigure}{0.49\textwidth}
\includegraphics[width=\linewidth]{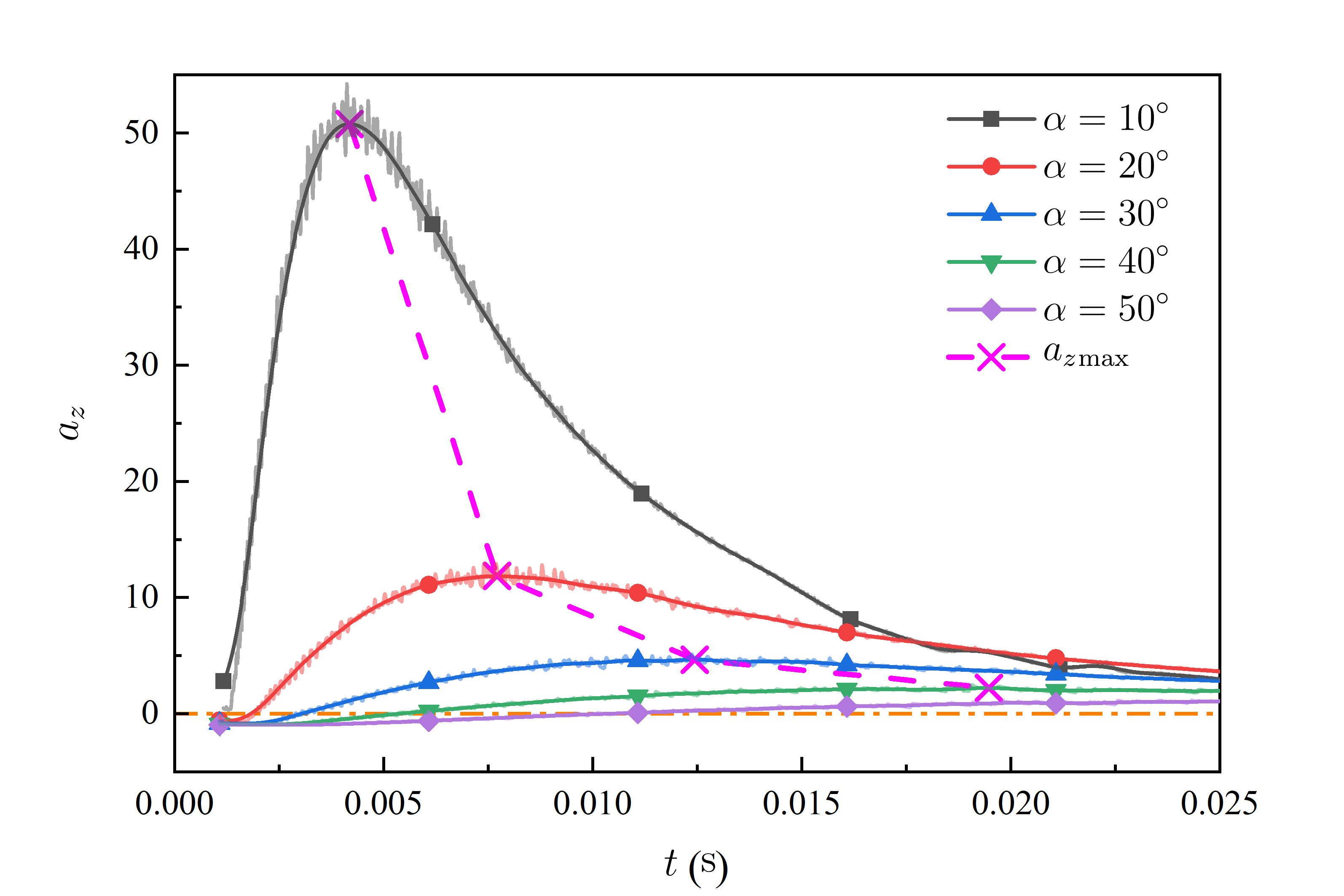} 
\caption{}
\label{fig:az_and_azmaxa}
\end{subfigure}
\begin{subfigure}{0.49\textwidth}
\includegraphics[width=\linewidth]{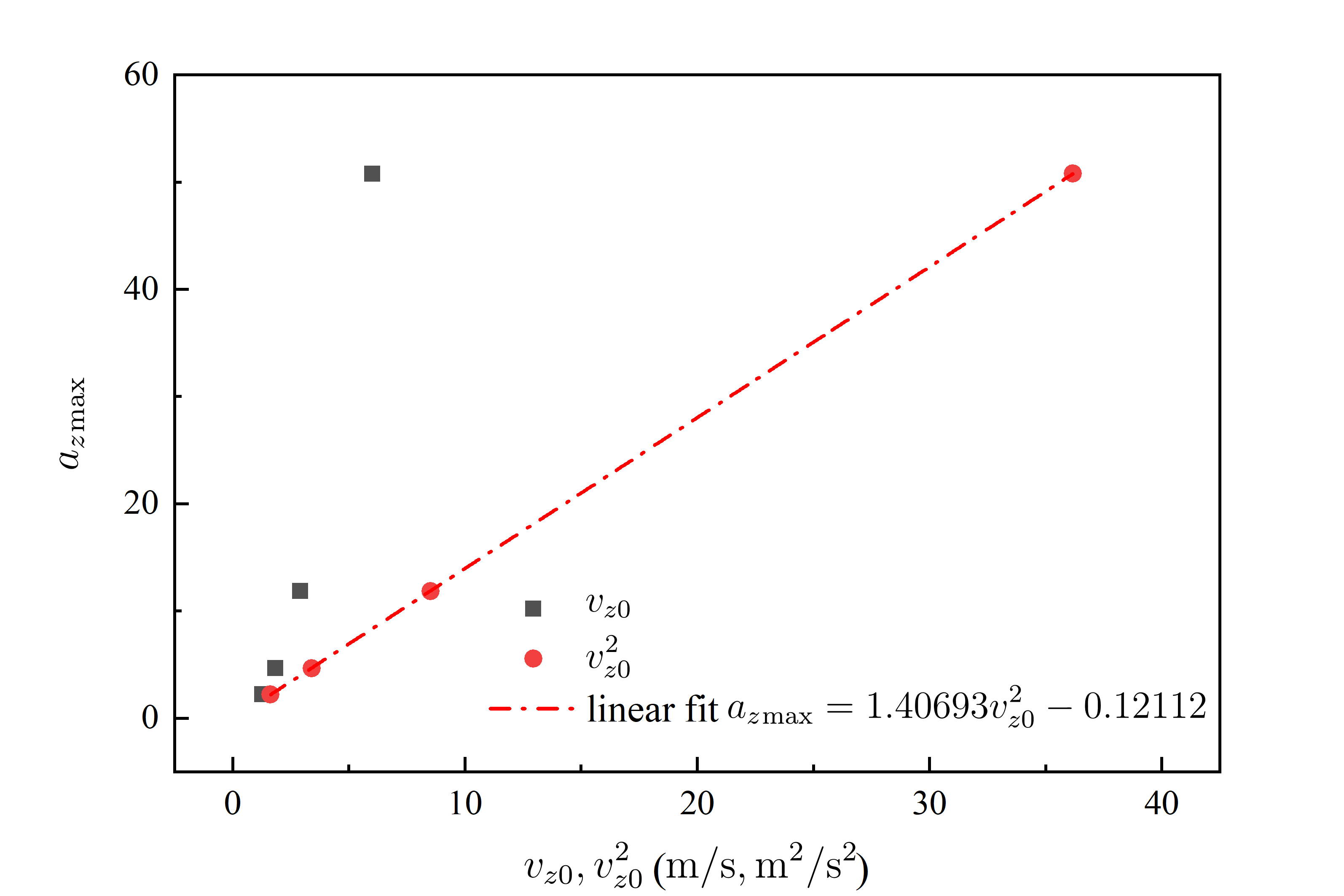}
\caption{}
\label{fig:az_and_azmaxb}
\end{subfigure}
\caption{Variation of \textcolor{black}{dimensionless} acceleration $z$ \textcolor{black}{with different velocity angle $\alpha$ and fixed horizontal velocity component for oblique water entry}: (a) versus time; (b) versus initial vertical velocity.}
\label{fig:az_and_azmax}
\end{figure}

\begin{figure}[hbt!]
\centering
\includegraphics[width=.5\textwidth]{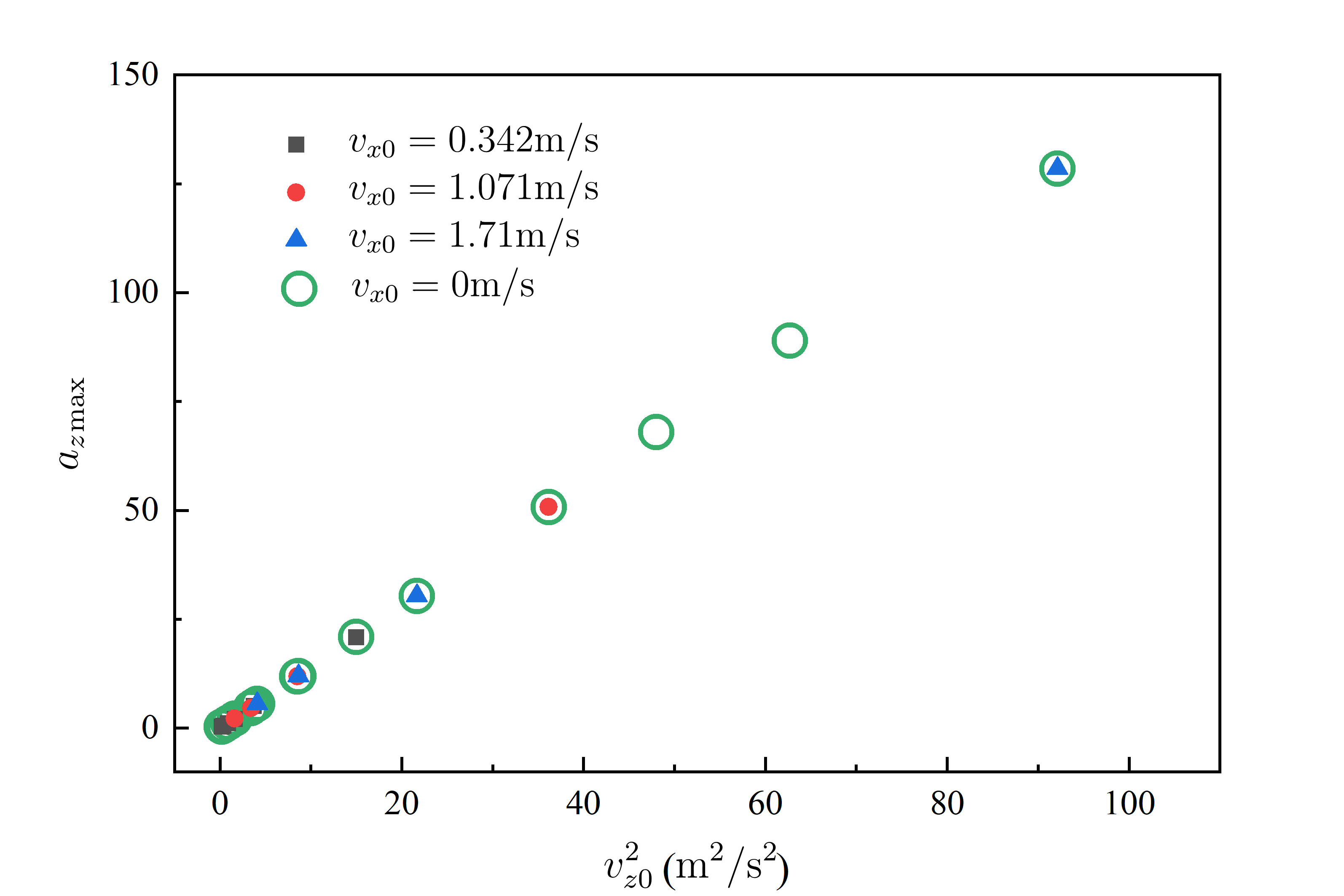}
\caption{Effect of the horizontal velocity on the relation between $a_{z\mathrm{max}}$ and $\upsilon_{z0}^2$ \textcolor{black}{for oblique water entry}.}
\label{fig:DifvxOnAzmaxvsVzz}
\end{figure}

Next, in order to better understand the effect of the variation of the vertical velocity, several simulations have been performed for constant $\upsilon_{x0}$ and $\alpha$ varying from 10$^\circ$ to 50$^\circ$, which corresponds to a reduction of the vertical velocity component. The time histories of \textcolor{black}{dimensionless acceleration in $z$-direction} $a_z$, defined as $a_z=(F_\mathrm{w}+F_\mathrm{a}-Mg)/Mg$, \textcolor{black}{where $F_\mathrm{w}$ and $F_\mathrm{a}$ denote the fluid force induced by water and air respectively,} are depicted in Fig.~\ref{fig:az_and_azmaxa}, along with several pink crosses marking the maximum value $a_{z\mathrm{max}}$. The data indicate that the increase in $\alpha$ causes a significant reduction of $a_z$ due to the corresponding reduction in the $\upsilon_{z0}$. Note that the positive values of $a_z$ denote upward acceleration. In particular, as $\upsilon_{z0}$ drops below a certain value, $a_z$ will experience a smooth trend in proximity to zero, known as ‘smooth entry’ \citep{vincent2018dynamic}. The data shown in Fig.~\ref{fig:az_and_azmaxb} indicate that $a_{z\mathrm{max}}$ is a linear function of $\upsilon_{z0}^2$, thus supporting the relationship formulated in the Eq.~\eqref{eq:amax}, except for the offset. Furthermore, other series of simulations have been conducted by varying the value of $\upsilon_{x0}$, including the case of zero horizontal velocity. Fig.~\ref{fig:DifvxOnAzmaxvsVzz} shows that all the data are aligned on the same straight line, thus confirming the validity of the relationship in the Eq.~\eqref{eq:amax}. Note that in the case of $\upsilon_{x0}$=0.342 m/s, $\alpha$ varies from 5$^\circ$ to 50$^\circ$. As highlighted in Table.~\ref{tab:compaatv}, a linear relation between $a_{z\mathrm{max}}$ and $\upsilon_{z0}^2$ exists, and only minor deviations can be observed in the slope $k$ compared with the theoretical estimate, derived from Eq.~\eqref{eq:amax}. However, there is an intercept value of $b$ for the numerical results which is presumably due to the gravity. On the other hand, the data shown in Fig.~\ref{fig:DifvxOnAzmaxvsVzz} and Table.~\ref{tab:compaatv} display a significant contribution of the vertical component of the velocity to the linear relation, independently of the value of $\upsilon_{x0}$.

\begin{table}[width=.9\linewidth, pos=hbt!]
\caption{Comparison between theoretical estimate and numerical results for the inclined water entry of a wedge}
\centering
\label{tab:compaatv}
\begin{tabular*}{\tblwidth}{@{}cccccccccc@{}}
\toprule
 & \multicolumn{3}{c}{$a_{z\mathrm{max}}$}& \multicolumn{3}{c}{$t^*, \mathrm{s}$}& \multicolumn{3}{c}{$\upsilon_z^*, \mathrm{m/s}$}\\
\midrule
 & $k$& err, \%& $b$& $k$& err, \%& $b$& $k$& err, \%& $b$\\
Theoretical value& 1.2807& -& -& 0.0197& -& -& 0.8333& -& -\\
$\upsilon_{x0}=0.342\mathrm{m/s}$& 1.3588& 6.09& -0.0509& 0.0285& 44.67& -0.0046& 0.8010& -3.87& 0.1121\\
$\upsilon_{x0}=1.071\mathrm{m/s}$& 1.4069& 9.85& -0.1211& 0.0228& 15.73& -0.0014& 0.8367& 0.41& 0.0142\\
$\upsilon_{x0}=1.710\mathrm{m/s}$& 1.3948& 8.91& 0.0364& 0.0185& -6.09& -& 0.8308& -0.30& 0.0306\\
\bottomrule
\end{tabular*}
\end{table}

\begin{figure}[hbt!]
\centering
\begin{subfigure}{0.49\textwidth}
\includegraphics[width=\linewidth]{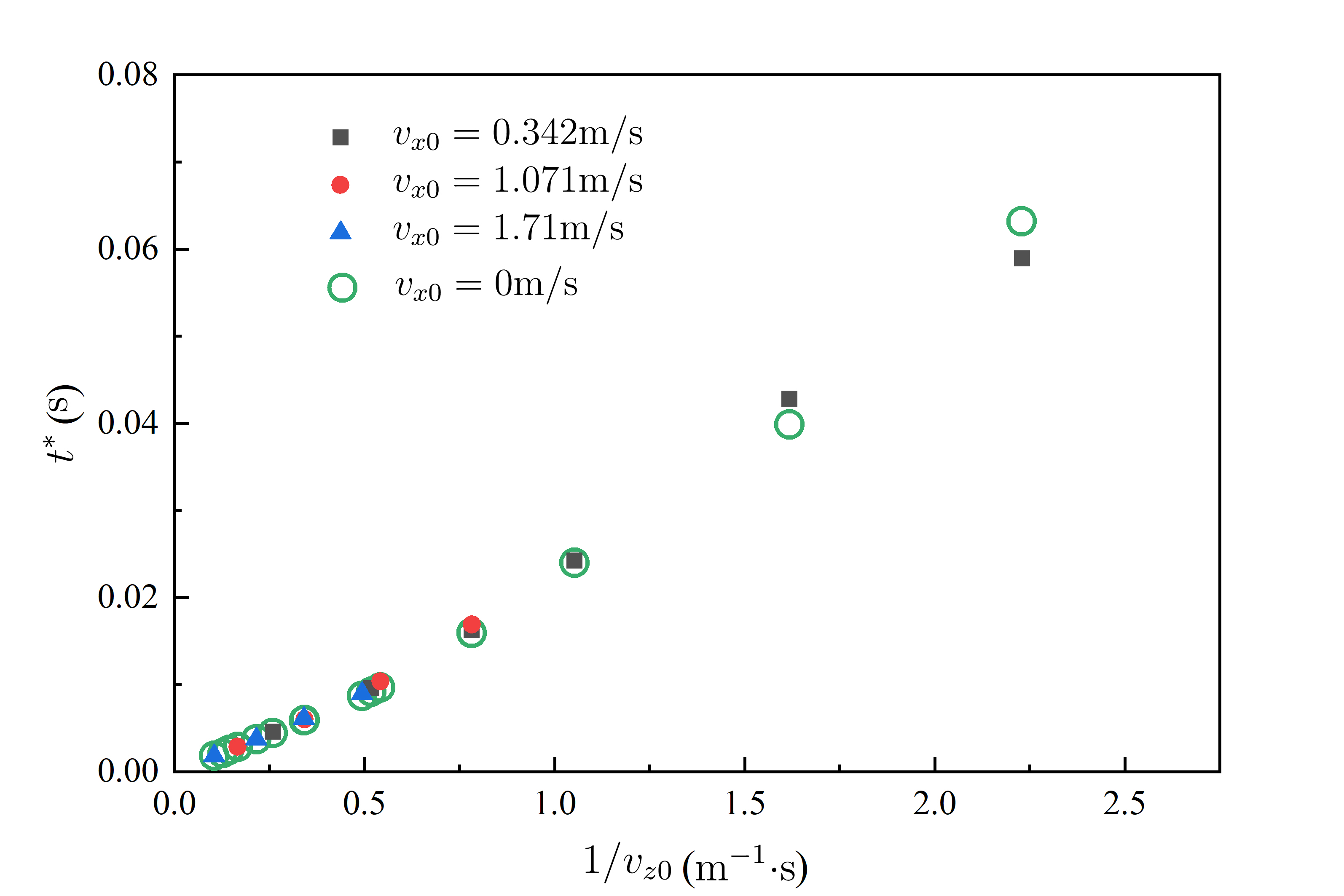} 
\caption{}
\label{fig:DifvxOnTZVzKappaa}
\end{subfigure}
\begin{subfigure}{0.49\textwidth}
\includegraphics[width=\linewidth]{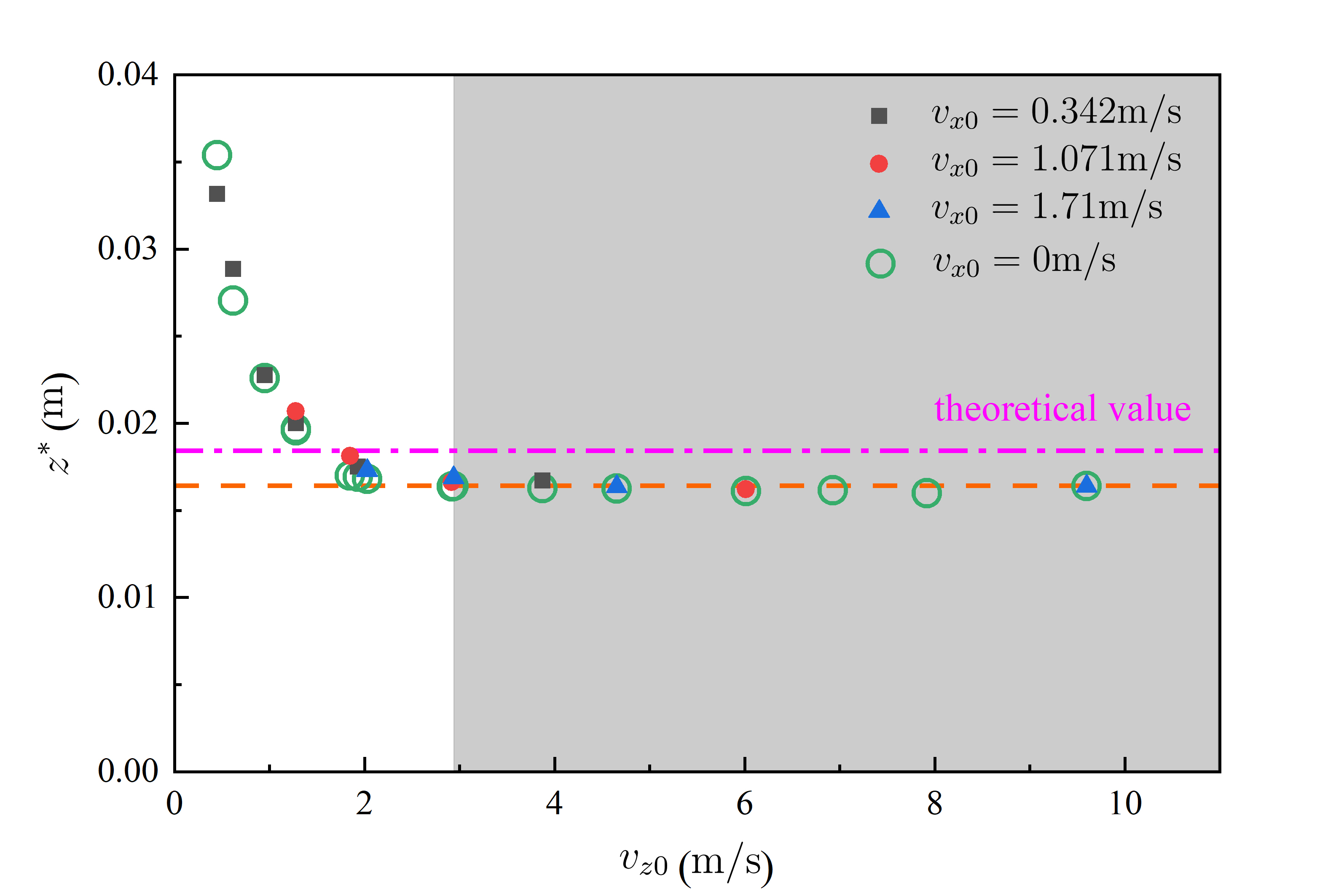}
\caption{}
\label{fig:DifvxOnTZVzKappab}
\end{subfigure}
\begin{subfigure}{0.49\textwidth}
\includegraphics[width=\linewidth]{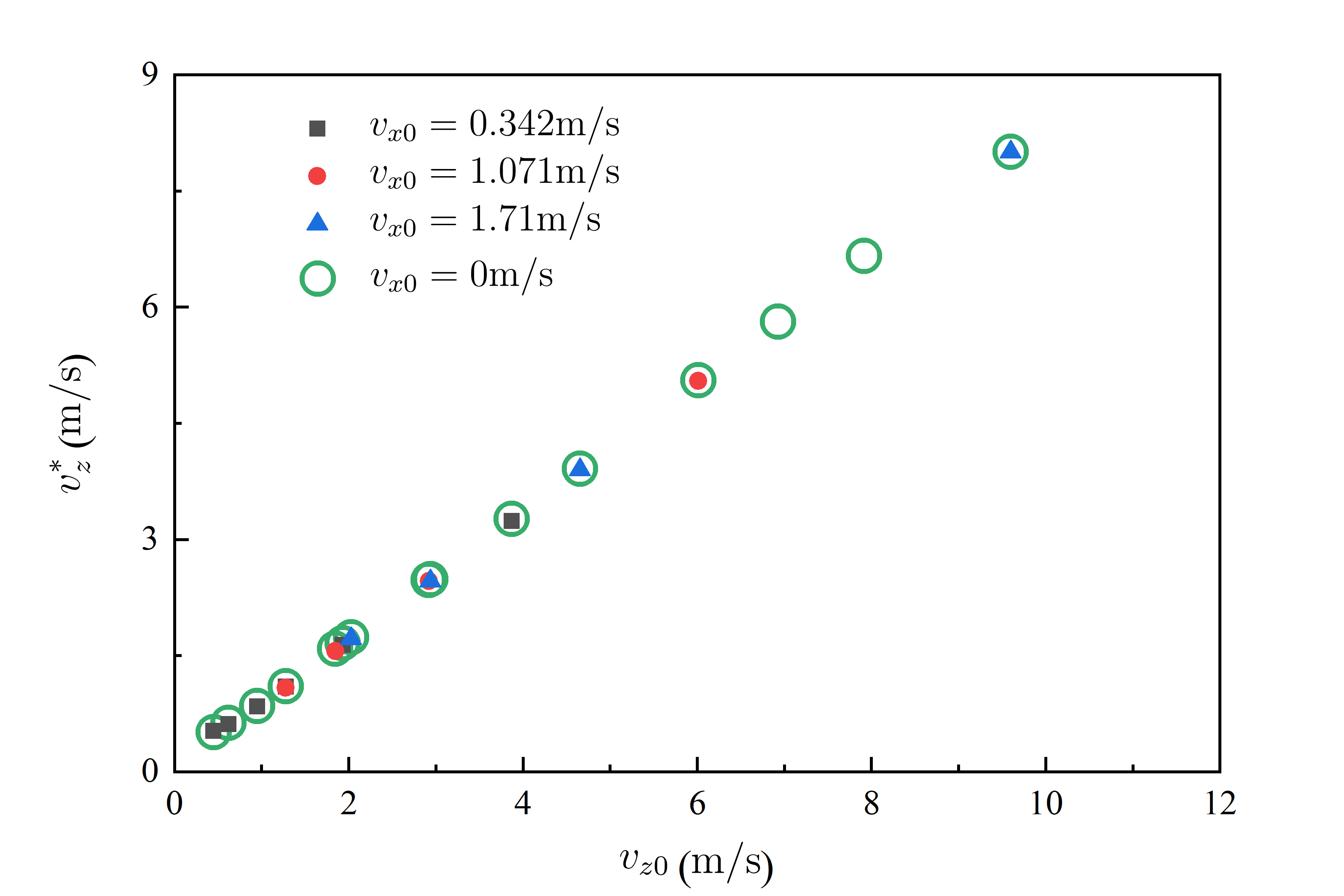} 
\caption{}
\label{fig:DifvxOnTZVzKappac}
\end{subfigure}
\begin{subfigure}{0.49\textwidth}
\includegraphics[width=\linewidth]{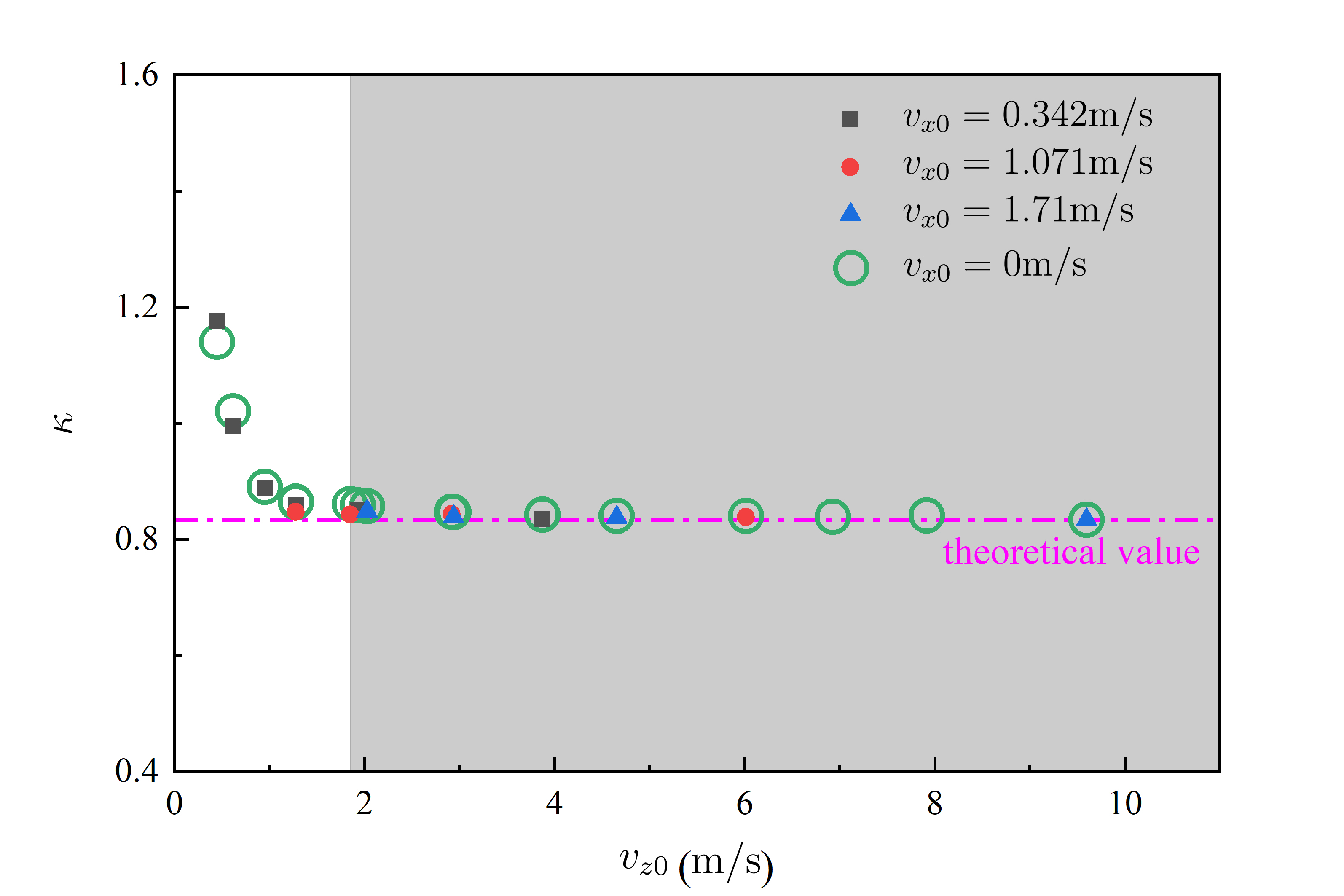}
\caption{}
\label{fig:DifvxOnTZVzKappad}
\end{subfigure}
\caption{Effect of initial vertical velocity on variable dynamic parameters \textcolor{black}{for the oblique water entry of a wedge}: (a) $t^*$; (b) $z^*$; (c) $\upsilon_z^*$; (d) $\kappa$.}
\label{fig:DifvxOnTZVzKappa}
\end{figure}

\begin{figure}[hbt!]
\centering
\includegraphics[width=0.98\textwidth]{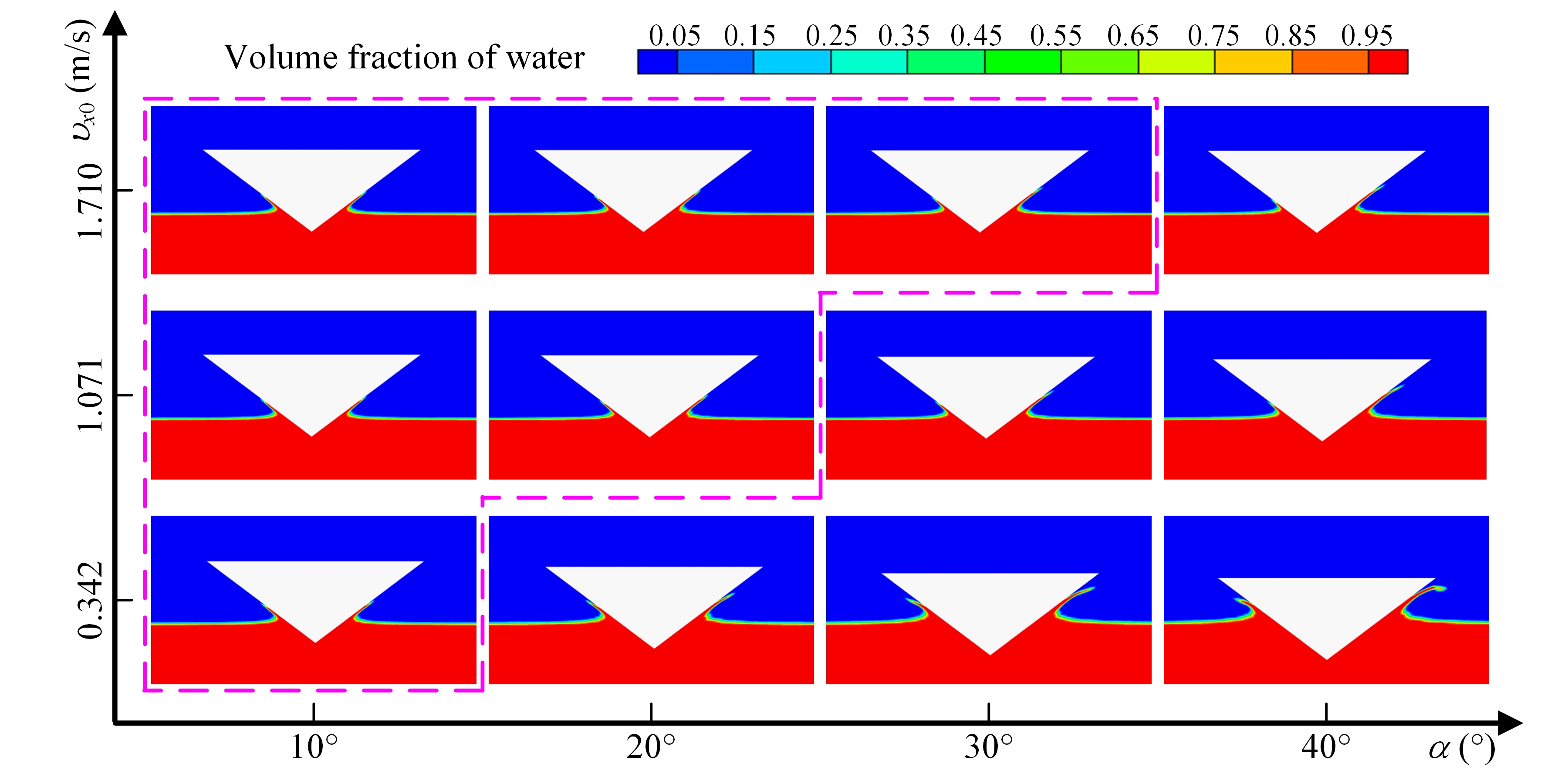}
\caption{Free surface deformation around the wedge at $t^*$ with different $\upsilon_{x0}$ and $\alpha$.}
\label{fig:ObliqueThreeVx0Level}
\end{figure}

In Fig.~\ref{fig:DifvxOnTZVzKappa}, the other four correlated variables are reported, viz., time $t^*$, penetration depth $z^*$, velocity $\upsilon_z^*$ and the ratio of velocity $\kappa$, defined as $\kappa=\upsilon_z^*/\upsilon_{z0}$, for the four cases introduced earlier. In Eq.~\eqref{eq:threepara_t}, a linear relation between $t^*$ and reciprocal of the initial vertical velocity $\upsilon_{z0}^{-1}$ was established that is similar to the solution in Fig.~\ref{fig:DifvxOnTZVzKappaa}, despite a small difference appears on $k$ among the three cases. As listed in Table \ref{tab:compaatv}, the error of the numerical values with respect to the theoretical estimate, varying from 44.67\% to -6.09\%, shows an obvious decreasing with the growth of $\upsilon_{x0}$. In fact, when reducing $\upsilon_{x0}$, the corresponding initial vertical velocity for lower $\alpha$ becomes smaller and, consequently, gravity effects increase causing larger differences with respect to the theoretical formulation which is derived without considering gravity. In Fig.~\ref{fig:DifvxOnTZVzKappab}, the values display a reduction of $z^*$ when increasing $\upsilon_{z0}$, where one can see that the greater is the $\upsilon_{z0}$, the closer $z^*$ will is to a asymptotic line slightly different from the theoretical result, however, $z^*$ should be constant in theory as it depends on $M$ and $\beta$ only (see Eq.~\eqref{eq:threepara}). The difference with respect to the theoretical line depends on the pile-up effect which is not taken into account in Von Karman’s momentum conservation and affects the evaluation of the hydrodynamic \textcolor{black}{behaviour}
\citep{mei1999on,iafrati2000hydroelastic}. Furthermore, the gray shaded area shows the range at which $z^*$ is close to the constant value and the lowest value of $\upsilon_{z0}$ is almost 2.95 m/s in this model, implying that the theoretical solution is nearly valid only when certain conditions on $\upsilon_{z0}$ are met. Fig.~\ref{fig:ObliqueThreeVx0Level} shows the water surface deformation around the wedge at $t^*$ for different cases with a cyan region, where it can be clearly noted that the displacements of the apex remain almost the same, despite different water jet zones form at the two sides. In the bottom-right picture, the spray seems to detach from the body and fall down. This is a consequence of the gravity. Moreover, as shown in all contours, it indicates that the maximum acceleration $a_{z\mathrm{max}}$ occurs before the wedge is completely submerged. 

Moving to the relationship between $\upsilon_z^*$ and $\upsilon_{z0}$, shown in Fig.~\ref{fig:DifvxOnTZVzKappac} and Table \ref{tab:compaatv}, a slight difference among the simulations and theory on $k$ can be observed, the error on $k$ being below 5\%. Furthermore, as shown in Fig.~\ref{fig:DifvxOnTZVzKappad}, the trend of $\kappa$ is similar to the one obtained from $z^*$, and a gray shaded region can be found where $\upsilon_z^*$ is 5/6 times $\upsilon_{z0}$ in agreement with the theoretical estimate. In other words, the value 5/6 about $\upsilon_z^*$ and $\upsilon_{z0}$ can only be set up when $\upsilon_{z0}$ is greater than 1.85m/s, which is smaller than the limitation 2.95m/s on $z^*$. As it can be seen in Fig.~\ref{fig:az_and_azmax}a), the acceleration experiences two phases, acceleration downwards and then upwards, before reaching the maximum. Since the wedge, with a deadrise angle $\beta$=37$^\circ$, undergoes a free fall motion, gravity plays a dominant role at the very early stage, leading to an accelerating period and an increase in the vertical velocity. Subsequently, with the increase of hydrodynamic force, the downward acceleration diminishes and gradually turns upwards. Thus, it can be concluded from Fig.~\ref{fig:az_and_azmax}a) that, for a given mass of the impacting body, the smaller is the initial vertical velocity, the longer is the accelerating time. Moreover, four distinctive points exceeding 1.0 are noticeable in Fig.~\ref{fig:DifvxOnTZVzKappad}, meaning that the vertical velocity of the body is larger than initial vertical velocity. Overall, it indicates that the accelerating phase not only lasts longer, but the effect of the accelerating phase become more dominant than the decelerating phase, as the initial vertical velocity decreases.

\begin{figure}[hbt!]
\centering
\includegraphics[width=.5\textwidth]{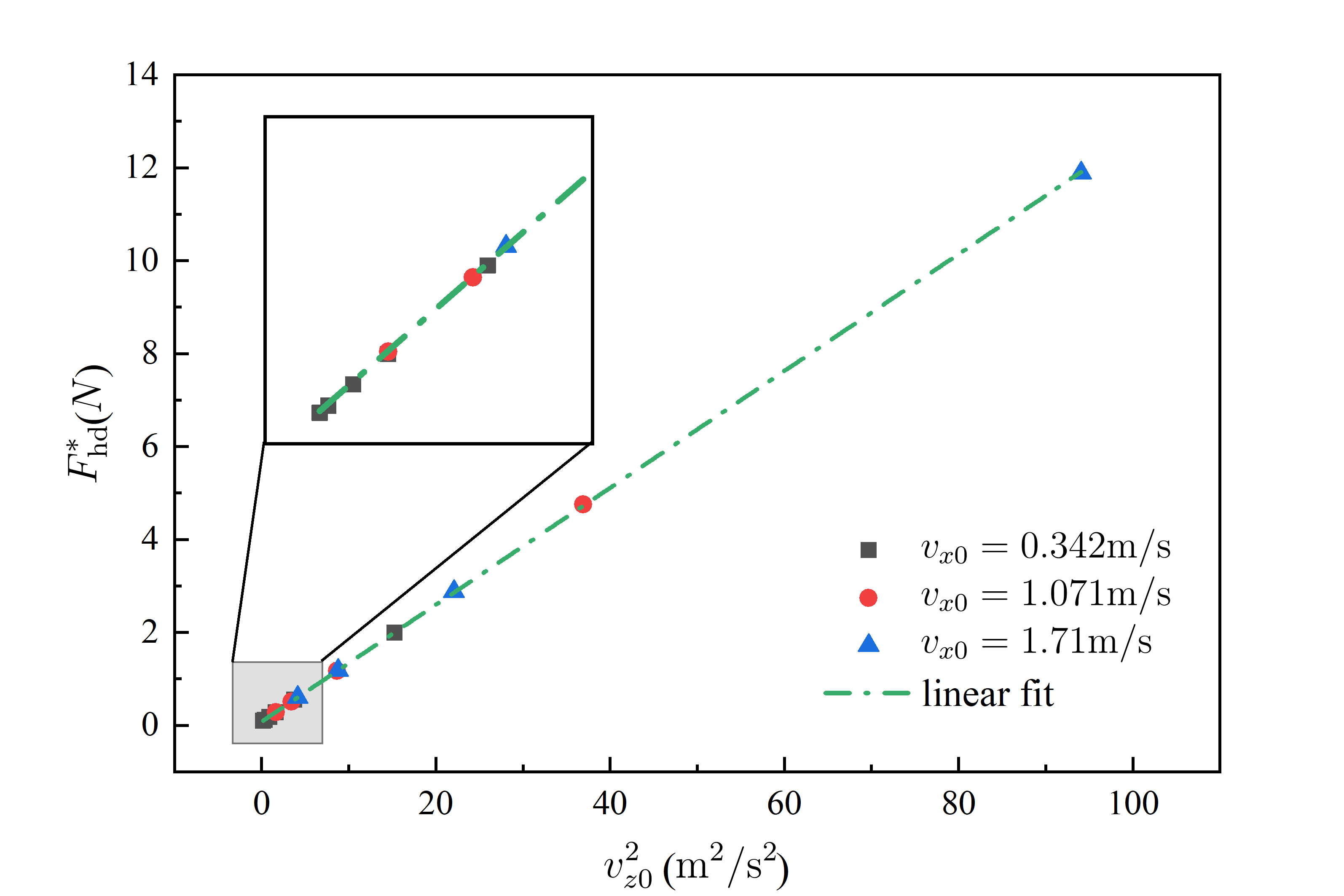}
\caption{Hydrodynamic forces versus the square of the initial vertical velocity \textcolor{black}{for the oblique water entry of a wedge with different initial velocity}.}
\label{fig:ObliqueDifvxOnFhd}
\end{figure}

It is worth noting that the momentum theorem (Eq.~\eqref{eq:momentum}) was obtained without gravity \citep{mei1999on}, whereas the gravitational field has been added into the numerical simulations. Nevertheless, following the investigation discussed above, the formulas \eqref{eq:amax}, \eqref{eq:threepara} and \eqref{eq:threepara_t} derived from Eq.~\eqref{eq:momentum} are still available when the initial vertical velocity becomes larger. In other words, gravity can be neglected with larger velocities, and it has been highlighted in \citep{zekri2021gravity}. Whereas, with slow impact speeds, the gravity should be considered in the model \citep{bertram2012practical}, as confirmed by the discrepancies occurred at the range of low velocities (see Fig.~\ref{fig:DifvxOnTZVzKappab} and \ref{fig:DifvxOnTZVzKappad}). Nonetheless, gravity seems to have no effects on the linear relation between $a_{z\mathrm{max}}$ and $\upsilon_{z0}^2$, except for the offset. The maximal vertical hydrodynamic force during impact is then introduced herein, defined as $F_\mathrm{hd}^*=M\cdot(a_\mathrm{max}\cdot g+g)-F_\mathrm{hs}^*$, where $F_\mathrm{hs}^*$ is the hydrostatic force approximately calculated by Archimedean principle. Results shown in Fig.~\ref{fig:ObliqueDifvxOnFhd} indicate that the linear relation still holds which is consistent with \citep{zekri2021gravity,bertram2012practical}, who found that ‘\textit{even when gravity is formally of the same order of magnitude as the fluid inertia, the effect of gravity on the hydrodynamic loads is still small and can be approximately neglected}’.

Based on the good collapse of the data from different initial horizontal velocity, it is believed that the initial vertical velocity plays a dominant role on the kinematic characteristics during wedge water entry with the given shape parameters, indicating that the effect of initial horizontal velocity on the relations can be ignored. For the analytical solutions based on Eq.~\eqref{eq:amax}, \eqref{eq:threepara} and Eq.~\eqref{eq:threepara_t} to be valid, there is a supplementary condition to the momentum theory which requires that the initial vertical velocity has to higher than a threshold value. Furthermore, the formula for the added mass, which is usually focused for the vertical water entry, is found to be valid for the oblique entry on the vertical direction as well.

\subsubsection{Effect of horizontal velocity}

\begin{figure}[hbt!]
\centering
\begin{subfigure}{0.49\textwidth}
\includegraphics[width=\linewidth]{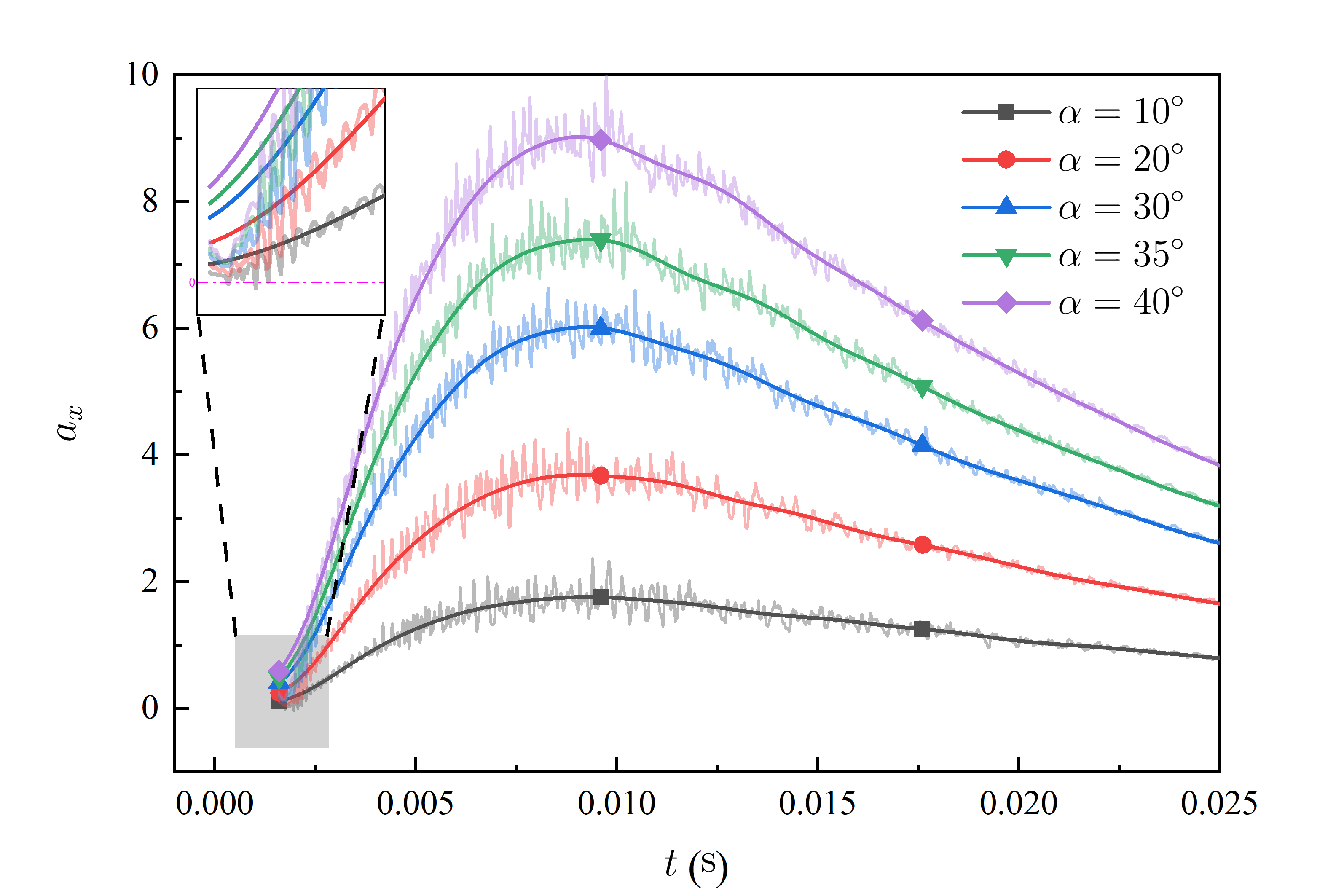} 
\caption{}
\label{fig:SameVzDifVx_axandaza}
\end{subfigure}
\begin{subfigure}{0.49\textwidth}
\includegraphics[width=\linewidth]{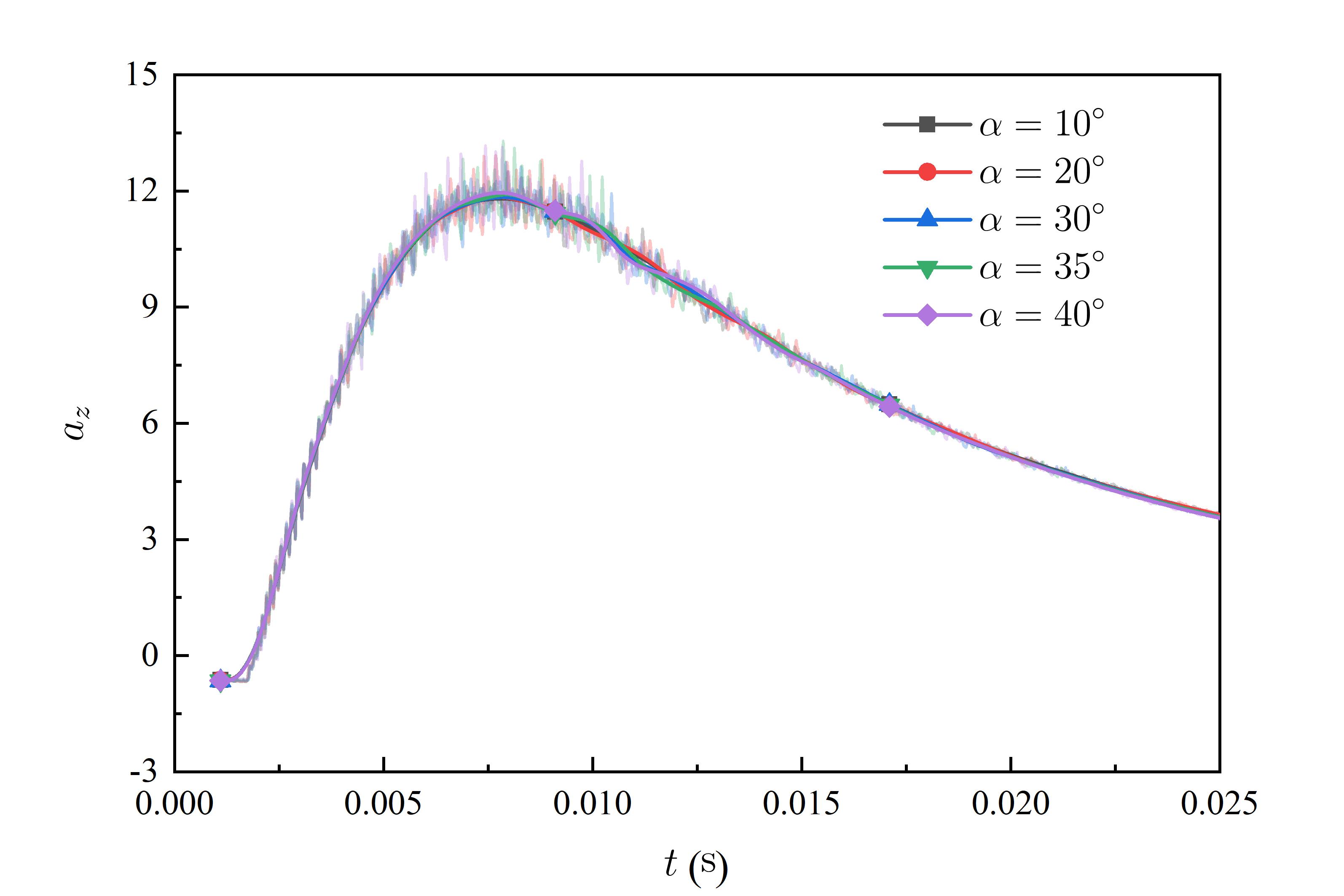}
\caption{}
\label{fig:SameVzDifVx_axandazb}
\end{subfigure}
\caption{Time histories of \textcolor{black}{dimensionless} acceleration \textcolor{black}{in both $x$- and $z$-directions} with different inclined angles \textcolor{black}{using the fixed vertical velocity component $\upsilon_{z0}$ = 2.943 m/s}: (a) $a_x$; (b) $a_z$.}
\label{fig:SameVzDifVx_axandaz}
\end{figure}

\begin{figure}[hbt!]
\centering
\includegraphics[width=.5\textwidth]{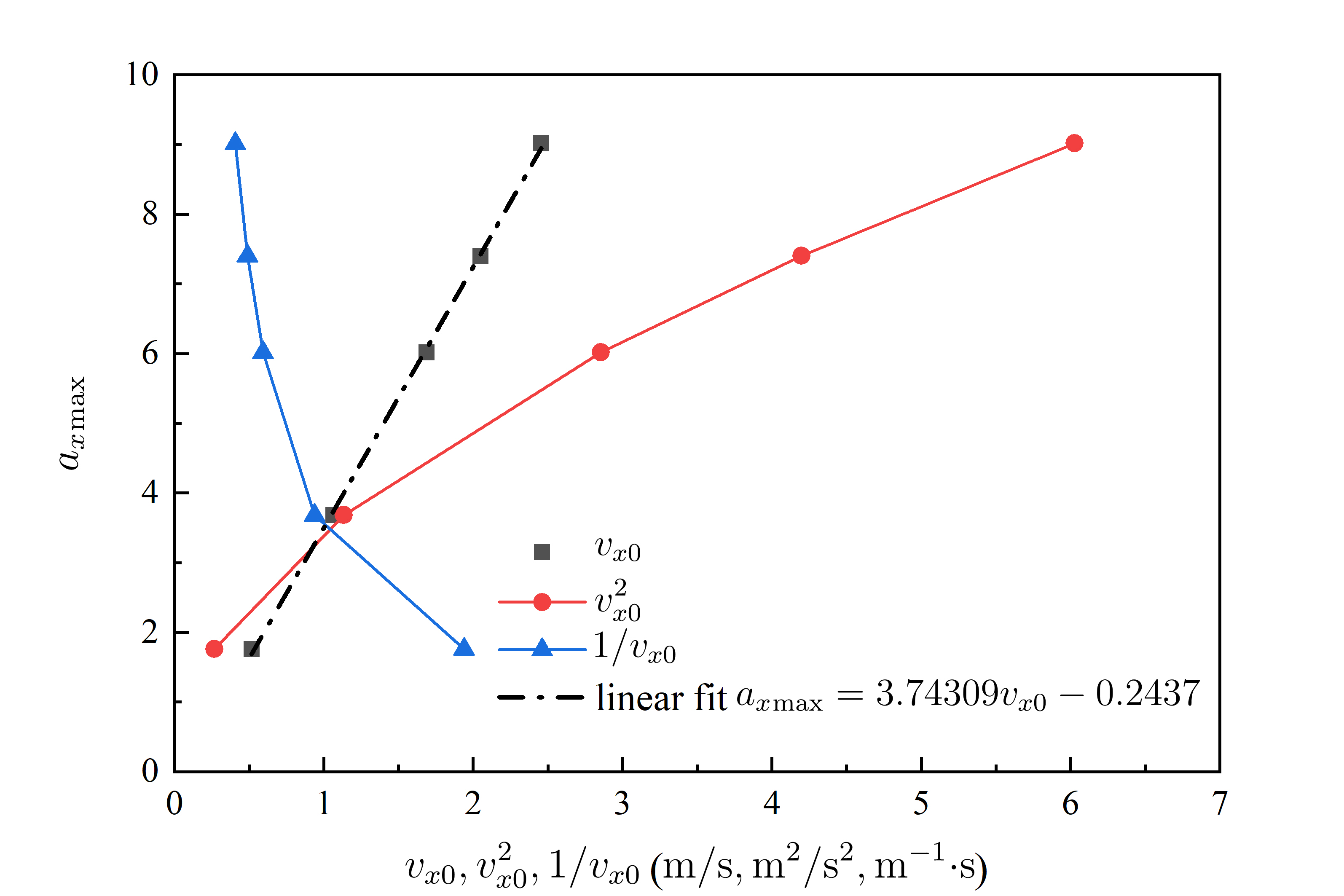}
\caption{Variation of maximum acceleration $x$ versus initial horizontal velocity.}
\label{fig:SameVzDifVx_axmax}
\end{figure}

\begin{figure}[hbt!]
\centering
\includegraphics[width=0.98\textwidth]{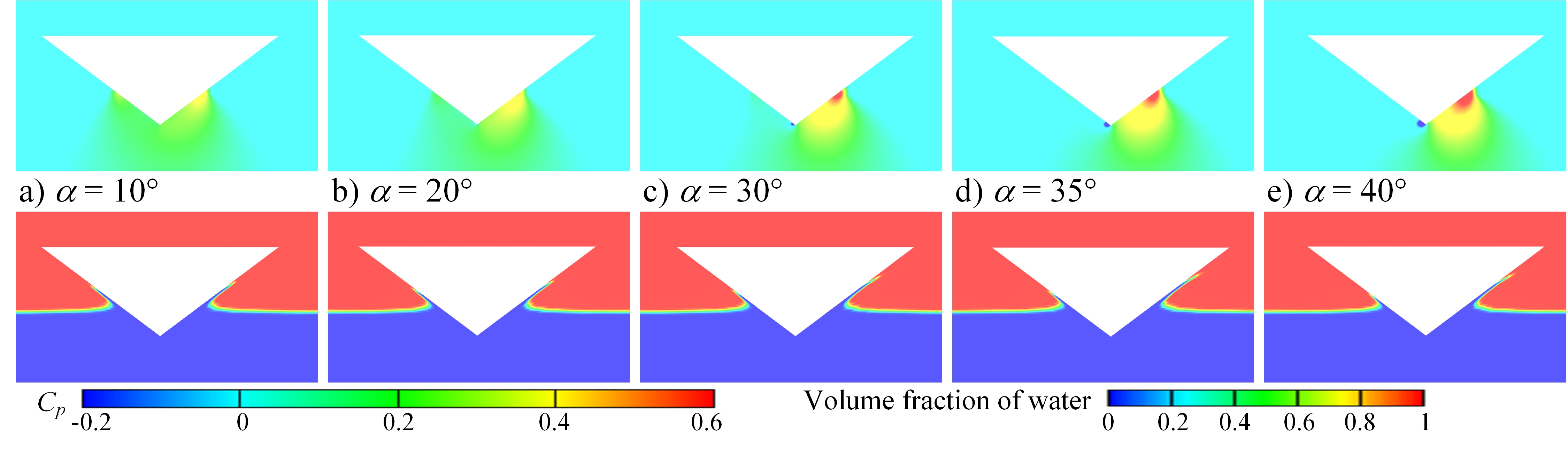}
\caption{Pressure distribution and water volume fraction for different velocity angle $\alpha$ when $a_x$ reaches its maximum.}
\label{fig:SameVzDifVxCpandVOF}
\end{figure}

\begin{figure}[hbt!]
\centering
\includegraphics[width=0.5\textwidth]{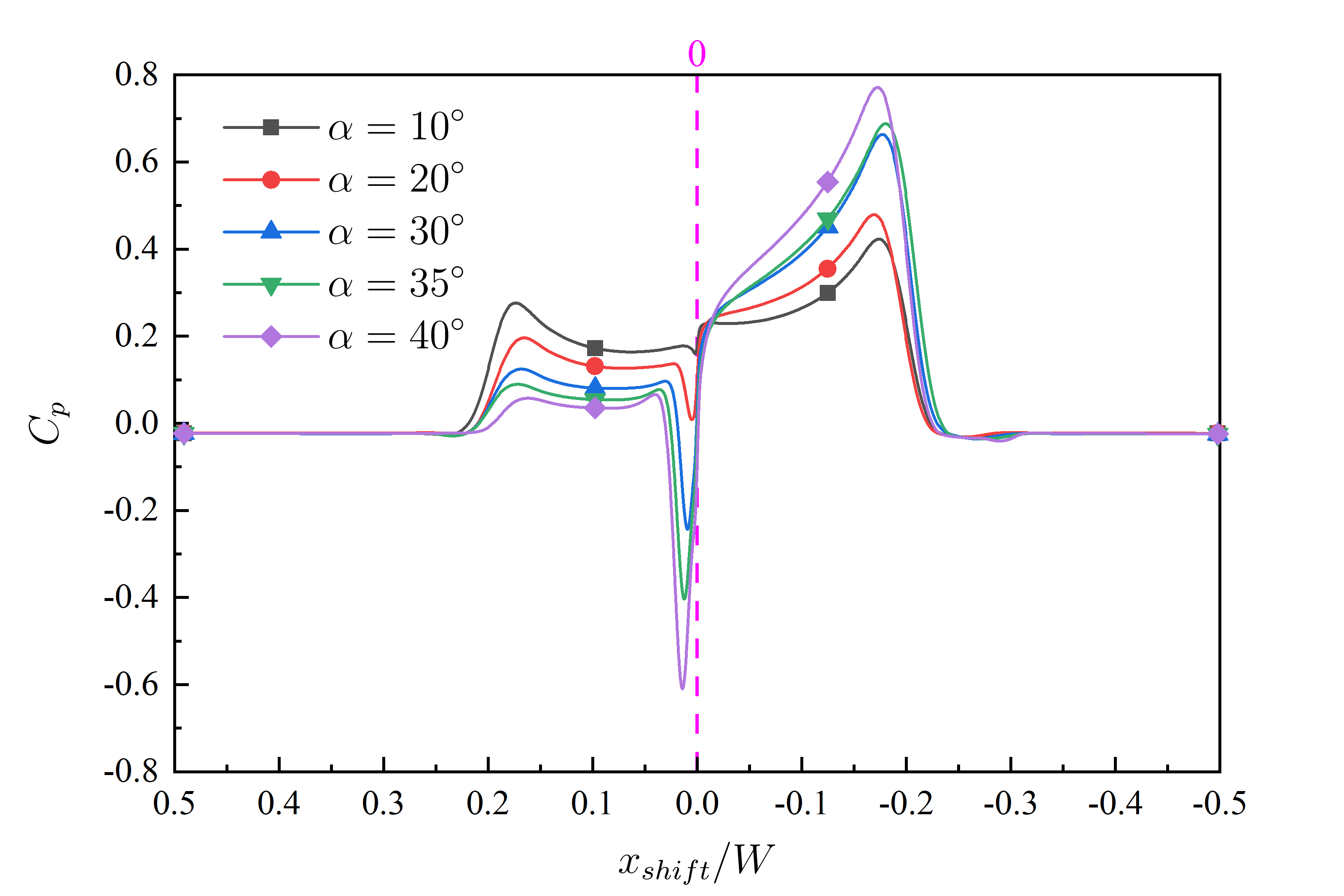}
\caption{Pressure coefficient along the normalized x-axis for different velocity angle $\alpha$ when $a_x$ reaches its maximum.}
\label{fig:SameVzDifVxCpline}
\end{figure}

In addition to the analysis of the effect of the initial vertical velocity on the load characteristics, for the oblique water entry of a wedge it is also significant to investigate the role played by the initial horizontal velocity. By assuming $\upsilon_{z0}$ constant and changing $\alpha$ to vary $\upsilon_{x0}$, similar to what done in the previous section, Fig.~\ref{fig:SameVzDifVx_axandaz} presents the time histories of $a_x$ and $a_z$ exerted on the wedge at various velocity angle $\alpha$, using the fixed vertical velocity component $\upsilon_{z0}$ = 2.943 m/s, derived from the previous case of $\upsilon_{x0}$ = 1.071 m/s and $\alpha$ = 20$^\circ$. As it can be seen, the value of $a_x$ exhibits an obvious decreasing trend when reducing $\alpha$ upon water impact, whereas no changes are observed in $a_z$, significantly differing from the situations of varying initial vertical velocity. Therefore, the data of $a_{x\mathrm{max}}$ are extracted and compared with three different functions of $\upsilon_{x0}$ as illustrated in Fig.~\ref{fig:SameVzDifVx_axmax}. It is interesting to note that the data fit well with a linear function, although the function is established between $a_{x\mathrm{max}}$ and $\upsilon_{x0}$, instead of $\upsilon_{x0}^2$, which is remarkably different from cases of varying $\upsilon_{z0}$. The pressure contour plots around the wedge with variable $\alpha$, when $a_{x\mathrm{max}}$ is achieved, are depicted in the upper side of Fig.~\ref{fig:SameVzDifVxCpandVOF} and Fig.~\ref{fig:SameVzDifVxCpline}, where the pressure coefficient $C_p$ is defined as $C_p=(p-p_0)/[0.5\rho(\upsilon_{z0}^2+\upsilon_{x0}^2)]$, and the value of $\upsilon_{x0}$ is referring to the initial horizontal velocity in the case of $\alpha$ = 40$^\circ$. It can be seen that a higher-pressure region occurs at the right-hand side of the wedge, whereas a zone with negative pressure is observed on the left, leading to the variation of $a_x$. It is therein evidenced that the pressure field varies significantly in the range $\alpha$ $\in$ [10$^\circ$, 40$^\circ$], when $a_x$ reaches the peak value. The comparison between Fig.~\ref{fig:SameVzDifVxCpandVOF} and Fig.~\ref{fig:SameVzDifVxCpline} indicates that the water jets originate from the pressure peak, and the low-pressure zone is close to the apex which is consistent with \citep{riccardi2004water,judge2004initial}. Furthermore, flow separation could be expected at the apex which can also lead to cavitation or ventilation due to horizontal-vertical impact velocity \citep{judge2004initial}, provided that fluid dynamic solution method is able to model cavitation and ventilation phenomena.

\begin{figure}[hbt!]
\centering
\begin{subfigure}{0.49\textwidth}
\includegraphics[width=\linewidth]{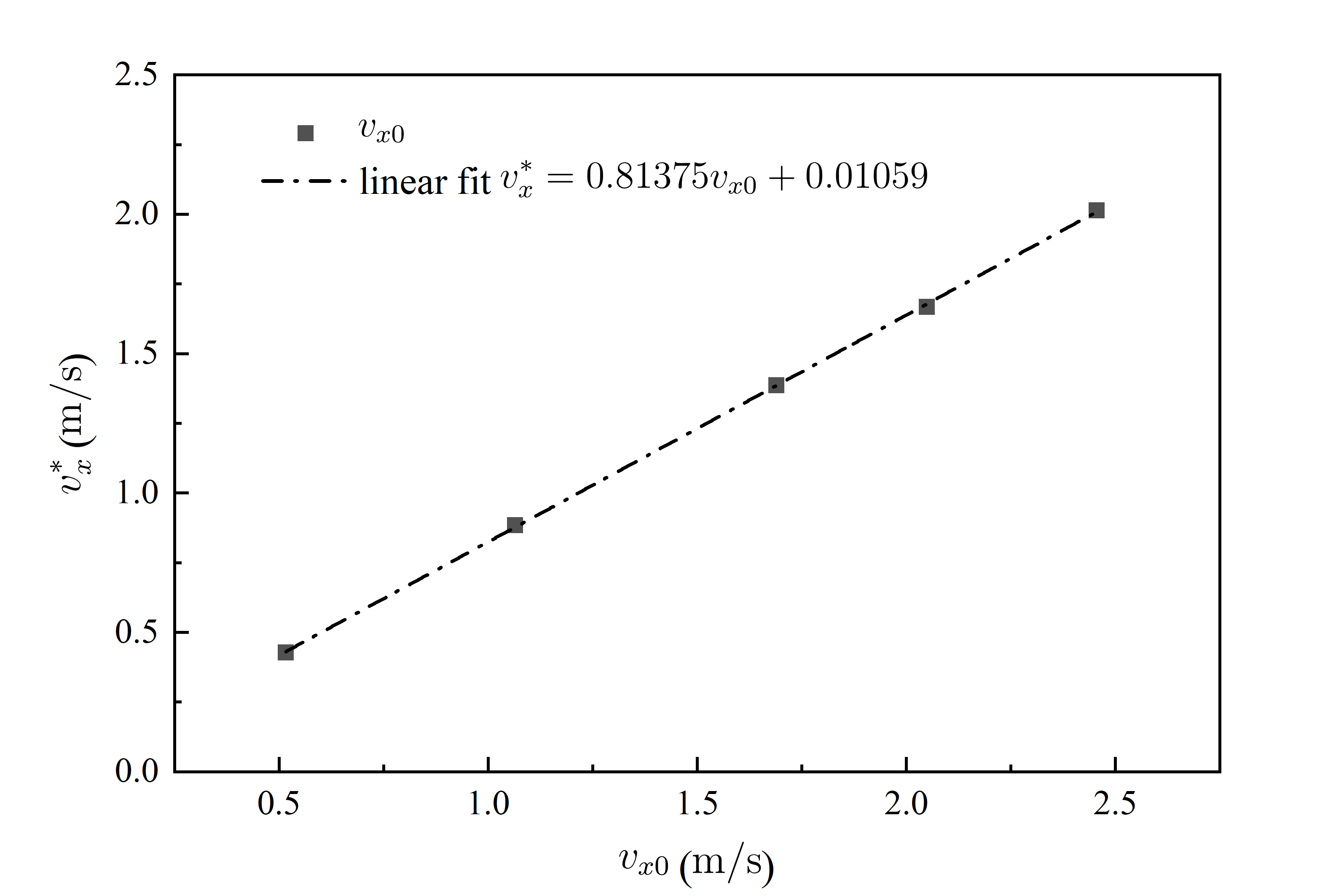} 
\caption{}
\label{fig:SameVzDifVxVxKTXZa}
\end{subfigure}
\begin{subfigure}{0.49\textwidth}
\includegraphics[width=\linewidth]{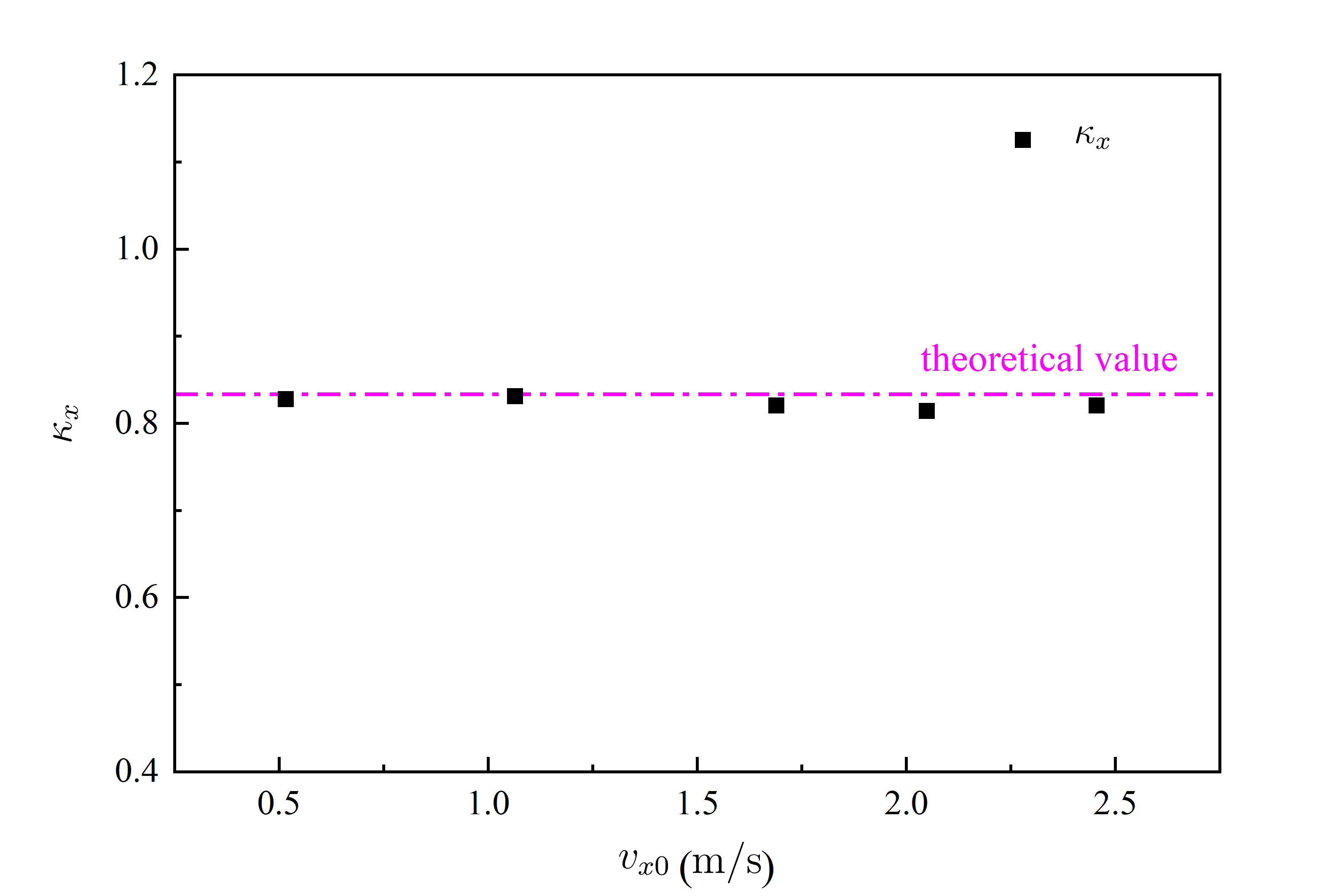}
\caption{}
\label{fig:SameVzDifVxVxKTXZb}
\end{subfigure}
\begin{subfigure}{0.49\textwidth}
\includegraphics[width=\linewidth]{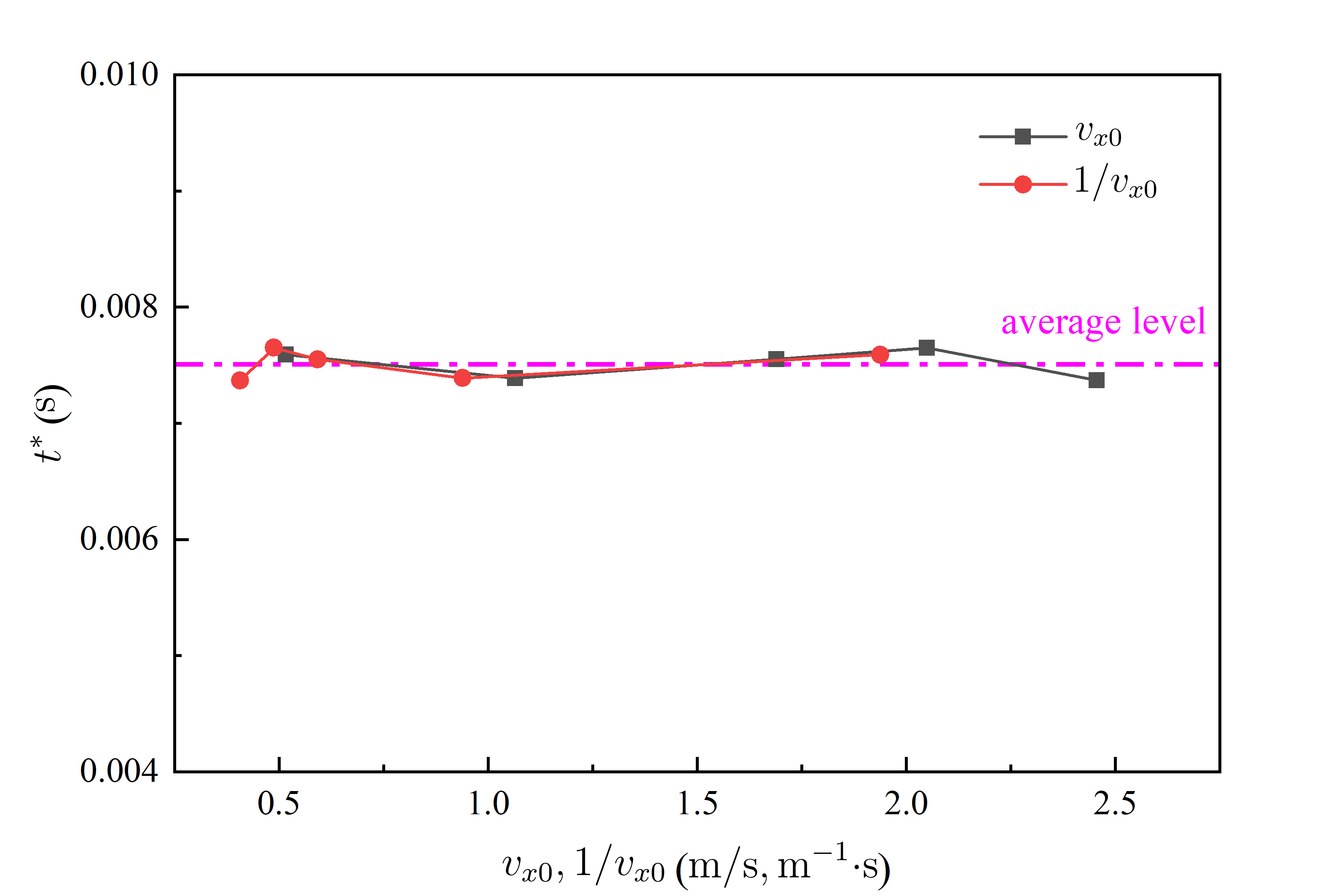} 
\caption{}
\label{fig:SameVzDifVxVxKTXZc}
\end{subfigure}
\begin{subfigure}{0.49\textwidth}
\includegraphics[width=\linewidth]{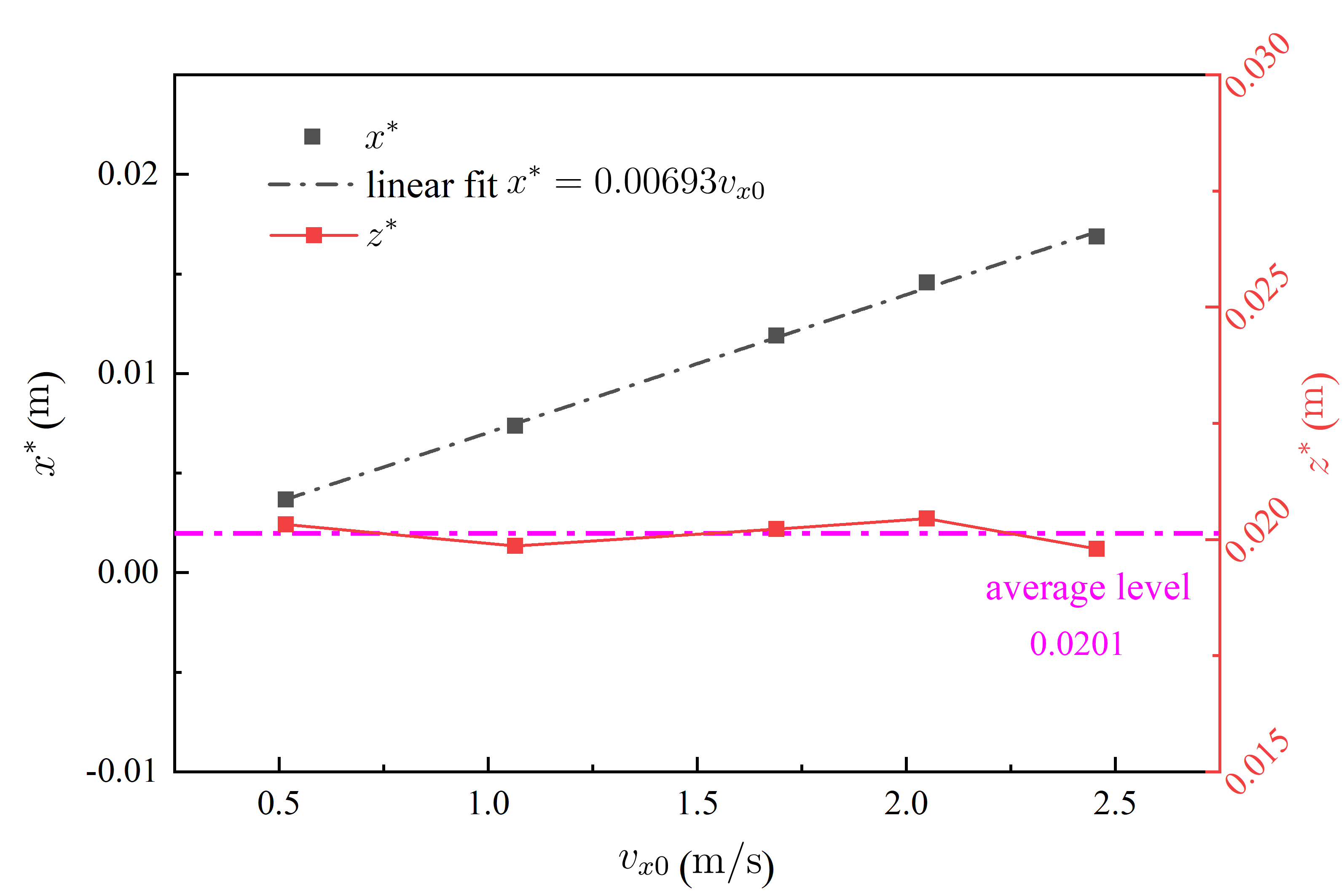}
\caption{}
\label{fig:SameVzDifVxVxKTXZd}
\end{subfigure}
\caption{Effect of initial horizontal velocity on variable dynamic parameters: (a) $\upsilon_x^*$; (b) $\kappa_x$; (c) $t^*$; (d) $x^*$ and $z^*$.}
\label{fig:SameVzDifVxVxKTXZ}
\end{figure}

In order to achieve a better comprehension of the effect of $\upsilon_{x0}$ on the impact dynamics, the value of the horizontal velocity component $\upsilon_{x}^*$, the ratio of velocity $\kappa_x$ and time $t^*$ at which $a_x$ reaches its maximal value are provided in Fig.~\ref{fig:SameVzDifVxVxKTXZ}. The results are shown for five distinct cases. It’s worth noticing that $\upsilon_{x}^*$ varies linearly with $\upsilon_{x0}$, as $\upsilon_{x}^*=0.81375\upsilon_{x0}+0.01059$. The parameter $k$ is numerically lower than the analytical one provided by Eq.~\eqref{eq:threepara}, as it is shown in Fig.~\ref{fig:SameVzDifVxVxKTXZb}. Nonetheless, the error of $k$ compared to theoretical estimate is -2.35\% with a root mean squared error (RMSE) on $\kappa$ is 0.0122, thus indicating the theory about $\upsilon_{z}^*-\upsilon_{z0}$ derived from vertical entry can be used. Moving to Fig.~\ref{fig:SameVzDifVxVxKTXZc}, the trend is quite different from the linear function displayed in Eq.~\eqref{eq:threepara_t} and Fig.~\ref{fig:DifvxOnTZVzKappaa} for the case of 2D wedge with various vertical velocities. The numerical values of $t^*$ are almost constant for both cases of $\upsilon_{x0}$ and $\upsilon_{x0}^{-1}$, and the standard deviation $\sigma$ of these data is $1.11\times10^{-4}$. The curves presented in Fig.~\ref{fig:SameVzDifVxVxKTXZd} demonstrate that $x^*$ is proportional to $\upsilon_{x0}$, expressed as $x^*=0.00693\upsilon_{x0}$, and the results of $z^*$ oscillate slightly around 0.0221 associated with $2.56\times10^{-4}$ in $\sigma$. The above result is confirmed by the lower part of Fig.~\ref{fig:SameVzDifVxCpandVOF}, where no substantial differences for vertical displacement are observed. In general, linear functions can be found on $a_{x\mathrm{max}}-\upsilon_{x0}$ and $\upsilon_x^{*}-\upsilon_{x0}$ (see Fig.~\ref{fig:SameVzDifVx_axmax} and Fig.~\ref{fig:SameVzDifVxVxKTXZa}), except for the relationship between the corresponding time $t^*$ and the initial horizontal velocity $\upsilon_{x0}$ where a constant trend is observed in Fig.~\ref{fig:SameVzDifVxVxKTXZc}.

\subsection{A cabin section in 3D}

\begin{figure}[hbt!]
\centering
\begin{subfigure}{0.49\textwidth}
\includegraphics[width=\linewidth]{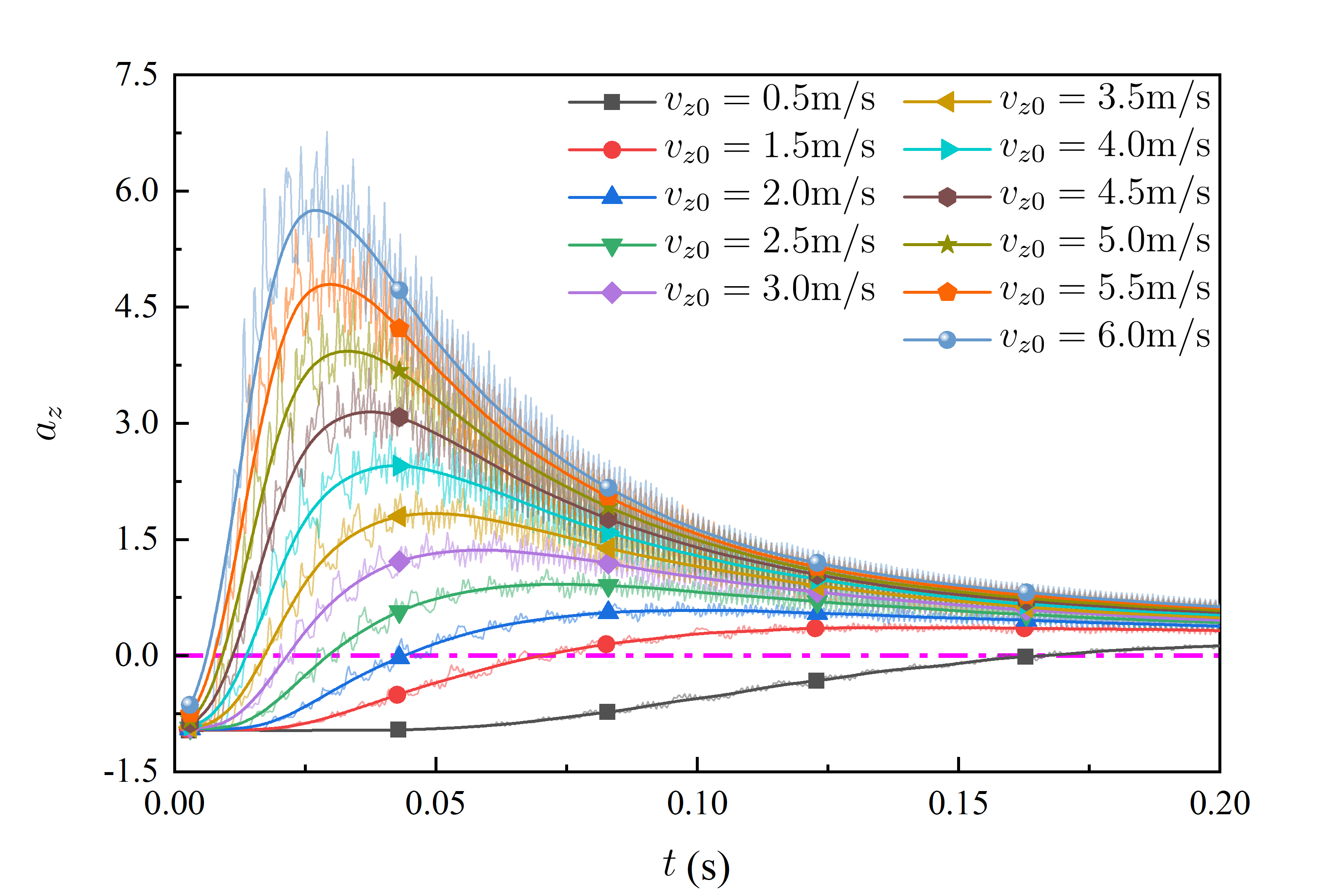} 
\caption{}
\label{fig:CabinSectionAzandCpa}
\end{subfigure}
\begin{subfigure}{0.49\textwidth}
\includegraphics[width=\linewidth]{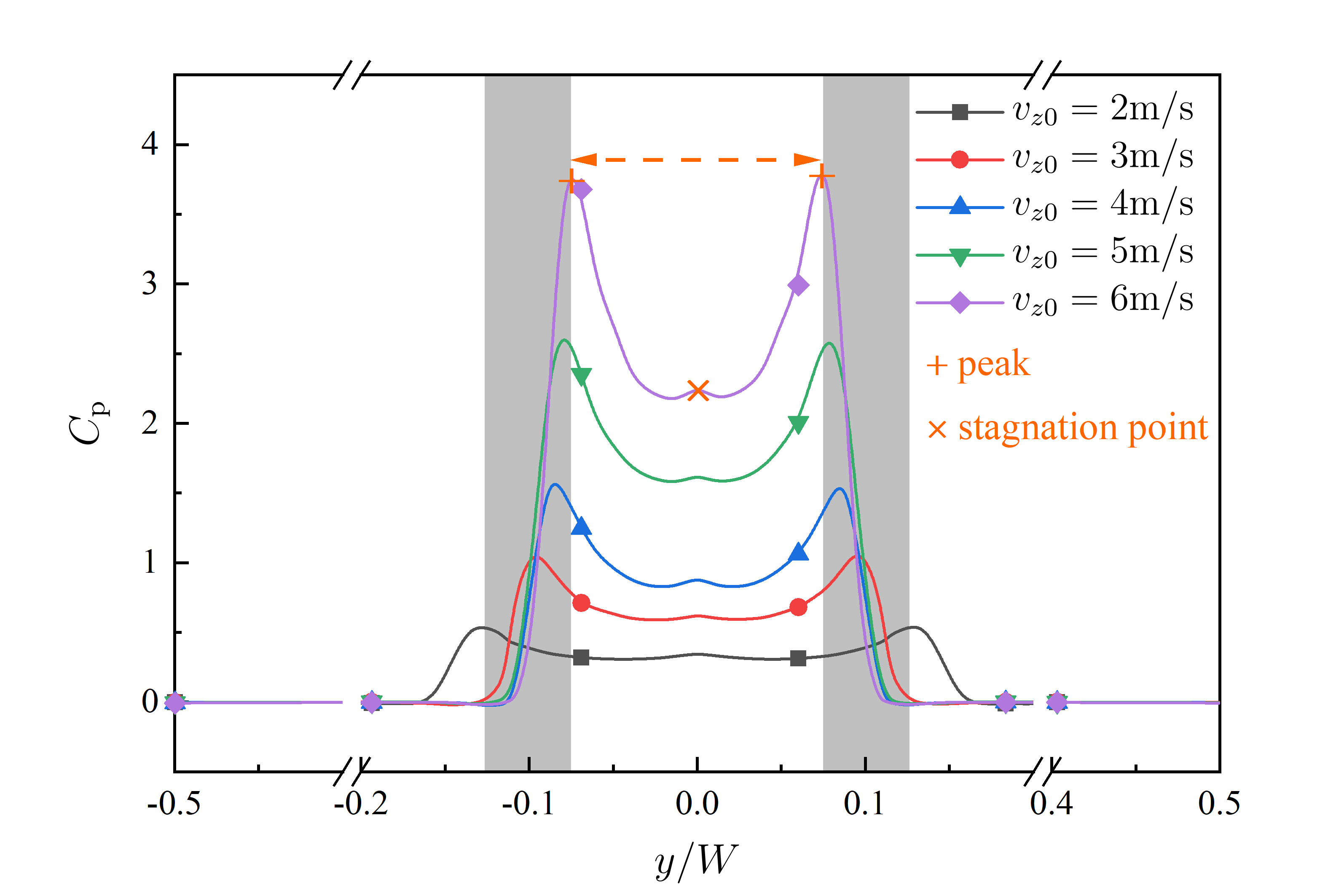}
\caption{}
\label{fig:CabinSectionAzandCpb}
\end{subfigure}
\caption{\textcolor{black}{In the case of a 3D cabin section:} (a) Time histories of \textcolor{black}{dimensionless} acceleration $a_z$ with different initial vertical velocity $\upsilon_{z0}$; (b) pressure coefficient at 0.5$L$ with different $\upsilon_{z0}$.}
\label{fig:CabinSectionAzandCp}
\end{figure}

The above results prove that it is possible to evaluate the load characteristics with the help of the linear relations, proposed in Eq.~\eqref{eq:amax}, \eqref{eq:threepara} and Eq.~\eqref{eq:threepara_t}, with large initial vertical velocity. This section presents the results of computational simulations of the vertical free fall of a cabin section (see Fig.~\ref{fig:CabinSection}), entering the free surface with various initial vertical velocity $\upsilon_{z0}$. Eleven cases with a series of $\upsilon_{z0}$ from 0.5 m/s to 6 m/s are simulated. Fig.~\ref{fig:CabinSectionAzandCpa} shows the evolution of $a_z$ acting on the cabin during the water entry. It is worth noting that the results have been filtered with a cutoff frequency 62.5 Hz. At the beginning of the impacting, the overall acceleration is negative indicating that gravity dominates and leads to an increase in the vertical velocity, while the hydrodynamic force only plays an auxiliary role at the onset of entry. As the body penetrates into the water, $a_z$ turns positive and reaches its peak value subsequently, which means the hydrodynamic force is dominant. Obviously, $a_z$ is linked with the initial impact velocity. The smaller $\upsilon_{z0}$ is, the smoother the trend of $a_z$ will be, until a point where the peak disappear. Such a behaviour can be also observed in Fig.~\ref{fig:CabinSectionAzandCpb}, where the pressure coefficient $C_p=(p-p_0)/(0.5\rho\upsilon_{z0}^2))$ is computed along the wetted part of the body at 0.5$L$ with $\upsilon_{z0}$ chosen as 6 m/s. As the initial impact velocity increases, the overall values of $C_p$ become higher for selected five cases with different initial vertical velocity, as shown in Fig.~\ref{fig:CabinSectionAzandCpb}, where three extreme values can be observed. One extreme value is at $y$=0, the apex of the body, so-called stagnation point, where the flow velocity is almost equal to zero, and the other two extreme values, marked with '+' in Fig.~\ref{fig:CabinSectionAzandCpb}, are inside the grey region. It can be seen that the distance between the peak points becomes narrower, and the difference is less pronounced as $\upsilon_{z0}$ grows.

\begin{figure}[hbt!]
\centering
\begin{subfigure}{0.49\textwidth}
\includegraphics[width=\linewidth]{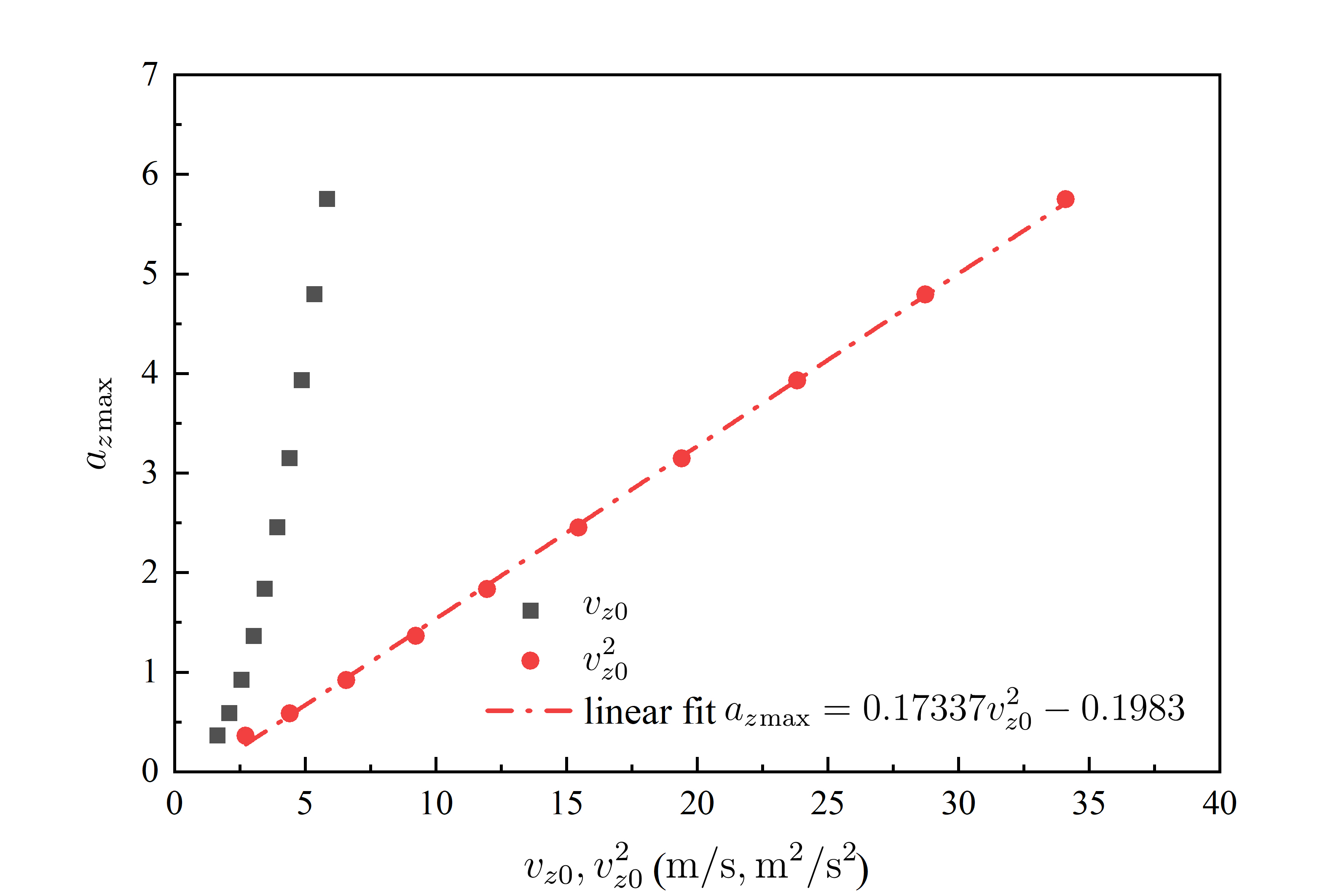} 
\caption{}
\label{fig:CabinSectionAzmaxandCpa}
\end{subfigure}
\begin{subfigure}{0.49\textwidth}
\includegraphics[width=\linewidth]{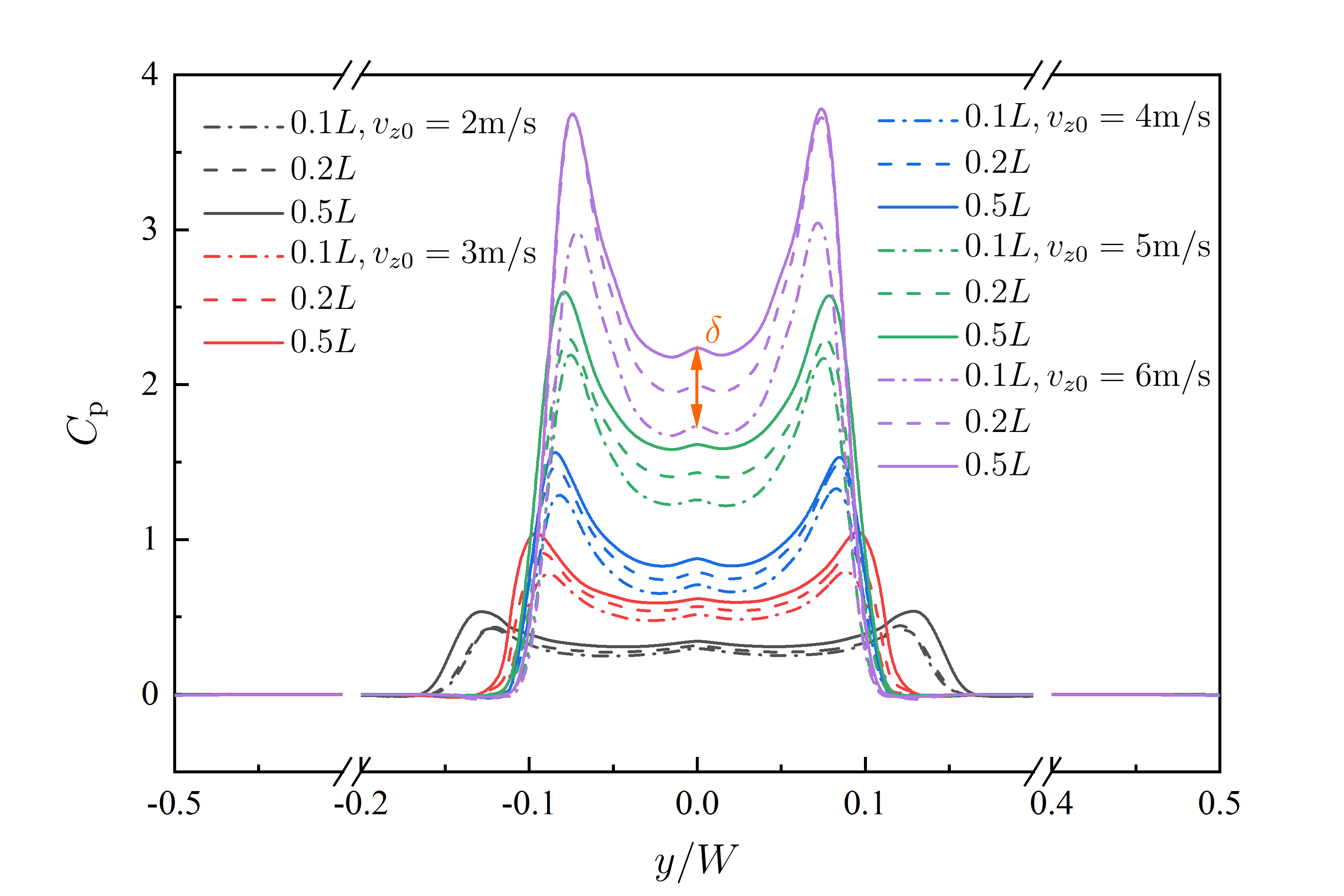}
\caption{}
\label{fig:CabinSectionAzmaxandCpb}
\end{subfigure}
\caption{\textcolor{black}{In the case of a 3D cabin section:} (a) Variation of $a_{z\mathrm{max}}$ versus $\upsilon_{z0}$ and $\upsilon_{z0}^2$; (b) pressure coefficient at three distinctive cross-setions for different $\upsilon_{z0}$.}
\label{fig:CabinSectionAzmaxandCp}
\end{figure}

Fig.~\ref{fig:CabinSectionAzmaxandCpa} demonstrates $a_{z\mathrm{max}}$ is still a linear function to $\upsilon_{z0}^2$, fitted by $a_{z\mathrm{max}} =0.1734\upsilon_{z0}^2-0.1983$, where $k$ is slightly lower than the theoretical one with -7.57\% error, as listed in Table~\ref{tab:compaatvCabin}. Fig.~\ref{fig:CabinSectionAzmaxandCpb} shows the results of $C_p$ at three distinctive cross-sections, viz., 0.1$L$, 0.2$L$ and 0.5$L$, for different values of $\upsilon_{z0}$, where the difference $\delta$ is caused by the three dimensional effects and introduces a difference between the numerical and the theoretical solution. The data show that the value of $\delta$ becomes larger as $\upsilon_{z0}$ rises, denoting more significant three-dimensional effects.

\begin{table}[width=.9\linewidth, pos=hbt!]
\caption{Comparison between theoretical estimate and numerical results for cabin section}
\centering
\label{tab:compaatvCabin}
\begin{tabular*}{\tblwidth}{@{}cccccccccc@{}}
\toprule
 & \multicolumn{3}{c}{$a_{z\mathrm{max}}$}& \multicolumn{3}{c}{$t^*, \mathrm{s}$}& \multicolumn{3}{c}{$\upsilon_z^*, \mathrm{m/s}$}\\
\midrule
 & $k$& err, \%& $b$& $k$& err, \%& $b$& $k$& err, \%& $b$\\
Theoretical value& 0.1876& -& -& 0.1343& -& -& 0.8333& -& -\\
Present study& 0.1734& -7.57& -0.1983& 0.2208& 64.41& -0.0215& 0.8135& -2.38& 0.2793\\
\bottomrule
\end{tabular*}
\end{table}

\begin{figure}[hbt!]
\centering
\begin{subfigure}{0.49\textwidth}
\includegraphics[width=\linewidth]{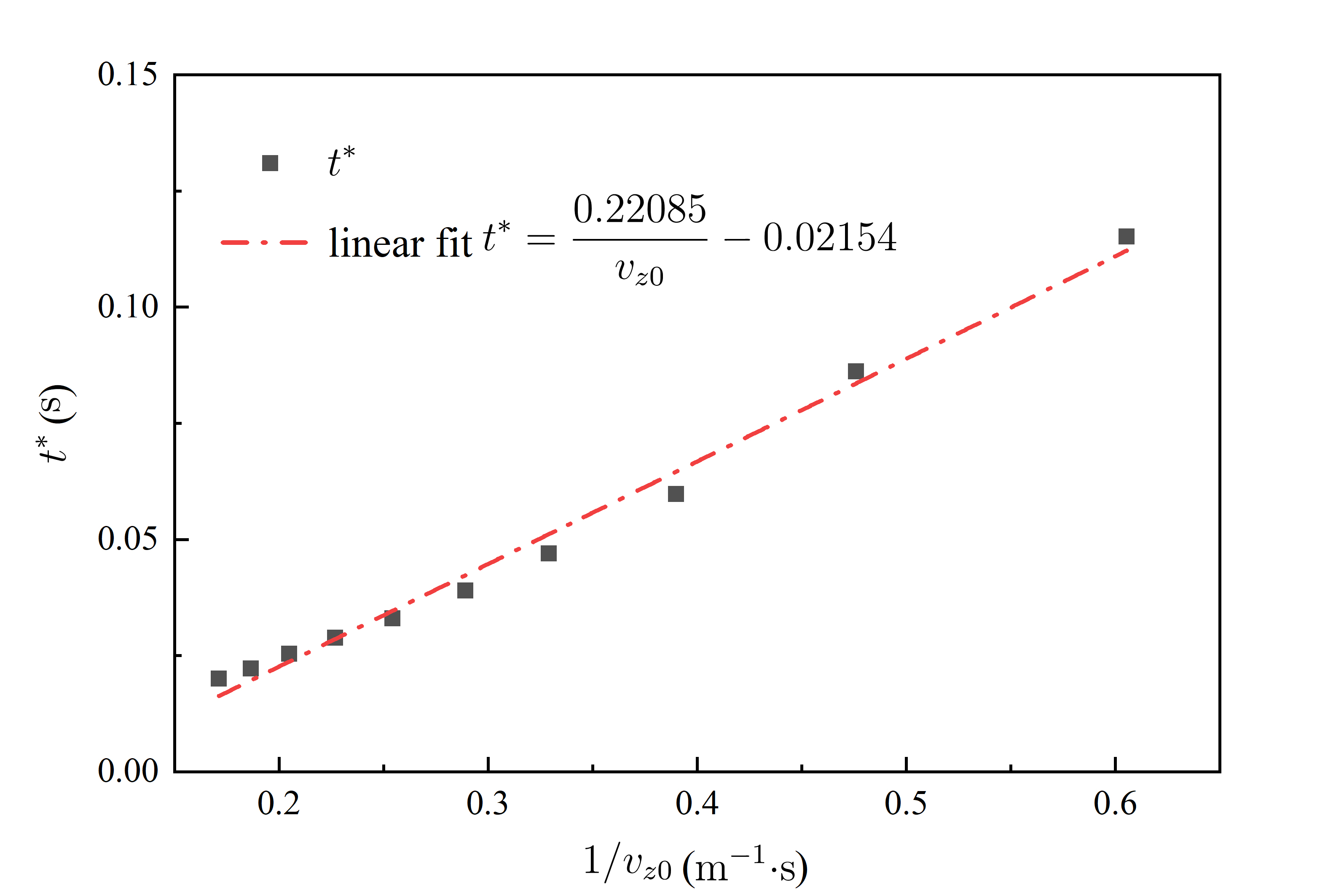} 
\caption{}
\label{fig:CabinTZVZKaa}
\end{subfigure}
\begin{subfigure}{0.49\textwidth}
\includegraphics[width=\linewidth]{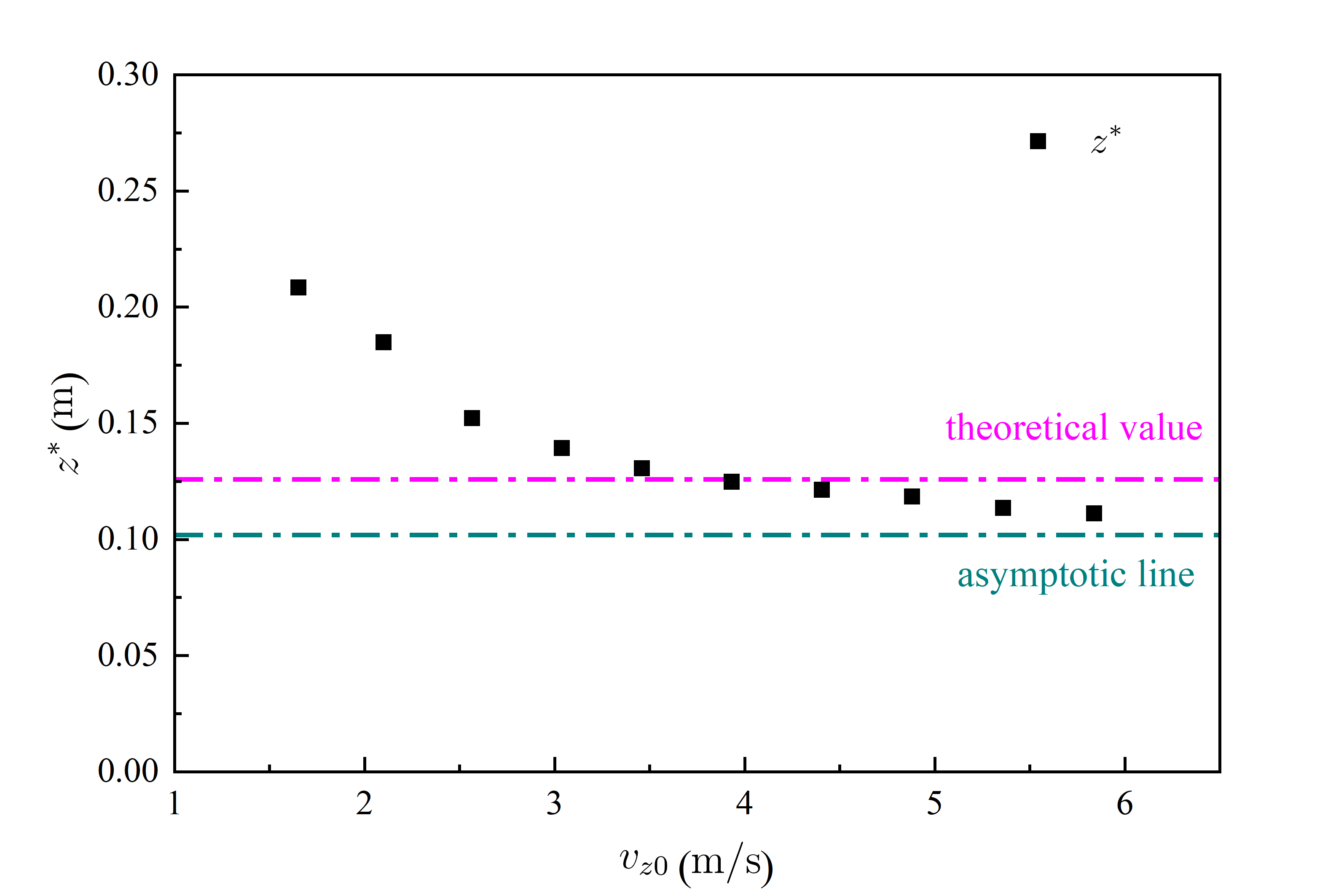}
\caption{}
\label{fig:CabinTZVZKab}
\end{subfigure}
\begin{subfigure}{0.49\textwidth}
\includegraphics[width=\linewidth]{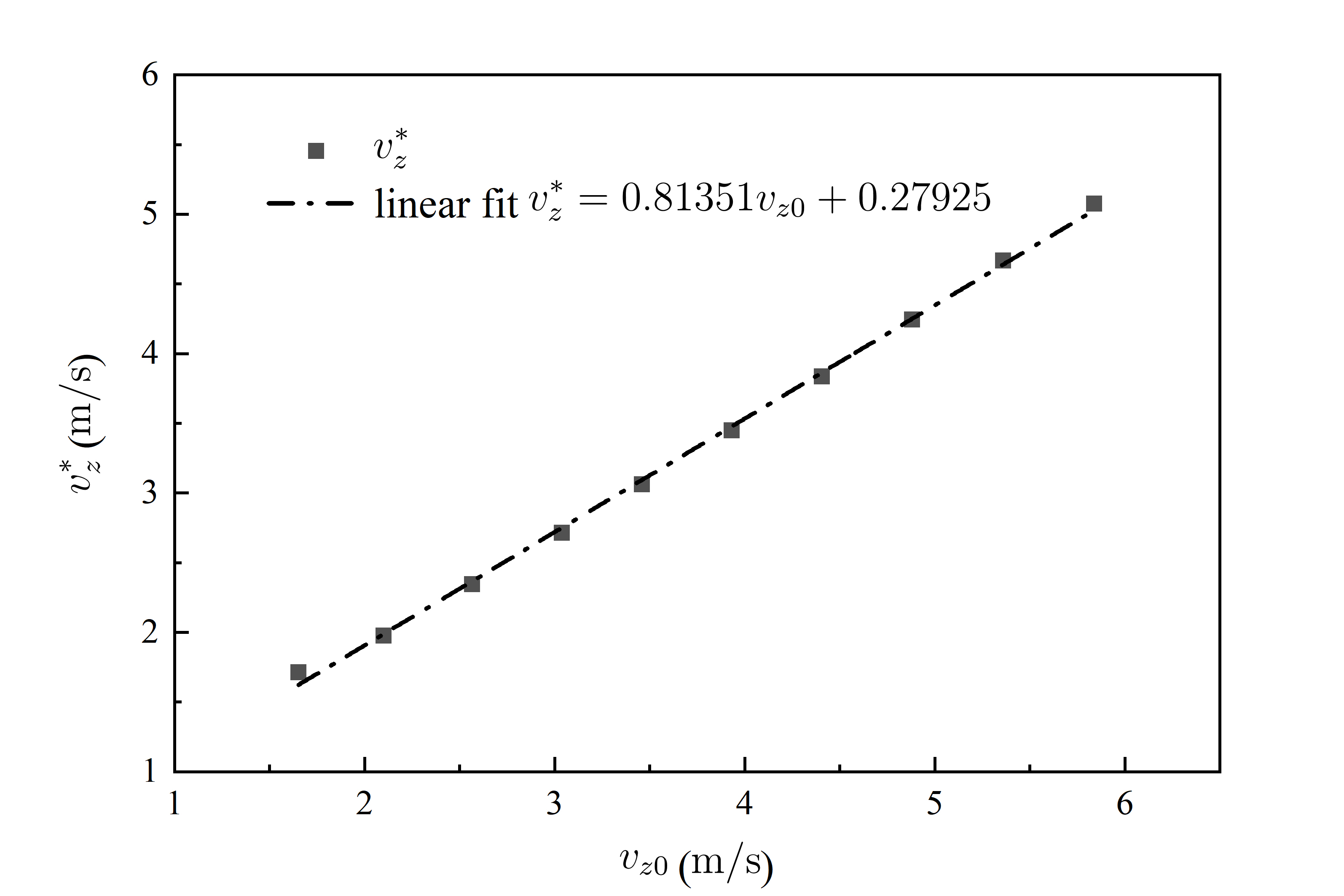} 
\caption{}
\label{fig:CabinTZVZKac}
\end{subfigure}
\begin{subfigure}{0.49\textwidth}
\includegraphics[width=\linewidth]{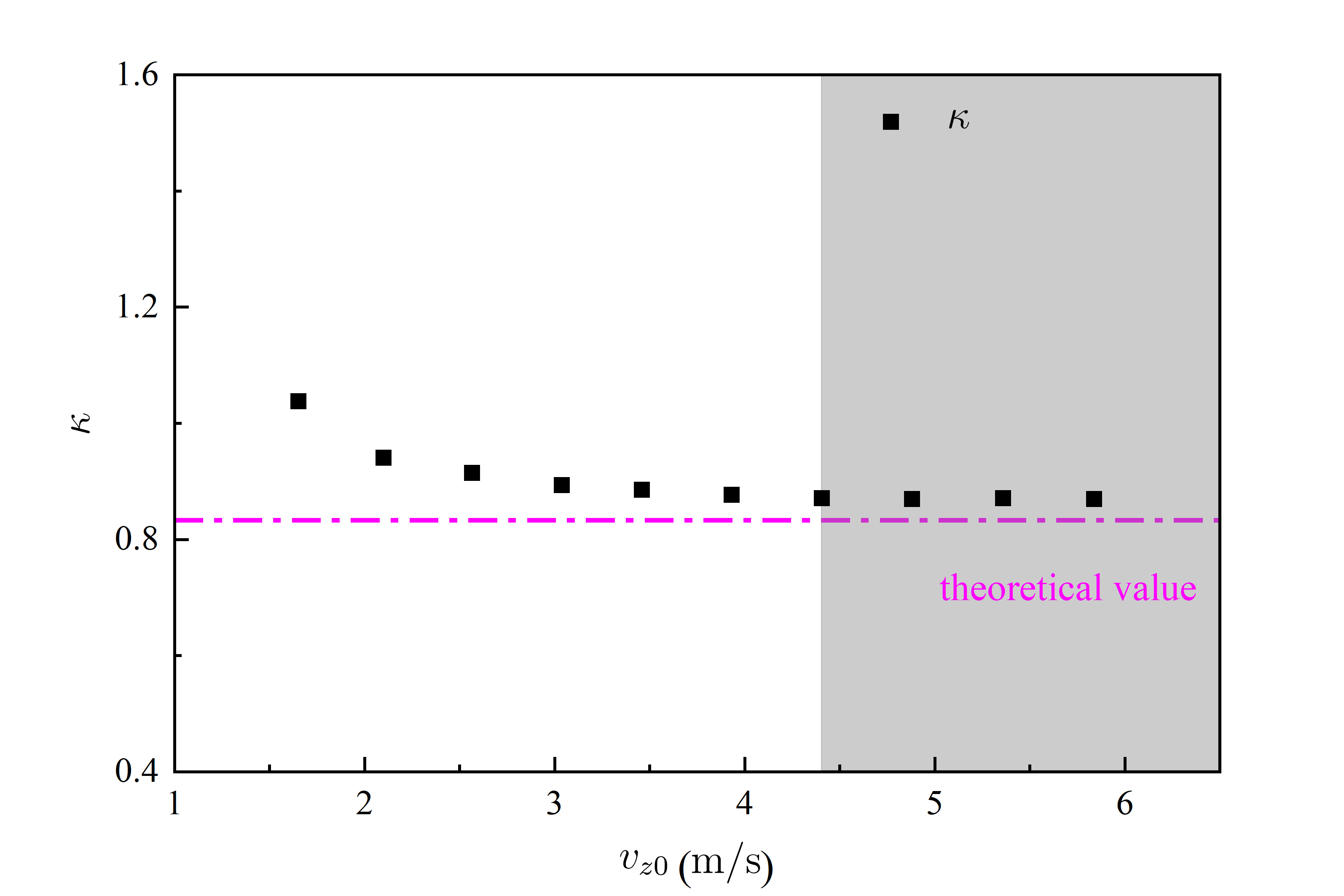}
\caption{}
\label{fig:CabinTZVZKad}
\end{subfigure}
\caption{Effect of the initial vertical velocity on the different parameters for the case of a 3D cabin section: (a) $t^*$; (b) $z^*$; (c) $\upsilon_z^*$; (d) $\kappa$.}
\label{fig:CabinTZVZKa}
\end{figure}

The relations about the other dynamic parameters in 3D cabin section are shown in Fig.~\ref{fig:CabinTZVZKa}. As can be seen, the corresponding time $t^*$ is a linear function of $\upsilon_{z0}^{-1}$ in Fig.~\ref{fig:CabinTZVZKaa}, although the numerical estimate of the parameter $k$ is 64.41\% different from the theoretical prediction, as also observed in the case of oblique entry of a symmetric wedge. Looking into the penetration depth $z^*$, there is a slight difference between the numerical results and the theoretical prediction, however, a new asymptotic line, lying below the theoretical one, appears and all data approach it asymptotically when increasing $\upsilon_{z0}$. It means that the maximum acceleration of the 3D cabin section occurs at a smaller depth due to the three-dimensional effects on slamming load \citep{wang2021assess} and pile-up effects. Another significant parameter to characterize the impact is the corresponding velocity $\upsilon_z^*$ as shown in Fig.~\ref{fig:CabinTZVZKac}, which displays a linear relation with $\upsilon_{z0}$. Specifically, as seen in  Fig.~\ref{fig:CabinTZVZKad}, the value of $\kappa$ approaches the theoretical line only for $\upsilon_{z0}$ greater than 4.5 m/s, whereas the large difference are observed for smaller initial impact velocities.

\begin{figure}[hbt!]
\centering
\begin{subfigure}{0.49\textwidth}
\includegraphics[width=\linewidth]{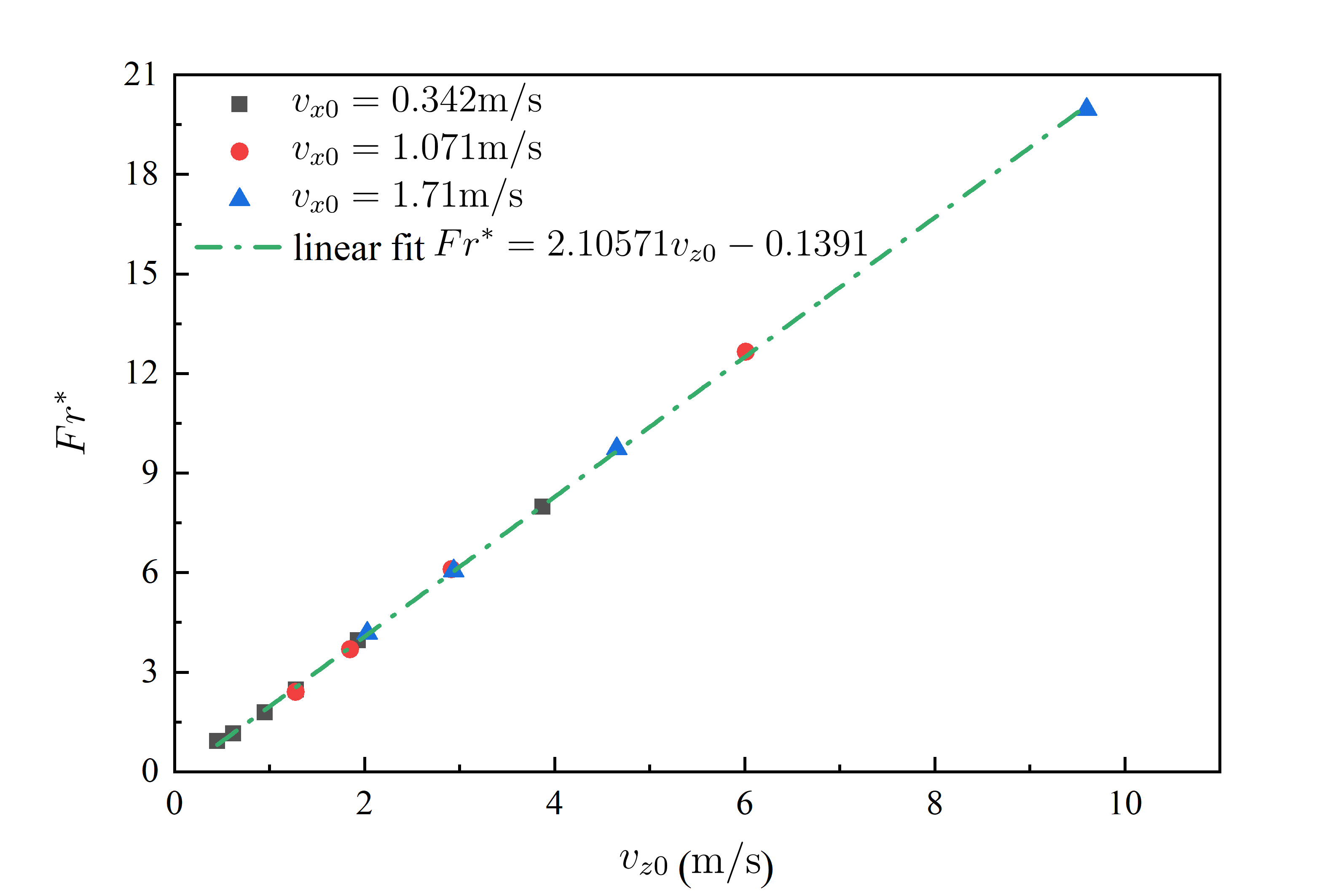} 
\caption{}
\label{fig:Frstara}
\end{subfigure}
\begin{subfigure}{0.49\textwidth}
\includegraphics[width=\linewidth]{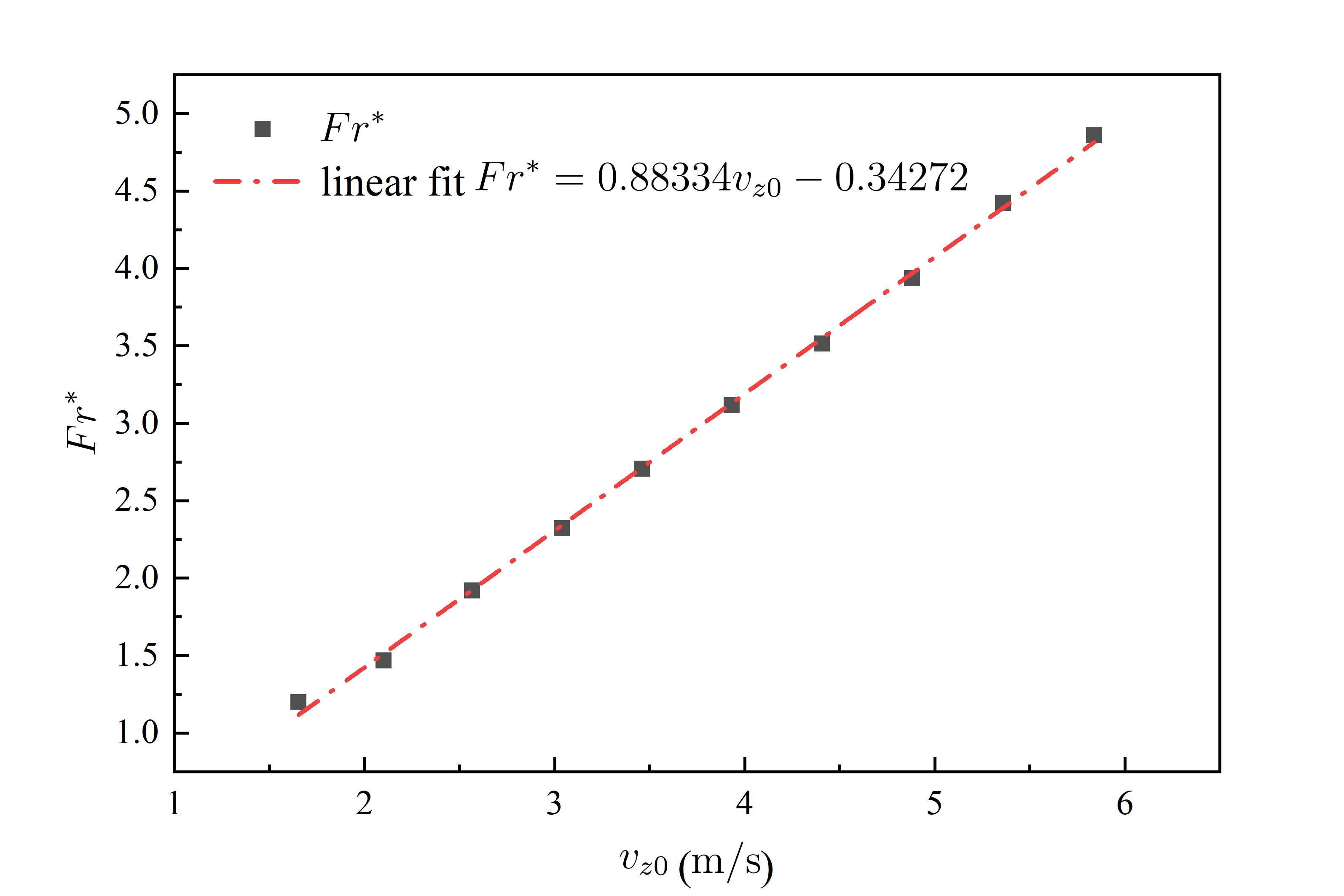}
\caption{}
\label{fig:Frstarb}
\end{subfigure}
\caption{The instantaneous Froude number as a linear function of the initial velocity: (a) 2D wedge; (b) 3D cabin section.}
\label{fig:Frstar}
\end{figure}

Subsequently, the instantaneous Froude number \citep{hulin2022gravity}, $Fr^*=\upsilon_z^*/\sqrt{gz^*}$, is introduced here to describe the combined relations between velocity and penetration depth, when the maximum value of acceleration is reached. As it can be seen in Fig.~\ref{fig:Frstar}, $Fr^*$ is found proportional to the initial vertical velocity $\upsilon_{z0}$, in the case of 2D wedge and 3D cabin section. The proportional relation can also be derived from Eq.~\eqref{eq:threepara}, where $z^*$ is independent of $\upsilon_{z0}$ and $\upsilon_z^*$ is considered as a linear function of $\upsilon_{z0}$. Being $a_{z\mathrm{max}}$ a linear function of $\upsilon_{z0}^2$, the relationship between maximum acceleration and the instantaneous Froude number $Fr^*$ can be easily established through Eq.~\eqref{eq:amax},~\eqref{eq:threepara}, as follows:

\begin{equation}
\label{eq:amaxandFr}
\dfrac{a_{z\mathrm{max}}}{(Fr^*)^2}=\dfrac{a_{z\mathrm{max}}\cdot gz^*}{(\upsilon_z^*)^2}=\dfrac{1}{3}
\longrightarrow a_{z\mathrm{max}} = \dfrac{1}{3} \cdot (Fr^*)^2
\end{equation}
The detailed results of 2D wedge and 3D cabin section are fitted and summarized in Table \ref{tab:FunctionAzmaxFr}. It can be seen that the numerical relations agree well with the theoretical prediction, although in 3D case the value of the slope, $k$, displays an obvious deviation associated with the three-dimensional effect. Moreover, Eq.~\eqref{eq:amaxandFr} can also be written as:

\begin{equation}
\label{eq:amaxvzandz}
\dfrac{a_{z\mathrm{max}}\cdot gz^*}{(\upsilon_z^*)^2}=\dfrac{1}{3}
\longrightarrow a_{z\mathrm{max}} = \dfrac{(\upsilon_z^*)^2}{3g\cdot z^*}
\end{equation}
providing a strong coupled relation among $a_{z\mathrm{max}}$, $\upsilon_z^*$ and $z^*$, instead of three separate expressions ( see Eq.~\eqref{eq:amax} and~\eqref{eq:threepara} ).

\begin{table}[width=.65\linewidth, pos=hbt!]
\caption{Function of $a_{z\mathrm{max}}$ and $Fr^*$ \textcolor{black}{derived from theoretical estimate and numerical results for a 2D wedge and a 3D cabin section}.}
\centering
\label{tab:FunctionAzmaxFr}
\begin{tabular*}{\tblwidth}{@{}ccc@{}}
\toprule
 & Expression& Error of $k$, \% \\
\midrule
Theoretical estimation& $a_{z\mathrm{max}}=\dfrac{1}{3}\cdot (Fr^*)^2$& -\\
2D wedge& $a_{z\mathrm{max}}=0.32142\cdot (Fr^*)^2+0.009169$& -3.57\\
3D cabin section& $a_{z\mathrm{max}}=0.24417\cdot (Fr^*)^2-0.05474$& -26.75\\
\bottomrule
\end{tabular*}
\end{table}

\subsection{V-shaped hull on amphibious aircraft}

Herein, the quantitative relations discussed above, (Eq.~\eqref{eq:amax}, \eqref{eq:threepara} and~\eqref{eq:threepara_t}), are employed to examine the effect of initial vertical velocity $\upsilon_{z0}$ on the load characteristics for the water landing of the V-shaped hull on amphibious aircraft (see Fig.~\ref{fig:SeaplaneConfig}). A set of numerical simulations is carried out with constant horizontal flight velocity $\upsilon_{x0}$= 37 m/s, which is determined as $\upsilon_{x0}=0.94\sqrt{2G/\rho S C_\mathrm{L}}$, where $G$ is the weight of aircraft , $S$ wing area  and $C_\mathrm{L}$ lift coefficient  regarding to the landing scenario \citep{lu2021effects}. The initial pitch $\theta_0$ is set as 7$^\circ$ which is considered as the suitable angle for landing event in the previous study \citep{lu2021effects}. Both the fixed and free pitch conditions have been simulated in the present study. Note that the wing components are taken into consideration in the present study. 

\begin{figure}[hbt!]
\centering
\begin{subfigure}{0.49\textwidth}
\includegraphics[width=\linewidth]{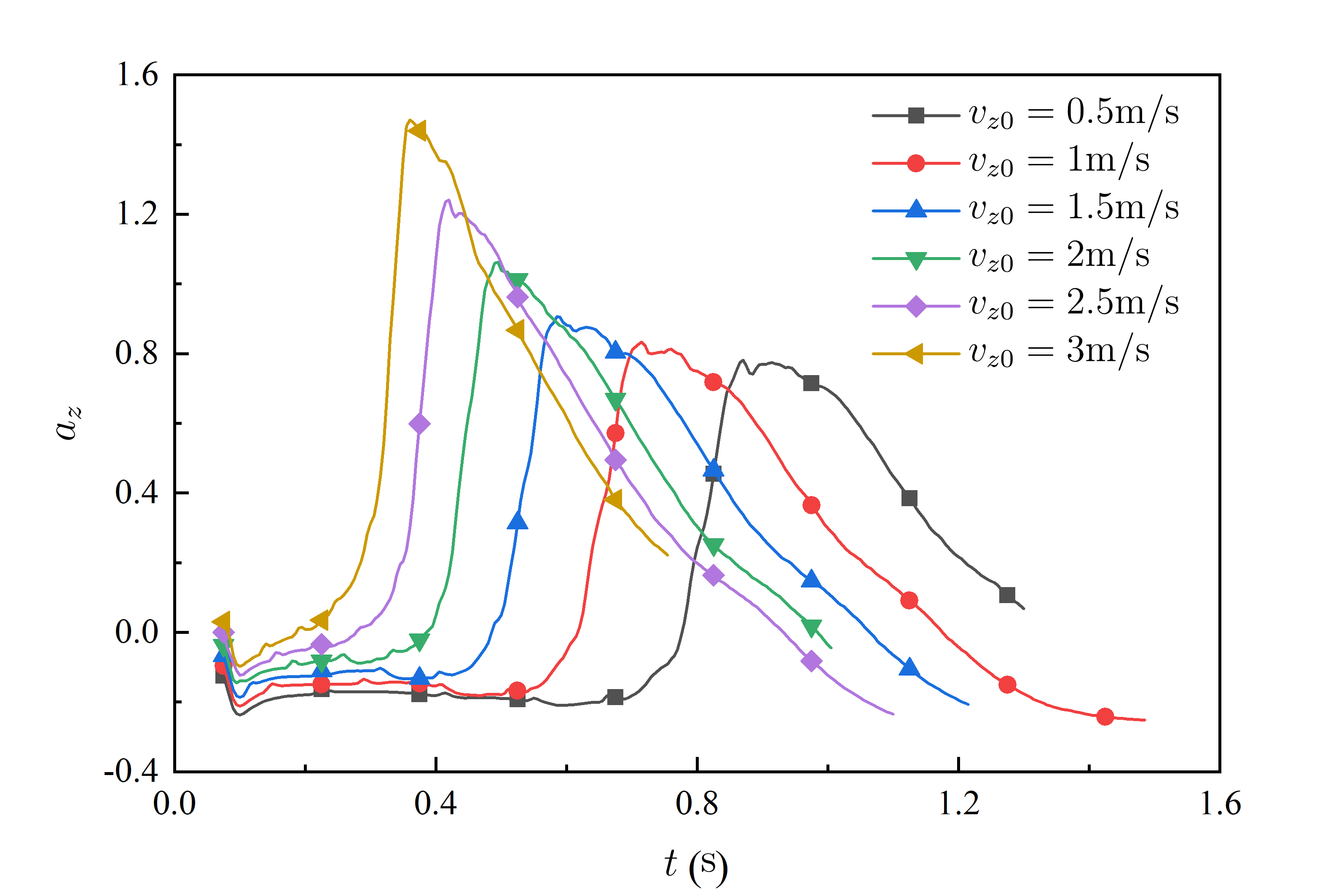} 
\caption{}
\label{fig:SeaFixUnfixAza}
\end{subfigure}
\begin{subfigure}{0.49\textwidth}
\includegraphics[width=\linewidth]{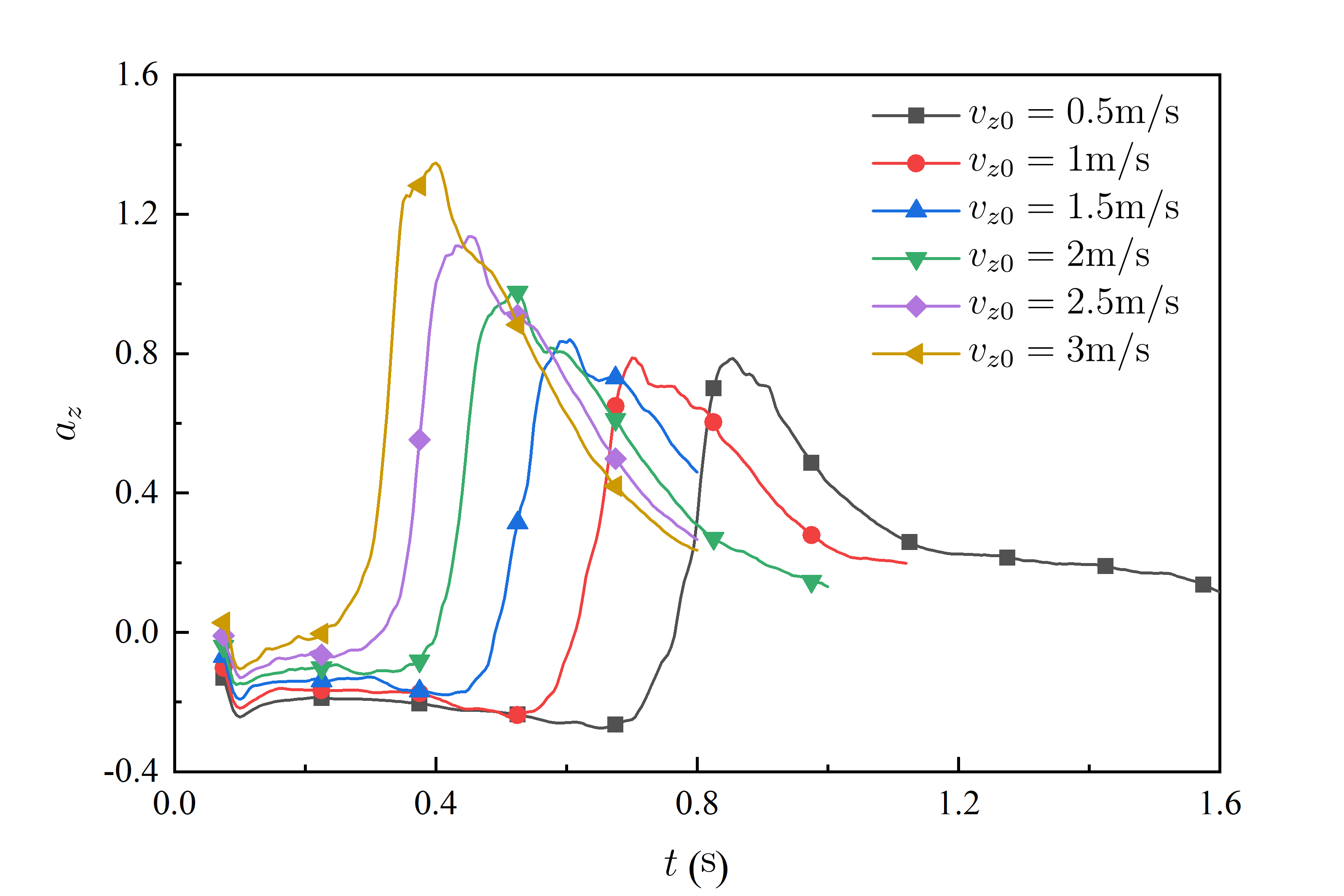}
\caption{}
\label{fig:SeaFixUnfixAzb}
\end{subfigure}
\caption{Comparison of fixed and free pitching condition on \textcolor{black}{dimensionless} acceleration \textcolor{black}{in $z$-direction} with different initial vertical velocity \textcolor{black}{for the amphibious aircraft}: (a) fixed pitch; (b) free pitch.}
\label{fig:SeaFixUnfixAz}
\end{figure}
\begin{figure}[hbt!]
\centering
\includegraphics[width=.5\textwidth]{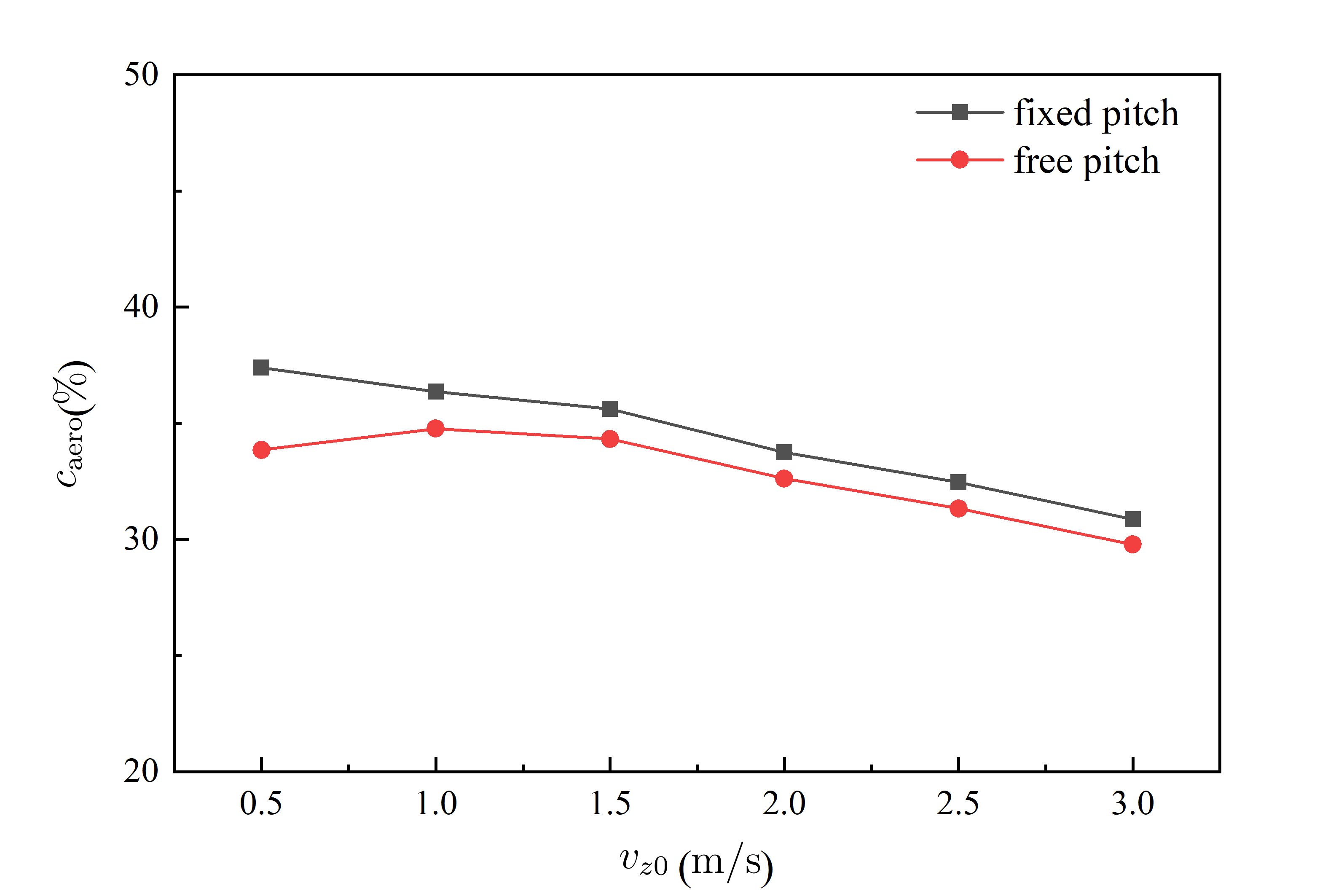}
\caption{Ratio of the aerodynamic to the fluid force as a function of $\upsilon_{z0}$.}
\label{fig:SeaAero}
\end{figure}

Results, shown in Fig.~\ref{fig:SeaFixUnfixAz}, indicate that $a_z$ decrease when reducing $\upsilon_{z0}$ in both conditions. It is worth noting that as $\upsilon_{z0}$ decreases below 1.5 m/s, the overall trend and the amplitudes of $a_z$ in each condition are quite similar, aside from the time lags. Differently from the conventional impact problem, the amphibian has aerodynamic devices, such as wings and tail wings, which introduce additional force components affecting the aircraft dynamics. Fig.~\ref{fig:SeaAero} shows the parameter $c_\mathrm{aero}$, which is the ratio between aerodynamic force to fluid force in the vertical direction derived for the different cases, when $a_z$ reaches the highest amplitude during the landing motion. It is shown that the parameter $c_\mathrm{aero}$ is always below 40\% and diminishes when increasing $\upsilon_{z0}$, thus indicating that the hydrodynamic force acting on the fuselage becomes larger as $\upsilon_{z0}$ grows, as expected.

\begin{figure}[hbt!]
\centering
\includegraphics[width=0.98\textwidth]{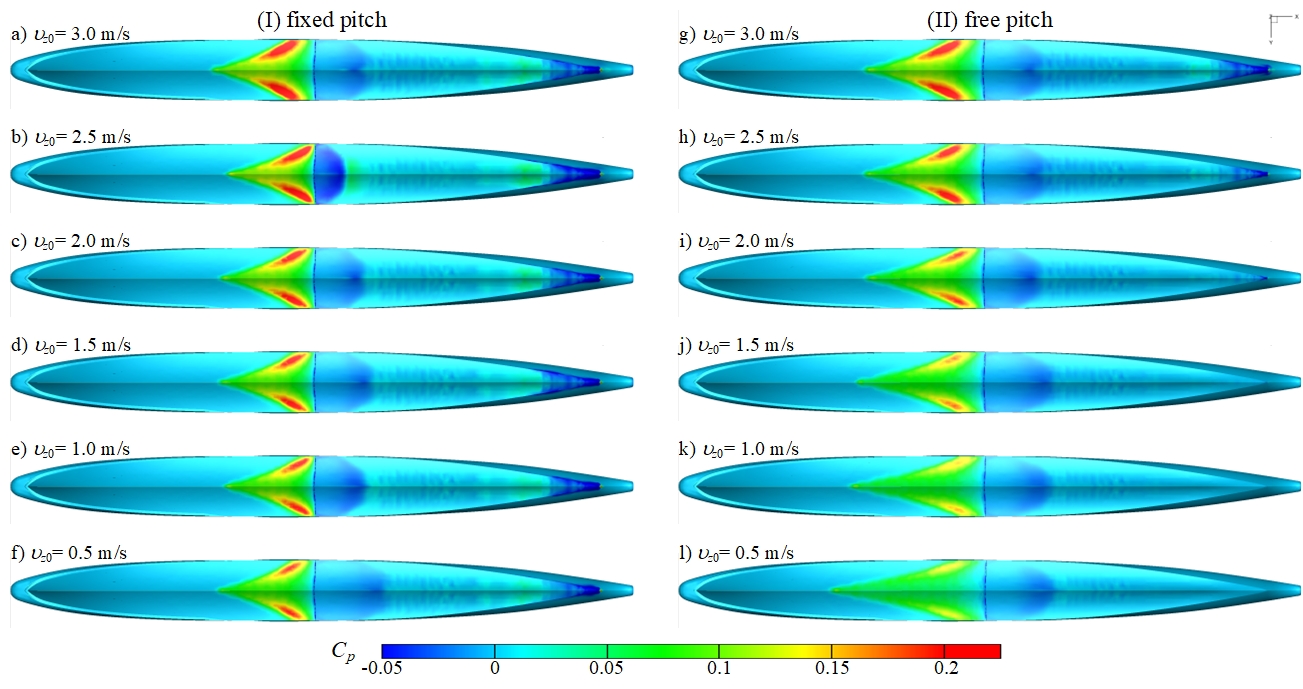}
\caption{Pressure distribution at the bottom of the aircraft for different $\upsilon_{z0}$ at $t^*$.}
\label{fig:SeaCp}
\end{figure}

Fig.~\ref{fig:SeaCp} illustrates the pressure distribution at the bottom of the aircraft when $a_z$ reaches its peak. The main fuselage portion striking with the free surface is the region over the forebody near the step. Note that the pressure coefficient displayed in the graph is defined as $C_p=(p-p_0)/(0.5\rho \upsilon_{x0}^2)$, where $\upsilon_{z0}$ is neglected being $\upsilon_{x0}$ = 37 m/s much greater than $\upsilon_{z0}$. The pressure peaks occur at the chine flare, after which the hydrodynamic decreases with the formation of a triangle-shaped region of positive pressure near the step. Correspondingly, negative pressure areas occur behind the step and the stern of the fuselage. The occurrence of negative pressures at the back of the fuselage is a consequence of the longitudinal curvature and it can be easily explained by exploiting a 2D+t concept in which the local cross section undergoes a water exit phase \citep{delbuono2021water}. The data also indicate that the high-pressure regions become smaller in size and reduce in magnitude when decreasing $\upsilon_{z0}$, which is coherent with the overall downtrend on the evolutions of $a_z$ revealed in Fig.~\ref{fig:SeaFixUnfixAz}.

\begin{figure}[hbt!]
\centering
\includegraphics[width=.5\textwidth]{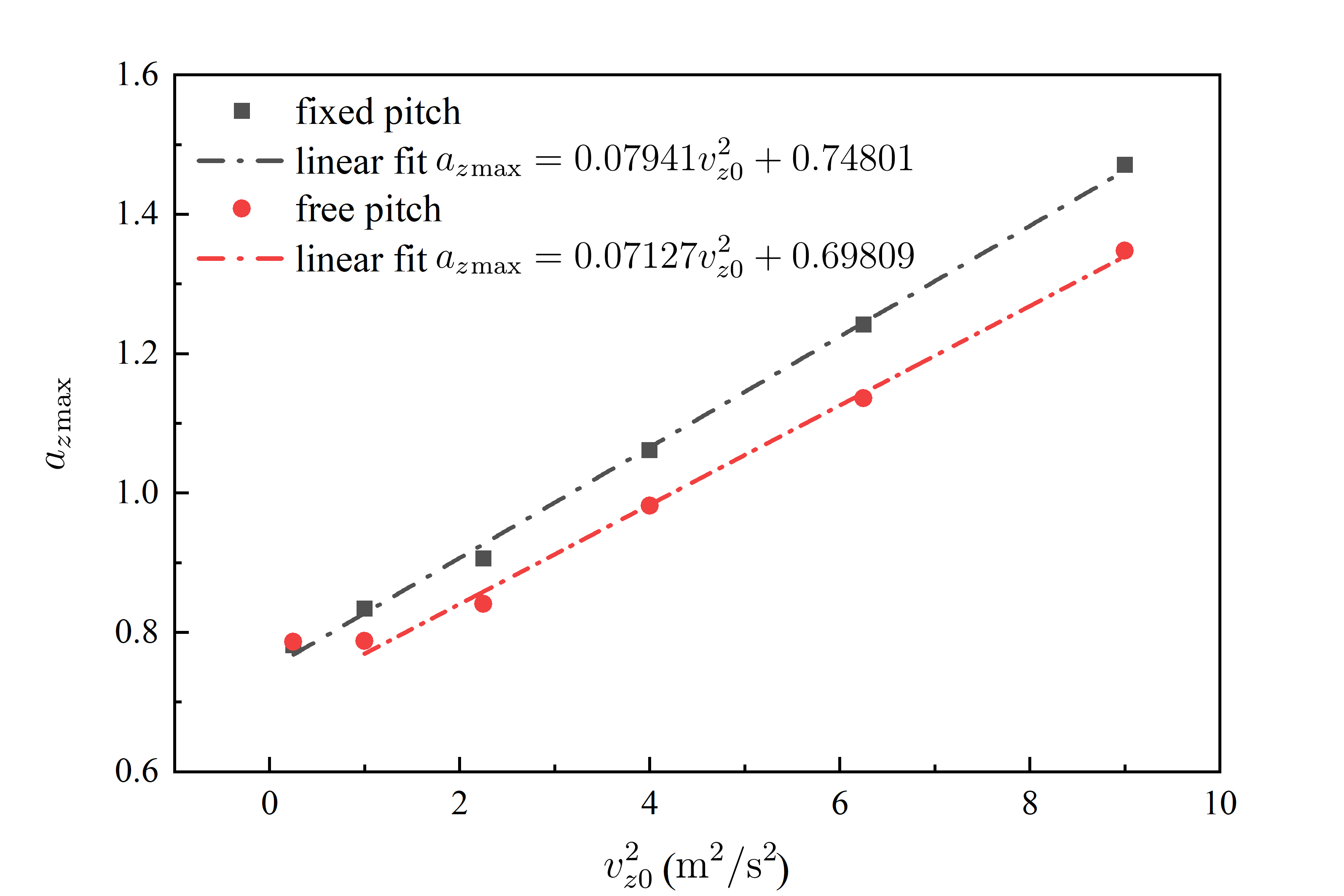}
\caption{Effect of the pitch motion, fixed and free, on the relation between $a_{z\mathrm{max}}$ and $\upsilon_{z0}^2$.}
\label{fig:SeaAzmax}
\end{figure}

\begin{figure}[hbt!]
\centering
\includegraphics[width=0.98\textwidth]{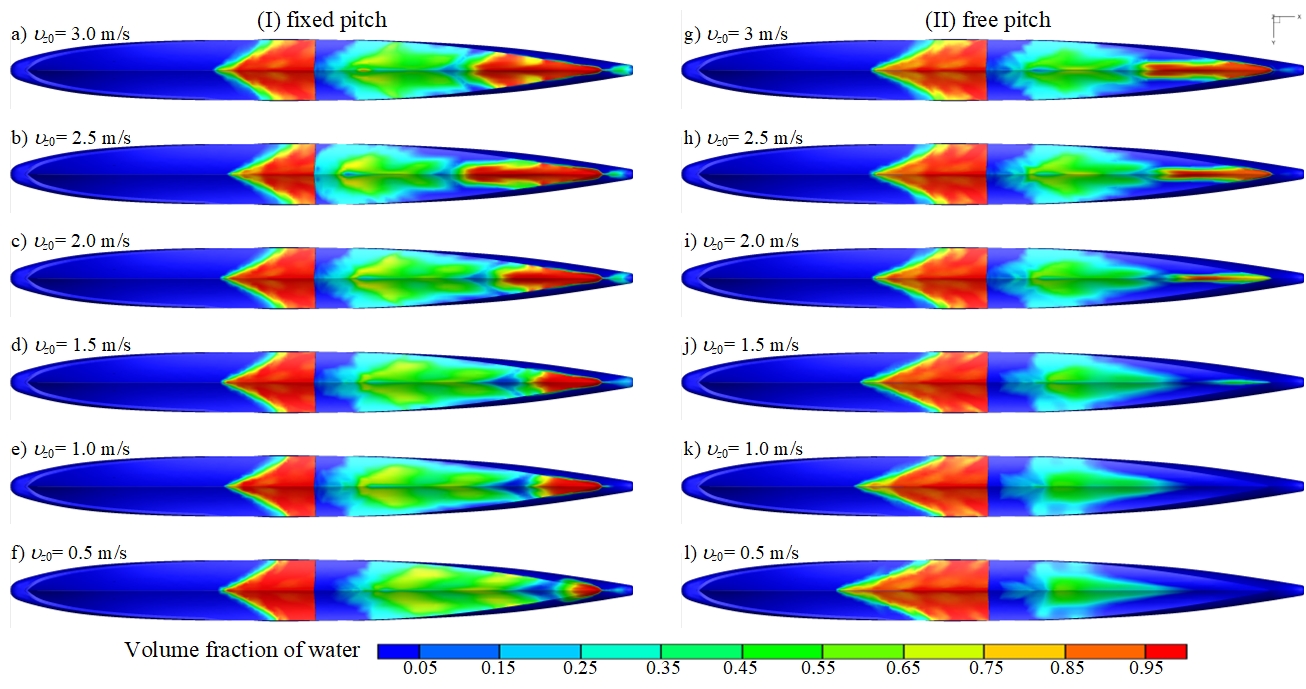}
\caption{Water volume fraction at the bottom of the aircraft for different $\upsilon_{z0}$ at $t^*$.}
\label{fig:SeaVOF}
\end{figure}

In order to achieve a better comprehension of the effect of the impact velocity on accelerations, the maximal values of $a_z$ are drawn as a function of the square of vertical velocity $\upsilon_{z0}^2$ in Fig.~\ref{fig:SeaAzmax}, although it is difficult to derive the slope $k$ from theoretical estimate Eq.~\eqref{eq:amax}. In the presence of a high horizontal speed, the pressure doesn’t depend much on the vertical velocity but rather on the horizontal velocity, pitch angle and pitch dynamics. Furthermore, there are the effects associated with the suction and the double-stepped planing phenomenon which are not accounted for in the theoretical model (see Fig.~\ref{fig:SeaVOF}). Instead, the linearity on $a_{z\mathrm{max}}-\upsilon_{z0}^2$ can be found as well, despite a little deviation appears when $\upsilon_{z0}$ is below 1m/s on the fixed pitching situation. The results of $a_{z\mathrm{max}}$ with fixed pitch is above that with free pitch. For instance, the high-pressure region presented in Fig.~\ref{fig:SeaCp}c) is larger than that in Fig.~\ref{fig:SeaCp}i).

\begin{figure}[hbt!]
\centering
\begin{subfigure}{0.49\textwidth}
\includegraphics[width=\linewidth]{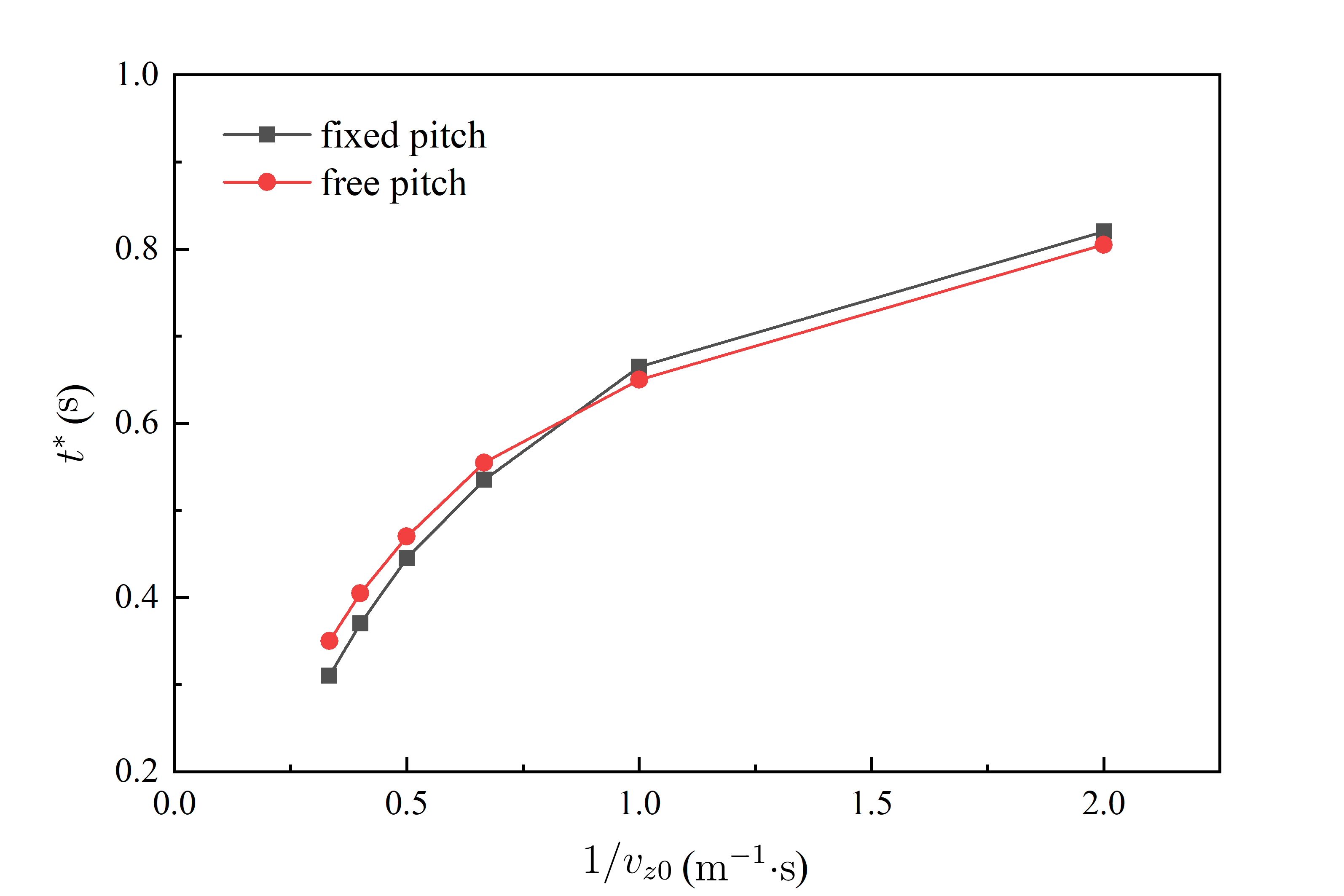} 
\caption{}
\label{fig:SeaTZVZKaa}
\end{subfigure}
\begin{subfigure}{0.49\textwidth}
\includegraphics[width=\linewidth]{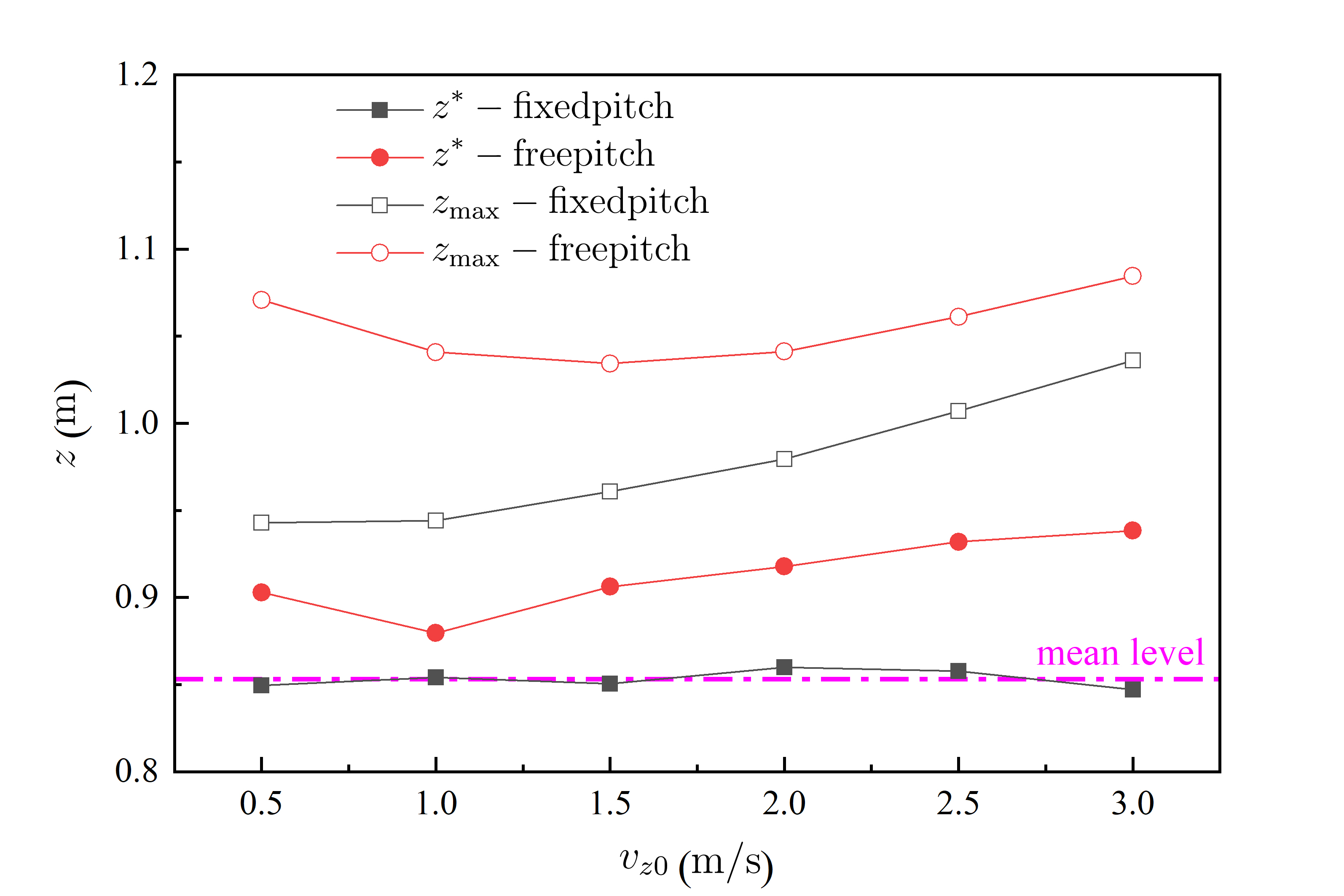}
\caption{}
\label{fig:SeaTZVZKab}
\end{subfigure}
\begin{subfigure}{0.49\textwidth}
\includegraphics[width=\linewidth]{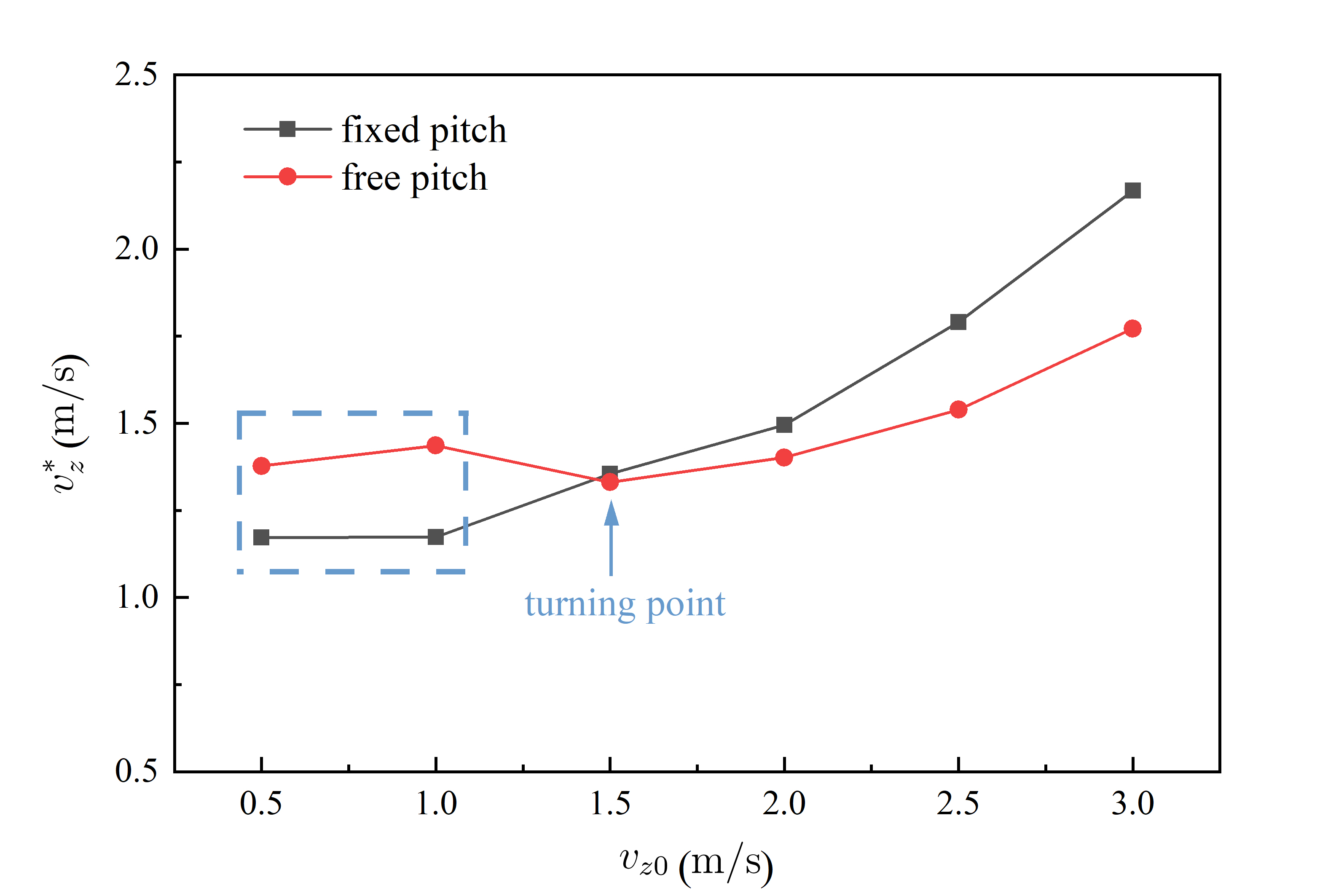} 
\caption{}
\label{fig:SeaTZVZKac}
\end{subfigure}
\begin{subfigure}{0.49\textwidth}
\includegraphics[width=\linewidth]{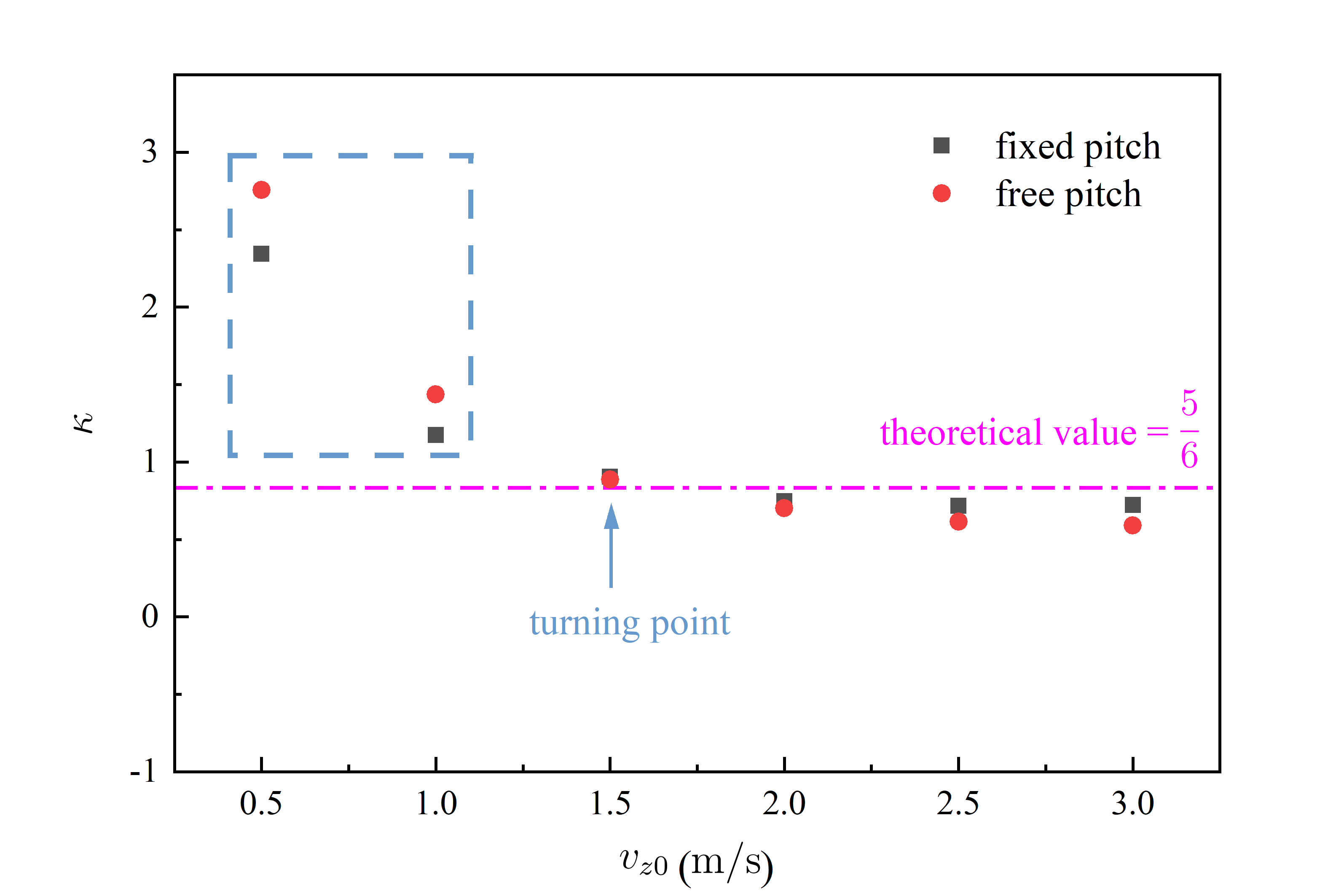}
\caption{}
\label{fig:SeaTZVZKad}
\end{subfigure}
\caption{Effect of initial vertical velocity on variable dynamic parameters \textcolor{black}{for the case of the amphibious aircraft}: (a) $t^*$; (b) $z^*$; (c) $\upsilon_z^*$; (d) $\kappa$.}
\label{fig:SeaTZVZKa}
\end{figure}

Owing to a considerable change on estimate of $a_{z\mathrm{max}}-\upsilon_{z0}^2$, it is necessary to check the effectiveness on Eq.~\eqref{eq:threepara} and \eqref{eq:threepara_t}. Fig.~\ref{fig:SeaTZVZKa} shows the variation of other four gauged factors by changing the initial impacting velocity $\upsilon_{z0}$, when the acceleration reaches its peak. As it can be seen in Fig.~\ref{fig:SeaTZVZKaa}, the larger the vertical velocity is, the shorter the time interval is, implying that the load distribution in time is smoother for lower $\upsilon_{z0}$. Moreover, there is no proportionality between $t^{*}-\upsilon_{z0}^{-1}$ in both cases of fixed and free pitch. Moving to penetration depth $z^*$ Fig.~\ref{fig:SeaTZVZKab}, a quite different evolution emerges between the fixed and free pitch conditions. In the fixed pitch condition small variations about the mean value occur, whereas, in the free pitch condition the depth $z^*$ grows as $\upsilon_{z0}$ increases gradually and approaching an asymptotic value. It is worth noticing that there is an inverse trend compared to the cases of wedge and cabin section. The depth, $z^*$, exhibits much smaller variations when an attitude control mode (fixed pitch) is exerted on the aircraft.

The results of maximal draught $z_\mathrm{max}$, reached by the hull, are also depicted in Fig.~\ref{fig:SeaTZVZKab}. Of course, $z_\mathrm{max}$ is above $z^*$, meaning that while $a_z$ attains its maximum, the aircraft continues to move downwards. It shows a monotonous increasing trend on the function of $z_\mathrm{max}$ to $\upsilon_{z0}$ for the case of fixed pitch, while a valley occurs in the free pitch motion. Turning to the behavior of the corresponding velocity $\upsilon_z^*$, it is interesting to see that the two cases share a quite similar evolution in $\upsilon_z^*-\upsilon_{z0}$, as presented in Fig.~\ref{fig:SeaTZVZKac} and \ref{fig:SeaTZVZKad}. Specifically, there is a turning point where $\upsilon_{z0}$ equals 1.5 m/s, whereas the trend is similar afterwards. On the left side of turning point, the results of fixed pitch are lower than that of free pitch. The blue dashed rectangle indicates the range at which $\upsilon_z^*$ is above $\upsilon_{z0}$, in other words, $\kappa>1$ (see Fig.~\ref{fig:SeaTZVZKad}), meaning that 
gravity plays a significant role when $\upsilon_{z0}$ is smaller than a certain value as mentioned on Sec.~\ref{sec:effectvertical}. It can be seen that the relation of $\upsilon_z^*-\upsilon_{z0}$ is not linear, quite different from the theoretical trend. Whereas, in Fig.~\ref{fig:SeaTZVZKad}, it is worth noting that all data approach the theoretical estimate, 5/6, which means $\kappa$ is still valid to some extent. Thus, the relations derived from Eq.~\eqref{eq:amax} and Eq.~\eqref{eq:threepara} are partly useful for the tendency prediction on $a_{z\mathrm{max}}$ and $\kappa$ through a simple analysis.

\begin{table}[width=0.95\linewidth, pos=hbt!]
\caption{Summary of theoretical quantitative relations compared with simulated results among three cases: simulated trend (black solid line); theoretical estimate (red dashed line); simulated asymptotic value (magenta dashed line)}
\centering
\label{tab:summary}
\begin{tabular*}{\tblwidth}{@{}cccccccccc@{}}
\toprule
\multirow{2}{*}{terms}& \multirow{2}{*}{theoretical relations}& \multicolumn{2}{c}{\includegraphics[width=.12\textwidth]{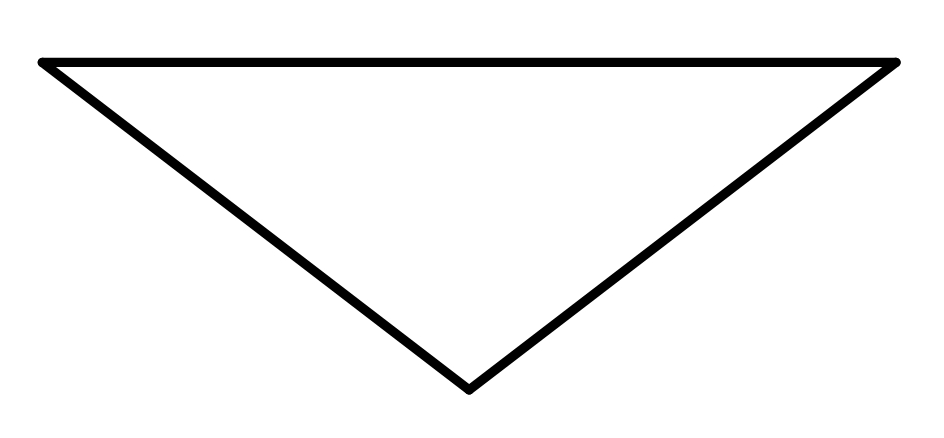}}& \multicolumn{2}{c}{\includegraphics[width=.1\textwidth]{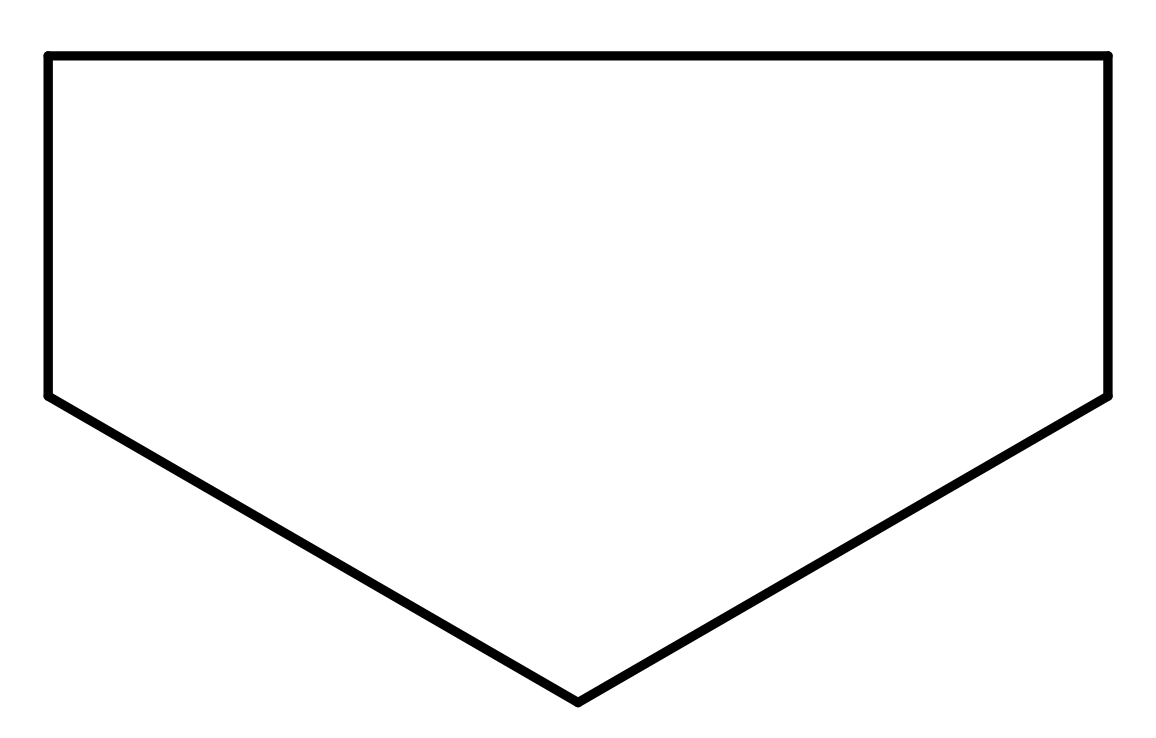}}& \multicolumn{4}{c}{\includegraphics[width=.3\textwidth]{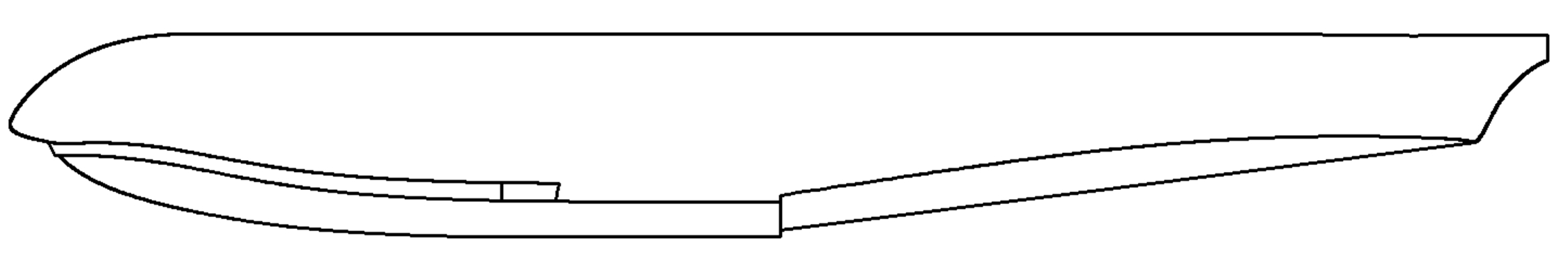}}\\\cline{7-10}
 & & & & & & \multicolumn{2}{c}{fixed pitch}& \multicolumn{2}{c}{free pitch}\\
\midrule
{$a_{z\mathrm{max}}-\upsilon_{z0}^2$}& linear& $\surd$& -& $\surd$& -& $\surd$& -& $\surd$& -\\
{$t^{*}-\upsilon_{z0}^{-1}$}& linear& $\surd$& -& $\surd$& -& $\times$& -& $\times$& -\\
{$z^{*}-\upsilon_{z0}$}& constant& $\times$& {\includegraphics[width=.1\textwidth]{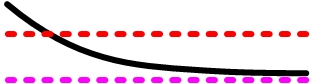}}& $\times$& {\includegraphics[width=.1\textwidth]{Fig/trend2.jpg}}& $\times$& constant& $\times$& {\includegraphics[width=.1\textwidth]{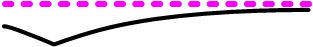}}\\
{$\upsilon_z^{*}-\upsilon_{z0}$}& linear& $\surd$& -& $\surd$& -& $\times$& -& $\times$& -\\
{$\kappa-\upsilon_{z0}$}& constant, 5/6& $\times$& {\includegraphics[width=.1\textwidth]{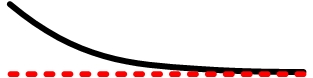}}& $\times$& {\includegraphics[width=.1\textwidth]{Fig/trend1.jpg}}& $\times$& {\includegraphics[width=.1\textwidth]{Fig/trend1.jpg}}& $\times$& {\includegraphics[width=.1\textwidth]{Fig/trend1.jpg}}\\
\bottomrule
\end{tabular*}
\end{table}

\section{Conclusion}
\label{sec:conclusion}
In the present study, the load characteristics of three models, such as a 2D symmetric wedge water entry, a 3D cabin section water entry and an amphibious aircraft landing on water have been investigated numerically. The effect of initial vertical velocity on the maximum acceleration, together with several relationships based on the transformation of momentum theorem, have been thoroughly analyzed. Contributions and findings are summarized in Table~\ref{tab:summary}, which can be described as:

1)	For the three V-shaped sectional area of bodies, such as 2D wedge, 3D cabin section and amphibious aircraft, impacting on water surface, the maximum vertical acceleration increases with the initial vertical velocity, and it is found herein that the value of maximal vertical acceleration is proportional to the square of the initial vertical impacting velocity. For oblique entry, the effect of horizontal velocity on acceleration has also been investigated and it is observed that the maximum horizontal acceleration is a linear function of initial horizontal velocity, rather than its square value.

2)	Another significant parameter, that these three models share the same trend, is the ratio of the corresponding velocity to the initial velocity, $\kappa$. Following the theoretical formulation, the value should be constant, 5/6, while the numerical results approach it in the case of large initial vertical velocity. It indicates that a threshold value of initial vertical velocity needs to be emphasized to make the theoretical result available. In other words, gravity can be neglected with larger velocities, however, with slow impact speeds, gravity should be considered in the model.

3)	For the relationship between penetration depth and the initial vertical velocity, the simulated results approach an asymptotic line (different from the theoretical estimate) with the increase of velocity in the 2D wedge case and the cabin section case. 
The difference between the numerical asymptotic line and the theoretical estimate is mainly caused by the water pile-up effect. For the 3D cabin section, the three-dimensional water flow in the spanwise direction could also be responsible for the difference.
Considering the complicated geometry of the hull, it is hard to determine the theoretical estimate. The numerical results of fixed pitch present a constant trend, whereas a constant value is not reached in the case of free pitch.

4)	Looking into other two linear relations, $t^{*}-\upsilon_{z0}^{-1}$ and $\upsilon_z^{*}-\upsilon_{z0}$, shown in Table~\ref{tab:summary}, they can be established upon the wedge and the cabin section compared with the theoretical results, while it is invalid for the hull. Besides, in the case of 2D wedge and 3D cabin section, the instantaneous Froude number, $Fr^*$, is displayed to describe the combined relations between velocity and penetration depth, when the maximum value of acceleration is satisfied. Due to the relationship of $Fr^{*}-\upsilon_{z0}$ and $a_{z\mathrm{max}}-\upsilon_{z0}^*$, the maximum acceleration $a_{z\mathrm{max}}$ is one third of the square of the instantaneous Froude number $Fr^*$. Moreover, a strong coupled relation among $a_{z\mathrm{max}}$, $\upsilon_z^*$ and $z^*$ is found, $a_{z\mathrm{max}}=(\upsilon_z^*)^2/(3g\cdot z^*)$.

\appendix
\section{Appendix}
\label{sec:appendix}
\setcounter{equation}{0}
\renewcommand\theequation{A\arabic{equation}}

In order to obtain the instantaneous acceleration shown in Eq.~\eqref{eq:a}, time derivative of the instantaneous velocity can be analytically computed as:
\begin{equation}
\label{eq:a_detail}
\begin{split}
a(t) &= \Dot{\upsilon}(t)=\dfrac{d}{dt} \left( \dfrac{2M\tan^2(\beta) \upsilon_0}{2M\tan^2(\beta)+\pi\rho z^2 (t)} \right)\\
&= 2M\tan^2(\beta)\upsilon_0 \cdot \dfrac{d}{dt} \left( \dfrac{1}{2M\tan^2(\beta)+\pi\rho z^2 (t)} \right)\\
&= 2M\tan^2(\beta)\upsilon_0 \cdot \left( - \dfrac{1}{(2M\tan^2(\beta)+\pi\rho z^2 (t))^2} \right) \cdot 2\pi\rho \cdot z(t) \cdot \Dot{z}(t)\\
&= - \dfrac{(2M\tan^2(\beta)\upsilon_0)^2}{(2M\tan^2(\beta)+\pi\rho z^2 (t))^2} \cdot \dfrac{2\pi\rho \cdot z(t) \cdot \Dot{z}(t)}{2M\tan^2(\beta)\upsilon_0}\\
&= - \upsilon^2 (t) \cdot \dfrac{2\pi\rho \cdot z(t) \cdot \upsilon(t)}{2M\tan^2(\beta)\upsilon_0}\\
&= - \dfrac{\pi\rho z(t)}{M \upsilon_0 \tan^2(\beta)} \cdot \upsilon^3(t)    
\end{split}
\end{equation}
The minus sign indicates that the direction of acceleration is opposite to the direction of velocity. Besides, we define that the positive value of acceleration is upwards, while the velocity and the penetration depth are positive downwards.

The acceleration reaches its peak value when $\Dot{a}(t)=0$, that is,

\begin{equation}
\label{eq:1}
\begin{split}
\Dot{a}(t) &= \dfrac{d}{dt} \left( - \dfrac{\pi\rho \cdot z(t)}{M\upsilon_0\tan^2(\beta)} \cdot \upsilon^3 (t) \right)\\
&= - \dfrac{\pi\rho}{M\upsilon_0\tan^2(\beta)} \cdot \dfrac{d}{dt}(z(t) \cdot \upsilon^3 (t))
=0
\end{split}
\end{equation}
Dividing out the constant term, Eq.~\eqref{eq:1} can be rewritten as:

\begin{equation}
\label{eq:2}
\dfrac{d}{dt}(z(t) \cdot \upsilon^3 (t))
=\upsilon^4 (t) + z(t) \cdot 3\upsilon^2 (t) \cdot a(t)=0
\longrightarrow \upsilon^2 (t) + 3z(t) \cdot a(t)=0
\end{equation}
Substituting Eq.~\eqref{eq:a_detail} into the second term of this last equation, the Eq.~\eqref{eq:2} can be expressed as:

\begin{equation}
\label{eq:3}
\upsilon^2 (t) - \dfrac{3\pi\rho \cdot z^2(t) \upsilon^3(t)}{M\upsilon_0\tan^2(\beta)}=0
\longrightarrow
1 - \dfrac{3\pi\rho \cdot z^2(t) \upsilon(t)}{M\upsilon_0\tan^2(\beta)}=0
\end{equation}
Then, a new expression of the instantaneous velocity can be obtained as:

\begin{equation}
\label{eq:5}
\upsilon(t)= \dfrac{M\upsilon_0\tan^2(\beta)}{3\pi\rho \cdot z^2(t)}
\end{equation}
Besides, the expression of the instantaneous velocity is also given by Eq.~\eqref{eq:upsilon}. Then, using Eq.~\eqref{eq:upsilon} and \eqref{eq:5}, we obtain the corresponding penetration depth $z^*$:

\begin{equation}
\label{eq:6}
\dfrac{2M\tan^2(\beta) \upsilon_0}{2M\tan^2(\beta)+\pi\rho z^2(t)}=
\dfrac{M\upsilon_0\tan^2(\beta)}{3\pi\rho \cdot z^2(t)}
\end{equation}

\begin{equation}
\label{eq:7}
z(t)=\sqrt{\dfrac{2M}{5\pi\rho}} \tan(\beta)=z^*
\end{equation}
Substituting Eq.~\eqref{eq:7} into Eq.~\eqref{eq:5}, the corresponding velocity $\upsilon^*$ can be expressed as:

\begin{equation}
\label{eq:8}
\upsilon= \dfrac{M\upsilon_0\tan^2(\beta)}{3\pi\rho \cdot \dfrac{2M\tan^2(\beta)}{5\pi\rho}}=\dfrac{5}{6}\upsilon_0
\end{equation}

Finally, by combing Eqs.~\eqref{eq:a_detail}, \eqref{eq:7} and \eqref{eq:8}, the maximal acceleration in the positive direction can be obtained as:

\begin{equation}
\label{eq:9}
\begin{split}
a^* &= \dfrac{\pi\rho \cdot z^*}{M \upsilon_0 \tan^2(\beta)} \cdot (\upsilon^*(t))^3\\
&= \dfrac{\pi\rho}{M \upsilon_0 \tan^2(\beta)} \cdot \sqrt{\dfrac{2M}{5\pi\rho}} \tan(\beta) \cdot \left( \dfrac{5\upsilon_0}{6} \right)^3\\
&=\upsilon_0^2 \left( \dfrac{5}{6} \right)^3 \dfrac{1}{\tan(\beta)} \sqrt{\dfrac{2\pi\rho}{5M}}
\end{split}
\end{equation}

\section*{Acknowledgments}

This work has been supported by China Scholarship Council (CSC, No. 202106830092) and the Project TORPEDO (inTerazione fluidO stRuttura in ProblEmi Di impattO) cooperated in the Institute of Marine Engineering of the National Research Council of Italy. The supports from Open Foundations of EDL Laboratory (EDL19092111) and the Aeronautical Science Foundation of China under grant no. 20182352015 are also acknowledged.








\printcredits

\bibliographystyle{cas-model2-names}

\bibliography{cas-refs}

\begin{thebibliography}{50}
\expandafter\ifx\csname natexlab\endcsname\relax\def\natexlab#1{#1}\fi
\providecommand{\url}[1]{\texttt{#1}}
\providecommand{\href}[2]{#2}
\providecommand{\path}[1]{#1}
\providecommand{\DOIprefix}{doi:}
\providecommand{\ArXivprefix}{arXiv:}
\providecommand{\URLprefix}{URL: }
\providecommand{\Pubmedprefix}{pmid:}
\providecommand{\doi}[1]{\href{http://dx.doi.org/#1}{\path{#1}}}
\providecommand{\Pubmed}[1]{\href{pmid:#1}{\path{#1}}}
\providecommand{\bibinfo}[2]{#2}
\ifx\xfnm\relax \def\xfnm[#1]{\unskip,\space#1}\fi
\bibitem[{Abraham et~al.(2014)Abraham, Gorman, Reseghetti, Sparrow, Stark and
  Shepard}]{abraham2014modeling}
\bibinfo{author}{Abraham, J.}, \bibinfo{author}{Gorman, J.},
  \bibinfo{author}{Reseghetti, F.}, \bibinfo{author}{Sparrow, E.},
  \bibinfo{author}{Stark, J.}, \bibinfo{author}{Shepard, T.},
  \bibinfo{year}{2014}.
\newblock \bibinfo{title}{Modeling and numerical simulation of the forces
  acting on a sphere during early-water entry}.
\newblock \bibinfo{journal}{Ocean Engineering} \bibinfo{volume}{76},
  \bibinfo{pages}{1--9}.
\newblock \DOIprefix\doi{10.1016/j.oceaneng.2013.11.015}.
\bibitem[{Benson and Bidwell(1945)}]{benson1945bibliography}
\bibinfo{author}{Benson, J.M.}, \bibinfo{author}{Bidwell, J.M.},
  \bibinfo{year}{1945}.
\newblock \bibinfo{title}{Bibliography and Review of Information Relating to
  the Hydrodynamics of Seaplanes}.
\newblock \bibinfo{type}{Technical Report} \bibinfo{number}{NACA ACR-L5G28}.
  NACA. \bibinfo{address}{Washington, D. C.}
\bibitem[{Bertram(2012)}]{bertram2012practical}
\bibinfo{author}{Bertram, V.}, \bibinfo{year}{2012}.
\newblock \bibinfo{title}{Practical Ship Hydrodynamics}. \bibinfo{edition}{2}
  ed.. \bibinfo{publisher}{Butterworth-Heinemann},
  \bibinfo{address}{Kidlington}. chapter~\bibinfo{chapter}{4}.
\newblock pp. \bibinfo{pages}{138--145}.
\newblock \DOIprefix\doi{10.1016/C2010-0-68326-X}.
\bibitem[{Bisagni and Pigazzini(2017)}]{bisagni2017modelling}
\bibinfo{author}{Bisagni, C.}, \bibinfo{author}{Pigazzini, M.S.},
  \bibinfo{year}{2017}.
\newblock \bibinfo{title}{Modelling strategies for numerical simulation of
  aircraft ditching}.
\newblock \bibinfo{journal}{International Journal of Crashworthiness}
  \bibinfo{volume}{23}, \bibinfo{pages}{377--394}.
\newblock \DOIprefix\doi{10.1080/13588265.2017.1328957}.
\bibitem[{Breton et~al.(2020)Breton, Tassin and
  Jacques}]{breton2020experimental}
\bibinfo{author}{Breton, T.}, \bibinfo{author}{Tassin, A.},
  \bibinfo{author}{Jacques, N.}, \bibinfo{year}{2020}.
\newblock \bibinfo{title}{Experimental investigation of the water entry and/or
  exit of axisymmetric bodies}.
\newblock \bibinfo{journal}{Journal of Fluid Mechanics} \bibinfo{volume}{901},
  \bibinfo{pages}{A37}.
\newblock \DOIprefix\doi{10.1017/jfm.2020.559}.
\bibitem[{Chen et~al.(2022)Chen, Xiao, Wu, Wang and Tong}]{chen2022numerical}
\bibinfo{author}{Chen, J.}, \bibinfo{author}{Xiao, T.}, \bibinfo{author}{Wu,
  B.}, \bibinfo{author}{Wang, F.}, \bibinfo{author}{Tong, M.},
  \bibinfo{year}{2022}.
\newblock \bibinfo{title}{Numerical study of wave effect on water entry of a
  three-dimensional symmetric wedge}.
\newblock \bibinfo{journal}{Ocean Engineering} \bibinfo{volume}{250},
  \bibinfo{pages}{110800}.
\newblock \DOIprefix\doi{10.1016/j.oceaneng.2022.110800}.
\bibitem[{Del~Buono et~al.(2021)Del~Buono, Bernardini, Tassin and
  Iafrati}]{delbuono2021water}
\bibinfo{author}{Del~Buono, A.}, \bibinfo{author}{Bernardini, G.},
  \bibinfo{author}{Tassin, A.}, \bibinfo{author}{Iafrati, A.},
  \bibinfo{year}{2021}.
\newblock \bibinfo{title}{Water entry and exit of 2d and axisymmetric bodies}.
\newblock \bibinfo{journal}{Journal of Fluids and Structures}
  \bibinfo{volume}{103}, \bibinfo{pages}{103269}.
\newblock \DOIprefix\doi{10.1007/s13272-017-0257-0}.
\bibitem[{Duan et~al.(2019)Duan, Sun, Chen, Wei and Yang}]{duan2019numerical}
\bibinfo{author}{Duan, X.}, \bibinfo{author}{Sun, W.}, \bibinfo{author}{Chen,
  C.}, \bibinfo{author}{Wei, M.}, \bibinfo{author}{Yang, Y.},
  \bibinfo{year}{2019}.
\newblock \bibinfo{title}{Numerical investigation of the porpoising motion of a
  seaplane planing on water with high speeds}.
\newblock \bibinfo{journal}{Aerospace Science and Technology}
  \bibinfo{volume}{84}, \bibinfo{pages}{980--994}.
\newblock \DOIprefix\doi{10.1016/j.ast.2018.11.037}.
\bibitem[{Gong et~al.(2009)Gong, Liu and Wang}]{gong2009water}
\bibinfo{author}{Gong, K.}, \bibinfo{author}{Liu, H.}, \bibinfo{author}{Wang,
  B.}, \bibinfo{year}{2009}.
\newblock \bibinfo{title}{Water entry of a wedge based on sph model with an
  improved boundary treatment}.
\newblock \bibinfo{journal}{Journal of Hydrodynamics} \bibinfo{volume}{21},
  \bibinfo{pages}{750--757}.
\newblock \DOIprefix\doi{10.1016/S1001-6058(08)60209-7}.
\bibitem[{Guo et~al.(2013)Guo, Liu, Qu and Wang}]{guo2013effect}
\bibinfo{author}{Guo, B.}, \bibinfo{author}{Liu, P.}, \bibinfo{author}{Qu, Q.},
  \bibinfo{author}{Wang, J.}, \bibinfo{year}{2013}.
\newblock \bibinfo{title}{Effect of pitch angle on initial stage of a transport
  airplane ditching}.
\newblock \bibinfo{journal}{Chinese Journal of Aeronautics}
  \bibinfo{volume}{26}, \bibinfo{pages}{17--26}.
\newblock \DOIprefix\doi{10.1016/j.cja.2012.12.024}.
\bibitem[{Hirt and Nichols(1981)}]{hirt1981volume}
\bibinfo{author}{Hirt, C.W.}, \bibinfo{author}{Nichols, B.D.},
  \bibinfo{year}{1981}.
\newblock \bibinfo{title}{Volume of fluid (vof) method for the dynamics of free
  boundaries}.
\newblock \bibinfo{journal}{Journal of Computational Physics}
  \bibinfo{volume}{39}, \bibinfo{pages}{201--225}.
\newblock \DOIprefix\doi{10.1016/0021-9991(81)90145-5}.
\bibitem[{Hughes et~al.(2013)Hughes, Vignjevic, Campbell, Vuyst, Djordjevic and
  Papagiannis}]{hughes2013from}
\bibinfo{author}{Hughes, K.}, \bibinfo{author}{Vignjevic, R.},
  \bibinfo{author}{Campbell, J.}, \bibinfo{author}{Vuyst, T.D.},
  \bibinfo{author}{Djordjevic, N.}, \bibinfo{author}{Papagiannis, L.},
  \bibinfo{year}{2013}.
\newblock \bibinfo{title}{From aerospace to offshore: Bridging the numerical
  simulation gaps-simulation advancements for fluid structure interaction
  problems}.
\newblock \bibinfo{journal}{International Journal of Impact Engineering}
  \bibinfo{volume}{61}, \bibinfo{pages}{48--63}.
\newblock \DOIprefix\doi{10.1016/j.ijimpeng.2013.05.001}.
\bibitem[{Hulin et~al.(2022)Hulin, Del~Buono, Tassin, Bernardini and
  Iafrati}]{hulin2022gravity}
\bibinfo{author}{Hulin, F.}, \bibinfo{author}{Del~Buono, A.},
  \bibinfo{author}{Tassin, A.}, \bibinfo{author}{Bernardini, G.},
  \bibinfo{author}{Iafrati, A.}, \bibinfo{year}{2022}.
\newblock \bibinfo{title}{Gravity effects in two-dimensional and axisymmetric
  water impact models}.
\newblock \bibinfo{journal}{Journal of Fluid Mechanics} \bibinfo{volume}{944},
  \bibinfo{pages}{A9}.
\newblock \DOIprefix\doi{10.1017/jfm.2022.448}.
\bibitem[{Iafrati(2016)}]{iafrati2016experimental}
\bibinfo{author}{Iafrati, A.}, \bibinfo{year}{2016}.
\newblock \bibinfo{title}{Experimental investigation of the water entry of a
  rectangular plate at high horizontal velocity}.
\newblock \bibinfo{journal}{Journal of Fluid Mechanics} \bibinfo{volume}{799},
  \bibinfo{pages}{637--672}.
\newblock \DOIprefix\doi{10.1017/jfm.2016.374}.
\bibitem[{Iafrati et~al.(2000)Iafrati, Carcaterra, Ciappi and
  Campana}]{iafrati2000hydroelastic}
\bibinfo{author}{Iafrati, A.}, \bibinfo{author}{Carcaterra, A.},
  \bibinfo{author}{Ciappi, E.}, \bibinfo{author}{Campana, E.F.},
  \bibinfo{year}{2000}.
\newblock \bibinfo{title}{Hydroelastic analysis of a simple oscillator
  impacting the free surface}.
\newblock \bibinfo{journal}{Journal of Ship Research} \bibinfo{volume}{44},
  \bibinfo{pages}{278--289}.
\newblock \DOIprefix\doi{10.5957/jsr.2000.44.4.278}.
\bibitem[{Iafrati and Grizzi(2019)}]{iafrati2019cavitation}
\bibinfo{author}{Iafrati, A.}, \bibinfo{author}{Grizzi, S.},
  \bibinfo{year}{2019}.
\newblock \bibinfo{title}{Cavitation and ventilation modalities during
  ditching}.
\newblock \bibinfo{journal}{Physics of Fluids} \bibinfo{volume}{31},
  \bibinfo{pages}{052101}.
\newblock \DOIprefix\doi{10.1063/1.5092559}.
\bibitem[{Judge et~al.(2004)Judge, Troesch and Perlin}]{judge2004initial}
\bibinfo{author}{Judge, C.}, \bibinfo{author}{Troesch, A.},
  \bibinfo{author}{Perlin, M.}, \bibinfo{year}{2004}.
\newblock \bibinfo{title}{Initial water impact of a wedge at vertical and
  oblique angles}.
\newblock \bibinfo{journal}{Journal of Engineering Mathematics}
  \bibinfo{volume}{48}, \bibinfo{pages}{279--303}.
\newblock \DOIprefix\doi{10.1023/B:engi.0000018187.33001.e1}.
\bibitem[{von Karman(1929)}]{karman1929impact}
\bibinfo{author}{von Karman, T.}, \bibinfo{year}{1929}.
\newblock \bibinfo{title}{The Impact on Seaplane Floats During Landing}.
\newblock \bibinfo{type}{Technical Report} \bibinfo{number}{NACA TN-321}. NACA.
  \bibinfo{address}{Washington, D. C.}
\bibitem[{Korobkin(2004)}]{korobkin2004analytical}
\bibinfo{author}{Korobkin, A.}, \bibinfo{year}{2004}.
\newblock \bibinfo{title}{Analytical models of water impact}.
\newblock \bibinfo{journal}{European Journal of Applied Mathematics}
  \bibinfo{volume}{15}, \bibinfo{pages}{821--838}.
\newblock \DOIprefix\doi{10.1017/S0956792504005765}.
\bibitem[{Korobkin and Scolan(2006)}]{korobkin2006three}
\bibinfo{author}{Korobkin, A.A.}, \bibinfo{author}{Scolan, Y.M.},
  \bibinfo{year}{2006}.
\newblock \bibinfo{title}{Three-dimensional theory of water impact. part 2.
  linearized wagner problem}.
\newblock \bibinfo{journal}{Journal of Fluid Mechanics} \bibinfo{volume}{549},
  \bibinfo{pages}{343--373}.
\newblock \DOIprefix\doi{10.1017/S0022112005008049}.
\bibitem[{Lu et~al.(2021)Lu, Xiao, Deng, Zhi, Zhu and Lu}]{lu2021effects}
\bibinfo{author}{Lu, Y.}, \bibinfo{author}{Xiao, T.}, \bibinfo{author}{Deng,
  S.}, \bibinfo{author}{Zhi, H.}, \bibinfo{author}{Zhu, Z.},
  \bibinfo{author}{Lu, Z.}, \bibinfo{year}{2021}.
\newblock \bibinfo{title}{Effects of initial conditions on landing performance
  of the amphibious aircraft}.
\newblock \bibinfo{journal}{Acta Aeronautica et Astronautica Sinica}
  \bibinfo{volume}{42}, \bibinfo{pages}{159--170}.
\newblock \DOIprefix\doi{10.7527/S1000-6893.2020.24483}. \bibinfo{note}{(in
  Chinese)}.
\bibitem[{Mei et~al.(1999)Mei, Liu and Yue}]{mei1999on}
\bibinfo{author}{Mei, X.}, \bibinfo{author}{Liu, Y.}, \bibinfo{author}{Yue,
  D.K.P.}, \bibinfo{year}{1999}.
\newblock \bibinfo{title}{On the water impact of general two-dimensional
  sections}.
\newblock \bibinfo{journal}{Applied Ocean Research} \bibinfo{volume}{21},
  \bibinfo{pages}{1--15}.
\newblock \DOIprefix\doi{10.1016/S0141-1187(98)00034-0}.
\bibitem[{Neuberg and Drimer(2017)}]{neuberg2017fatigue}
\bibinfo{author}{Neuberg, O.}, \bibinfo{author}{Drimer, N.},
  \bibinfo{year}{2017}.
\newblock \bibinfo{title}{Fatigue limit state design of fast boats}.
\newblock \bibinfo{journal}{Marine Structures} \bibinfo{volume}{55},
  \bibinfo{pages}{17--36}.
\newblock \DOIprefix\doi{10.1016/j.marstruc.2017.05.002}.
\bibitem[{Panciroli et~al.(2013)Panciroli, Abrate and
  Minak}]{panciroli2013dynamic}
\bibinfo{author}{Panciroli, R.}, \bibinfo{author}{Abrate, S.},
  \bibinfo{author}{Minak, G.}, \bibinfo{year}{2013}.
\newblock \bibinfo{title}{Dynamic response of flexible wedges entering the
  water}.
\newblock \bibinfo{journal}{Composite Structures} \bibinfo{volume}{99},
  \bibinfo{pages}{163--171}.
\newblock \DOIprefix\doi{10.1016/j.compstruct.2012.11.042}.
\bibitem[{Qiu and Song(2013)}]{qiu2013efficient}
\bibinfo{author}{Qiu, L.}, \bibinfo{author}{Song, W.}, \bibinfo{year}{2013}.
\newblock \bibinfo{title}{Efficient decoupled hydrodynamic and aerodynamic
  analysis of amphibious aircraft water takeoff process}.
\newblock \bibinfo{journal}{Journal of Aircraft} \bibinfo{volume}{50},
  \bibinfo{pages}{1369--1379}.
\newblock \DOIprefix\doi{10.2514/1.C031846}.
\bibitem[{Qu et~al.(2016)Qu, Liu, Liu, Guo and Agarwal}]{qu2016numerical}
\bibinfo{author}{Qu, Q.}, \bibinfo{author}{Liu, C.}, \bibinfo{author}{Liu, P.},
  \bibinfo{author}{Guo, B.}, \bibinfo{author}{Agarwal, R.K.},
  \bibinfo{year}{2016}.
\newblock \bibinfo{title}{Numerical simulation of water-landing performance of
  a regional aircraft}.
\newblock \bibinfo{journal}{Journal of Aircraft} \bibinfo{volume}{53},
  \bibinfo{pages}{1680--1689}.
\newblock \DOIprefix\doi{10.2514/1.C033686}.
\bibitem[{Riccardi and Iafrati(2004)}]{riccardi2004water}
\bibinfo{author}{Riccardi, G.}, \bibinfo{author}{Iafrati, A.},
  \bibinfo{year}{2004}.
\newblock \bibinfo{title}{Water impact of an asymmetric floating wedge}.
\newblock \bibinfo{journal}{Journal of Engineering Mathematics}
  \bibinfo{volume}{49}, \bibinfo{pages}{19--39}.
\newblock \DOIprefix\doi{10.1023/B:ENGI.0000014885.89822.f5}.
\bibitem[{Russo et~al.(2018)Russo, Jalalisendi, Falcucci and
  Porfiri}]{russo2018experimental}
\bibinfo{author}{Russo, S.}, \bibinfo{author}{Jalalisendi, M.},
  \bibinfo{author}{Falcucci, G.}, \bibinfo{author}{Porfiri, M.},
  \bibinfo{year}{2018}.
\newblock \bibinfo{title}{Experimental characterization of oblique and
  asymmetric water entry}.
\newblock \bibinfo{journal}{Experimental Thermal and Fluid Science}
  \bibinfo{volume}{92}, \bibinfo{pages}{141--161}.
\newblock \DOIprefix\doi{10.1016/j.expthermflusci.2017.10.028}.
\bibitem[{Scolan and Korobkin(2001)}]{scolan2001three}
\bibinfo{author}{Scolan, Y.M.}, \bibinfo{author}{Korobkin, A.A.},
  \bibinfo{year}{2001}.
\newblock \bibinfo{title}{Three-dimensional theory of water impact. part 1.
  inverse wagner problem}.
\newblock \bibinfo{journal}{Journal of Fluid Mechanics} \bibinfo{volume}{440},
  \bibinfo{pages}{293--326}.
\newblock \DOIprefix\doi{10.1017/S002211200100475X}.
\bibitem[{Sheng et~al.(2022)Sheng, Yu, Wang and Chen}]{sheng2022acfd}
\bibinfo{author}{Sheng, C.}, \bibinfo{author}{Yu, P.}, \bibinfo{author}{Wang,
  T.}, \bibinfo{author}{Chen, N.}, \bibinfo{year}{2022}.
\newblock \bibinfo{title}{A cfd based kriging model for predicting the impact
  force on the sphere bottom during the early-water entry}.
\newblock \bibinfo{journal}{Ocean Engineering} \bibinfo{volume}{243},
  \bibinfo{pages}{110304}.
\newblock \DOIprefix\doi{10.1016/j.oceaneng.2021.110304}.
\bibitem[{Siemann et~al.(2017)Siemann, Schwinn, Scherer and
  Kohlgruber}]{siemann2017advances}
\bibinfo{author}{Siemann, M.H.}, \bibinfo{author}{Schwinn, D.B.},
  \bibinfo{author}{Scherer, J.}, \bibinfo{author}{Kohlgruber, D.},
  \bibinfo{year}{2017}.
\newblock \bibinfo{title}{Advances in numerical ditching simulation of flexible
  aircraft models}.
\newblock \bibinfo{journal}{International Journal of Crashworthiness}
  \bibinfo{volume}{23}, \bibinfo{pages}{236--251}.
\newblock \DOIprefix\doi{10.1080/13588265.2017.1359462}.
\bibitem[{Siemann and Langrand(2017)}]{siemann2017coupled}
\bibinfo{author}{Siemann, M.N.}, \bibinfo{author}{Langrand, B.},
  \bibinfo{year}{2017}.
\newblock \bibinfo{title}{Coupled fluid-structure computational methods for
  aircraft ditching simulation: Comparison of ale-fe and sph-fe approaches}.
\newblock \bibinfo{journal}{Computers and Structures} \bibinfo{volume}{188},
  \bibinfo{pages}{95--108}.
\newblock \DOIprefix\doi{10.1016/j.compstruc.2017.04.004}.
\bibitem[{Terziev et~al.(2022)Terziev, Tezdogan and Incecik}]{terziev2022scale}
\bibinfo{author}{Terziev, M.}, \bibinfo{author}{Tezdogan, T.},
  \bibinfo{author}{Incecik, A.}, \bibinfo{year}{2022}.
\newblock \bibinfo{title}{Scale effects and full-scale ship hydrodynamics: A
  review}.
\newblock \bibinfo{journal}{Ocean Engineering} \bibinfo{volume}{245},
  \bibinfo{pages}{110496}.
\newblock \DOIprefix\doi{10.1016/j.oceaneng.2021.110496}.
\bibitem[{Vincent et~al.(2018)Vincent, Xiao, Yohann, Jung and
  Kanso}]{vincent2018dynamic}
\bibinfo{author}{Vincent, L.}, \bibinfo{author}{Xiao, T.},
  \bibinfo{author}{Yohann, D.}, \bibinfo{author}{Jung, S.},
  \bibinfo{author}{Kanso, E.}, \bibinfo{year}{2018}.
\newblock \bibinfo{title}{Dynamics of water entry}.
\newblock \bibinfo{journal}{Journal of Fluid Mechanics} \bibinfo{volume}{846},
  \bibinfo{pages}{508--535}.
\newblock \DOIprefix\doi{10.1017/jfm.2018.273}.
\bibitem[{Wagner(1932)}]{wagner1932phenomena}
\bibinfo{author}{Wagner, H.}, \bibinfo{year}{1932}.
\newblock \bibinfo{title}{Phenomena associated with impacts and sliding on
  liquid surface}.
\newblock \bibinfo{journal}{Journal of Applied Mathematics and Mechanics}
  \bibinfo{volume}{12}, \bibinfo{pages}{193--215}.
\newblock \bibinfo{note}{(in German)}.
\bibitem[{Wang et~al.(2015)Wang, Lugni and Faltinsen}]{wang2015experimental}
\bibinfo{author}{Wang, J.}, \bibinfo{author}{Lugni, C.},
  \bibinfo{author}{Faltinsen, O.M.}, \bibinfo{year}{2015}.
\newblock \bibinfo{title}{Experimental and numerical investigation of a
  freefall wedge vertically entering the water surface}.
\newblock \bibinfo{journal}{Applied Ocean Research} \bibinfo{volume}{51},
  \bibinfo{pages}{181--203}.
\newblock \DOIprefix\doi{10.1016/j.apor.2015.04.003}.
\bibitem[{Wang et~al.(2021a)Wang, Gadelho, Islam and Soares}]{wang2021cfd}
\bibinfo{author}{Wang, S.}, \bibinfo{author}{Gadelho, J.},
  \bibinfo{author}{Islam, H.}, \bibinfo{author}{Soares, C.G.},
  \bibinfo{year}{2021}a.
\newblock \bibinfo{title}{Cfd modelling and grid uncertainty analysis of the
  free-falling water entry of 2d rigid bodies}.
\newblock \bibinfo{journal}{Applied Ocean Research} \bibinfo{volume}{115},
  \bibinfo{pages}{102813}.
\newblock \DOIprefix\doi{10.1016/j.apor.2021.102813}.
\bibitem[{Wang et~al.(2021b)Wang, Xiang and Soares}]{wang2021assess}
\bibinfo{author}{Wang, S.}, \bibinfo{author}{Xiang, G.},
  \bibinfo{author}{Soares, C.G.}, \bibinfo{year}{2021}b.
\newblock \bibinfo{title}{Assessment of three-dimensional effects on slamming
  load predictions using openfoam}.
\newblock \bibinfo{journal}{Applied Ocean Research} \bibinfo{volume}{112},
  \bibinfo{pages}{102646}.
\newblock \DOIprefix\doi{10.1016/j.apor.2021.102646}.
\bibitem[{Wen et~al.(2020)Wen, Liu, Qu and Hu}]{wen2020impact}
\bibinfo{author}{Wen, X.}, \bibinfo{author}{Liu, P.}, \bibinfo{author}{Qu, Q.},
  \bibinfo{author}{Hu, T.}, \bibinfo{year}{2020}.
\newblock \bibinfo{title}{Impact of wedge bodies on wedge-shaped water surface
  with varying speed}.
\newblock \bibinfo{journal}{Journal of Fluids and Structures}
  \bibinfo{volume}{92}, \bibinfo{pages}{102831}.
\newblock \DOIprefix\doi{10.1016/j.jfluidstructs.2019.102831}.
\bibitem[{Woodgate et~al.(2019)Woodgate, Barakos, Scrase and
  Neville}]{woodgate2019simulation}
\bibinfo{author}{Woodgate, M.A.}, \bibinfo{author}{Barakos, G.N.},
  \bibinfo{author}{Scrase, N.}, \bibinfo{author}{Neville, T.},
  \bibinfo{year}{2019}.
\newblock \bibinfo{title}{Simulation of helicopter ditching using smoothed
  particle hydrodynamics}.
\newblock \bibinfo{journal}{Aerospace Science and Technology}
  \bibinfo{volume}{85}, \bibinfo{pages}{277–292}.
\newblock \DOIprefix\doi{10.1016/j.ast.2018.12.016}.
\bibitem[{Wu and Sun(2014)}]{wu2014similarity}
\bibinfo{author}{Wu, G.X.}, \bibinfo{author}{Sun, S.L.}, \bibinfo{year}{2014}.
\newblock \bibinfo{title}{Similarity solution for oblique water entry of an
  expanding paraboloid}.
\newblock \bibinfo{journal}{Journal of Fluid Mechanics} \bibinfo{volume}{745},
  \bibinfo{pages}{398--408}.
\newblock \DOIprefix\doi{10.1017/jfm.2014.111}.
\bibitem[{Xiao et~al.(2021a)Xiao, Lu, Deng, Zhi, Zhu and
  Chen}]{xiao2021hydrodynamic}
\bibinfo{author}{Xiao, T.}, \bibinfo{author}{Lu, Y.}, \bibinfo{author}{Deng,
  S.}, \bibinfo{author}{Zhi, H.}, \bibinfo{author}{Zhu, Z.},
  \bibinfo{author}{Chen, J.}, \bibinfo{year}{2021}a.
\newblock \bibinfo{title}{Hydrodynamic characteristics of a helicopter ditching
  on different positions of wavy water}.
\newblock \bibinfo{journal}{Journal of Aircraft} \bibinfo{volume}{58},
  \bibinfo{pages}{1--12}.
\newblock \DOIprefix\doi{10.2514/1.C036186}.
\bibitem[{Xiao et~al.(2021b)Xiao, Lu and Deng}]{xiao2021effect}
\bibinfo{author}{Xiao, T.}, \bibinfo{author}{Lu, Z.}, \bibinfo{author}{Deng,
  S.}, \bibinfo{year}{2021}b.
\newblock \bibinfo{title}{Effect of initial pitching angle on helicopter
  ditching characteristics using sph method}.
\newblock \bibinfo{journal}{Journal of Aircraft} \bibinfo{volume}{58},
  \bibinfo{pages}{167--181}.
\newblock \DOIprefix\doi{10.2514/1.C035898}.
\bibitem[{Xiao et~al.(2017)Xiao, Qin, Lu, Sun, Tong and
  Wang}]{xiao2017development}
\bibinfo{author}{Xiao, T.}, \bibinfo{author}{Qin, N.}, \bibinfo{author}{Lu,
  Z.}, \bibinfo{author}{Sun, X.}, \bibinfo{author}{Tong, M.},
  \bibinfo{author}{Wang, Z.}, \bibinfo{year}{2017}.
\newblock \bibinfo{title}{Development of a smoothed particle hydrodynamics
  method and its application to aircraft ditching simulations}.
\newblock \bibinfo{journal}{Aerospace Science and Technology}
  \bibinfo{volume}{166}, \bibinfo{pages}{28--43}.
\newblock \DOIprefix\doi{10.1016/j.ast.2017.02.022}.
\bibitem[{Yang et~al.(2020)Yang, Ma, Wen and Yang}]{yang2020crashworthy}
\bibinfo{author}{Yang, X.}, \bibinfo{author}{Ma, J.}, \bibinfo{author}{Wen,
  D.}, \bibinfo{author}{Yang, J.}, \bibinfo{year}{2020}.
\newblock \bibinfo{title}{Crashworthy design and energy absorption mechanisms
  for helicopter structures: A systematic literature review}.
\newblock \bibinfo{journal}{Progress in Aerospace Sciences}
  \bibinfo{volume}{114}, \bibinfo{pages}{100618}.
\newblock \DOIprefix\doi{10.1016/j.paerosci.2020.100618}.
\bibitem[{Yang and Xu(2018)}]{yang2018numerical}
\bibinfo{author}{Yang, X.B.}, \bibinfo{author}{Xu, G.D.}, \bibinfo{year}{2018}.
\newblock \bibinfo{title}{Numerical simulation of the oblique water entry of
  wedges with vortex shedding}.
\newblock \bibinfo{journal}{Brodogradnja} \bibinfo{volume}{69},
  \bibinfo{pages}{69--83}.
\newblock \DOIprefix\doi{10.21278/brod69406}.
\bibitem[{Yu et~al.(2019)Yu, Shen, Zhen, Tang and Wang}]{yu2019parametric}
\bibinfo{author}{Yu, P.}, \bibinfo{author}{Shen, C.}, \bibinfo{author}{Zhen,
  C.}, \bibinfo{author}{Tang, H.}, \bibinfo{author}{Wang, T.},
  \bibinfo{year}{2019}.
\newblock \bibinfo{title}{Parametric study on the free-fall water entry of a
  sphere by using the rans method}.
\newblock \bibinfo{journal}{Journal of Marine Science and Engineering}
  \bibinfo{volume}{7}, \bibinfo{pages}{122}.
\newblock \DOIprefix\doi{10.3390/jmse7050122}.
\bibitem[{Zekri et~al.(2021)Zekri, Korobkin and Cooker}]{zekri2021gravity}
\bibinfo{author}{Zekri, H.J.}, \bibinfo{author}{Korobkin, A.A.},
  \bibinfo{author}{Cooker, M.J.}, \bibinfo{year}{2021}.
\newblock \bibinfo{title}{Gravity effect on water entry during an early stage}.
\newblock \bibinfo{journal}{Journal of Fluid Mechanics} \bibinfo{volume}{916},
  \bibinfo{pages}{A10}.
\newblock \DOIprefix\doi{10.1017/jfm.2021.190}.
\bibitem[{Zhao and Faltinsen(1993)}]{zhao1993water}
\bibinfo{author}{Zhao, R.}, \bibinfo{author}{Faltinsen, O.M.},
  \bibinfo{year}{1993}.
\newblock \bibinfo{title}{Water entry of two-dimensional bodies}.
\newblock \bibinfo{journal}{Journal of Fluid Mechanics} \bibinfo{volume}{246},
  \bibinfo{pages}{593--612}.
\newblock \DOIprefix\doi{10.1017/S002211209300028X}.
\bibitem[{Zheng et~al.(2021)Zheng, Qu, Liu, Wen and Zhang}]{zheng2021numerical}
\bibinfo{author}{Zheng, Y.}, \bibinfo{author}{Qu, Q.}, \bibinfo{author}{Liu,
  P.}, \bibinfo{author}{Wen, X.}, \bibinfo{author}{Zhang, Z.},
  \bibinfo{year}{2021}.
\newblock \bibinfo{title}{Numerical analysis of the porpoising motion of a
  blended wing body aircraft during ditching}.
\newblock \bibinfo{journal}{Aerospace Science and Technology}
  \bibinfo{volume}{119}, \bibinfo{pages}{10731}.
\newblock \DOIprefix\doi{10.1016/j.ast.2021.107131}.

\end{thebibliography}



\end{document}